\documentclass[fleqn,usenatbib]{mnras}

\usepackage{newtxtext,newtxmath}
\usepackage{chemformula}
\usepackage{subcaption}

\usepackage[T1]{fontenc}

\usepackage{graphicx}	
\usepackage{amsmath}	
\usepackage{siunitx}
\usepackage[version=4]{mhchem}
\usepackage{xspace}
\usepackage{anyfontsize}

\usepackage{orcidlink}
\usepackage{adjustbox}
\usepackage{booktabs}
\usepackage{pdflscape}

\newcommand{\Rjup}{R$_{\mathrm{J}}$}

\newcommand{\degree}{$^{\circ}$}

\newcommand{\myplanet}{HAT-P-30\,b\xspace}
\newcommand{\eureka}{\texttt{Eureka!}\xspace}
\newcommand{\tiberius}{\texttt{Tiberius}\xspace}

\newcommand{\prt}{\texttt{petitRADTRANS}\xspace}
\newcommand{\bear}{\texttt{BeAR}\xspace}

\newcommand{\new}[1]{#1}

\newcommand{\newer}[1]{#1}

\defcitealias{penzlinBOWIEALIGNHowFormation2024}{Penzlin \& Booth et al. (2024)}

\defcitealias{thejwsttransitingexoplanetcommunityearlyreleasescienceteamIdentificationCarbonDioxide2023}{JWST Transiting Exoplanet Community Early Release Science Team 2023}

\defcitealias{molliere2022interpreting}{Molli\'ere et al. (2022)}

\title[BOWIE-ALIGN: the misaligned \myplanet]{BOWIE-ALIGN: Sub-solar C/O ratio and metallicity atmosphere of the misaligned hot Jupiter \myplanet}

\author[A. B. Claringbold et al.]{Alastair B. Claringbold$^{\orcidlink{0000-0003-1309-5558},1,2}$\thanks{E-mail: alastair.claringbold@warwick.ac.uk},
Chloe E. Fisher$^{{\orcidlink{0000-0003-0652-2902}},3}$,
James Kirk $^{\orcidlink{0000-0002-4207-6615},4}$,
Eva-Maria Ahrer $^{\orcidlink{0000-0003-0973-8426},5}$,
\newauthor
Anna B. T. Penzlin$^{\orcidlink{0000-0002-8873-6826},6,4}$,
Daniel P. Thorngren$^{{\orcidlink{0000-0002-5113-8558}},7}$,
Mercedes L\'opez-Morales$^{\orcidlink{0000-0003-3204-8183},8}$,
Peter J. Wheatley$^{\orcidlink{0000-0003-1452-2240},1,2}$,
\newauthor
Lili Alderson$^{\orcidlink{0000-0001-8703-7751},9}$,
Richard~A. Booth $^{\orcidlink{0000-0002-0364-937X},10}$,
Duncan A. Christie$^{\orcidlink{0000-0002-4997-0847},5}$,
Charlotte Fairman$^{\orcidlink{0000-0001-9665-5260},11}$,
\newauthor
Nathan J. Mayne$^{{\orcidlink{0000-0001-6707-4563}},12}$,
Mason McCormack$^{\orcidlink{0000-0002-1463-9847},13}$,
Annabella Meech$^{\orcidlink{0000-0002-7500-7173},14}$,
James E. Owen$^{\orcidlink{0000-0002-4856-7837},4}$
Vatsal Panwar$^{\orcidlink{0000-0002-2513-4465},1,2}$,
\newauthor
Denis E. Sergeev$^{{\orcidlink{0000-0001-8832-5288}},11}$,
Daniel Valentine$^{{\orcidlink{0000-0002-2643-6836}},11}$
Hannah R. Wakeford$^{\orcidlink{0000-0003-4328-3867}~11}$,
Maria Zamyatina$^{{\orcidlink{0000-0002-9705-0535}},12}$
\\
$^{1}$Department of Physics, University of Warwick, Gibbet Hill Road, Coventry CV4 7AL, UK\\
$^{2}$Centre for Exoplanets and Habitability, University of Warwick, Gibbet Hill Road, Coventry CV4 7AL, UK\\
$^{3}$Department of Physics, University of Oxford, Denys Wilkinson Building, Keble Road, Oxford, OX1 3RH, United Kingdom\\
$^{4}$Department of Physics, Imperial College London, Prince Consort Road, SW7 2AZ, London, UK\\
$^{5}$Max Planck Institute for Astronomy (MPIA), K\"{o}nigstuhl 17, 69117 Heidelberg, Germany \\
$^{6}$Ludwig-Maximilians-Universit{\"a}t M{\"u}nchen, Universit{\"a}ts-Sternwarte, Scheinerstr.~1, 81679 M{\"u}nchen, Germany\\
$^{7}$Department of Physics and Astronomy, Johns Hopkins University, Baltimore, MD 21218, USA\\
$^{8}$Space Telescope Science Institute, 3700 San Martin Drive, Baltimore MD 21218, USA\\
$^{9}$Department of Astronomy, Cornell University, 122 Sciences Drive, Ithaca, NY 14853, USA \\
$^{10}$School of Physics and Astronomy, University of Leeds, Leeds, LS2 9JT\\
$^{11}$School of Physics, HH Wills Physics Laboratory, Universty of Bristol, Tyndall Avenue, Bristol BS8 1TL, UK\\
$^{12}$Department of Physics and Astronomy, Faculty of Environment, Science and Economy, University of Exeter, Exeter EX4 4QL, UK\\
$^{13}$Department of Astronomy \& Astrophysics, University of Chicago, Chicago, IL 60637, USA\\
$^{14}$Center for Astrophysics | Harvard \& Smithsonian, 60 Garden St, Cambridge, MA 02138, USA\\
}

\date{Accepted XXX. Received YYY; in original form ZZZ}

\pubyear{\the\year{}}

\begin{document}
\label{firstpage}
\pagerange{\pageref{firstpage}--\pageref{lastpage}}
\maketitle

\begin{abstract}

\noindent We present the JWST NIRSpec/G395H transmission spectrum of the misaligned hot Jupiter HAT-P-30b from 2.8--5.2 $\mu$m as part of the BOWIE-ALIGN survey, a comparative survey designed to probe the link between planet formation and atmospheric composition in samples of misaligned and aligned hot Jupiters orbiting F-type stars.
Through independent data reductions and retrieval analyses, we find evidence for absorption features of \ch{H2O} and \ch{CO2} in the atmosphere of \myplanet. Our retrieved abundances are consistent with equilibrium chemistry, from which we infer a sub-solar C/O ratio (0.16--0.45), and sub-solar and sub-stellar metallicity (0.2--0.8$\times$solar, compared to a stellar metallicity of 1.1--1.6$\times$solar), with muted spectral features.
This 
composition challenges formation models of continuous migration and accretion within a steady disc of stellar metallicity, and
 could be the result of low C/O ratio gas accretion within the water ice line, low metallicity accretion due to the trapping of volatiles further out in the disc, or the combined accretion of low metallicity gas and carbon-poor solids. 

\end{abstract}




\section{Introduction}

The idea that the composition of an exoplanet atmosphere may conceal clues as to its formation location and migration history has been a key driver for the characterization of exoplanets. With the unparalleled precision and near-infrared wavelength coverage afforded by JWST, it is now possible to accurately measure relative elemental abundances in exoplanet atmospheres \citep[e.g.,][]{JWST2023,Ahrer2023,Alderson2023,Feinstein2023,Rustamkulov2023}. In principle, we can use the C/O ratio to infer where a planet formed relative to the ice lines of \ch{H2O}, \ch{CO2}, and CO in the protoplanetary disc \citep{Oberg2011}.

In practice, relating the atmospheric composition for a single planet to its formation history challenges this simple picture, with numerous physical processes obscuring the relationship between formation models and observed atmospheres. This is because, fundamentally, modelling planet formation includes a high dimensionality of poorly constrained parameters, including temperature profile, dust-to-gas ratio, and composition of the protoplanetary disc \citep[e.g.,][]{molliere2022interpreting}.
This can be additionally complicated by including physical processes important during formation, such as relative accretion of gas and solid \citep{Espinoza2017}, the evolution of ice lines \citep{Morbidelli2016,Owen2020}, the drift of volatile-carrying solids \citep{Booth2017,Schneider2021}, and the trapping of volatiles within ice \citep{ligterink2024mind}, as well as the complexity arising due to migration, as planets can accrete in multiple environments through their evolution. Observations of protoplanetary discs have also revealed them to be diverse objects, with their radial composition varying between host stars  \citep{law2021molecules,Law2021}. Furthermore, directly relating a planet's atmospheric composition to its bulk elemental ratios is non-trivial, with potential impacts from cloud formation \citep{Helling2016} or enrichment of an atmosphere uncoupled from the interior \citep{Muller2024}.

As demonstrated by \citetalias{penzlinBOWIEALIGNHowFormation2024}, while any single planet is a poor tracer of planet formation, we can use populations of exoplanets with similar dynamical properties implying shared formation histories 
to test whether atmospheric composition traces planet formation. In particular, \citetalias{penzlinBOWIEALIGNHowFormation2024} show how planets which underwent disc-migration should be chemically distinct from planets that underwent disc-free migration, with the precise nature of this distinction depending upon the behaviour of accreted silicates and the dominant carbon-bearing species within the disc. We can leverage the two distinct populations of aligned hot Jupiters, that likely migrated through the disc, and misaligned hot Jupiters, that likely underwent high-eccentricity migration after disc dispersal \citep{Rasio1996,Wu2003,2016Munoz}. Comparing these two dynamically distinct populations we can test 
the predictions of \citetalias{penzlinBOWIEALIGNHowFormation2024}.

This is precisely the goal of the BOWIE-ALIGN survey (A spectral Light Investigation into gas Giant origiNs; JWST GO 3838; PIs: Kirk \& Ahrer), by spectroscopically characterizing eight planets: four aligned hot Jupiters believed to have migrated through the disc, and four misaligned hot Jupiters believed to have migrated via high-eccentricity migration \citep{kirkBOWIEALIGNJWSTComparative2024}. All eight planets orbit stars above the Kraft Break ($T_{\mathrm{eff}}\gtrsim6100$ K), an observed shift in rotation rates attributed to much thinner surface convective zones than cooler stars causing less efficient magnetic breaking \citep{Kraft1967,beyerKraftBreakSharply2024}. Realignment of hot Jupiters due to tidal interactions with the star is unlikely to occur above the Kraft break, fitting the observation that hot Jupiters around cooler stars tend to have low stellar obliquities \citep{winnHotStarsHot2010,albrechtOBLIQUITIESHOTJUPITER2012}. This ensures the aligned planets in the BOWIE-ALIGN sample are not aligned due to re-alignment of the star, and are instead likely aligned due to migration through the disc.

The transmission spectra of three planets from the BOWIE-ALIGN sample have been published to date. The misaligned planet WASP-15\,b was found to host a super-stellar metallicity atmosphere with a solar C/O ratio and evidence of \ch{SO2} absorption \citep{kirk2025bowie}. This combination of atmospheric properties was tentatively attributed to late planetesimal accretion. Meanwhile the aligned planet TrES-4\,b was found to instead host a sub-stellar metallicity atmosphere with a sub-solar C/O ratio, challenging traditional planet formation models by suggesting either low C/O ratio gas accretion, or a combination of gas and carbon-poor solid accretion \citep{meech2025bowie}. Another aligned planet, KELT-7\,b, was found to have very weak spectral features due to either a high cloud deck or low atmospheric metallicity, limiting the ability to place constraints on the C/O ratio and metallicity (Ahrer et al. \textit{submitted}).

In this work we present the JWST NIRSpec/G395H transmission spectrum of \myplanet, a misaligned hot Jupiter in the BOWIE-ALIGN programme, with a measured obliquity of $70.5^{+2.9}_{-2.8} $\,\degree \citep{cegla2023exploring}. \myplanet, also known as WASP-51\,b, has a mass of $0.746 \pm 0.021$\,M$_\mathrm{Jup}$ \citep{Bonomo2017}, a radius of $1.42 \pm 0.03$\,R$_\mathrm{Jup}$, equilibrium temperature of $1630 \pm 42$\,K \citep{Blazek2022}, and an orbital period of $2.8106013\pm0.0000006$\,days \citep{Ivshina2022}. We include a full summary of our adopted system parameters in Table \ref{tab:planetary-parameters}.

\begin{table}
    \centering
    \caption{System parameters for \myplanet. References are for: [1] \citet{Johnson2011},  [2] \citet{Bonomo2017}, [3] \citet{Blazek2022}, [4] \citet[][]{Ivshina2022}, and [5] \citet{cegla2023exploring}.}
    \begin{tabular}{l c c }
    \hline
        Parameter & Value &   Reference\\ \hline
        \textit{Stellar parameters}\\
        Mass, M$_\textrm{*}$ (M$_\odot$) & $1.242 \pm 0.041$ & [2]\\
        Radius, R$_\textrm{*}$ (R$_\odot$) & $1.215 \pm0.051$ & [2] \\ 
        Effective Temperature, $T_{\textrm{eff}}$ (K) & $6304 \pm 88$ & [1]\\
        Age (Gyr) & $1.0^{+0.8}_{-0.5} $  & [2] \\
        Surface gravity, log $g$ (cgs) & $4.36 \pm 0.04$  & [2]\\
        Metallicity [Fe/H] (dex) & $0.13\pm0.08$ &  [2] \\
        \hline
        \textit{Planetary parameters}\\
        Mass, M$_\mathrm{p}$ (M$_\mathrm{Jup}$) & $0.746^{+0.020}_{-0.021} $ & [2] \\   
        Radius, R$_\mathrm{p}$ (R$_\mathrm{Jup}$) & $1.417 \pm 0.033$ & [3] \\  
        Equ. Temperature, T$_\mathrm{eq}$ (K) & $1630 \pm 42$ & [3] \\   
        Gravity, ($\mathrm{ms^{-2}}$) & $9.2 \pm 0.5$ &  \\ 
        Orbital period, P (days) & $2.8106013\pm0.0000006$ & [4] \\
        Obliquity, $\lambda$ (degrees) & $70.5^{+2.9}_{-2.8} $ & [5] \\
        \hline
    \end{tabular}
    
    \label{tab:planetary-parameters}
\end{table}

We describe the JWST observations in Section \ref{sec:observations}, and the data reduction and light curve fitting in Section \ref{sec:data_reduction}. To derive the atmospheric composition of \myplanet, we outline an atmospheric retrieval analysis to interpret the spectrum in Section \ref{sec:retrievals}. In Section \ref{sec:interior}, we use interior structure models to place constraints on the atmospheric composition and infer the bulk metallicity of \myplanet based on the system parameters. We include a discussion of the atmosphere of \myplanet and its relevance to its formation history in Section \ref{sec:discussion}, and present our conclusions in Section \ref{sec:conclusions}.


\begin{figure*}
    \centering
    \includegraphics[width=0.95\linewidth]{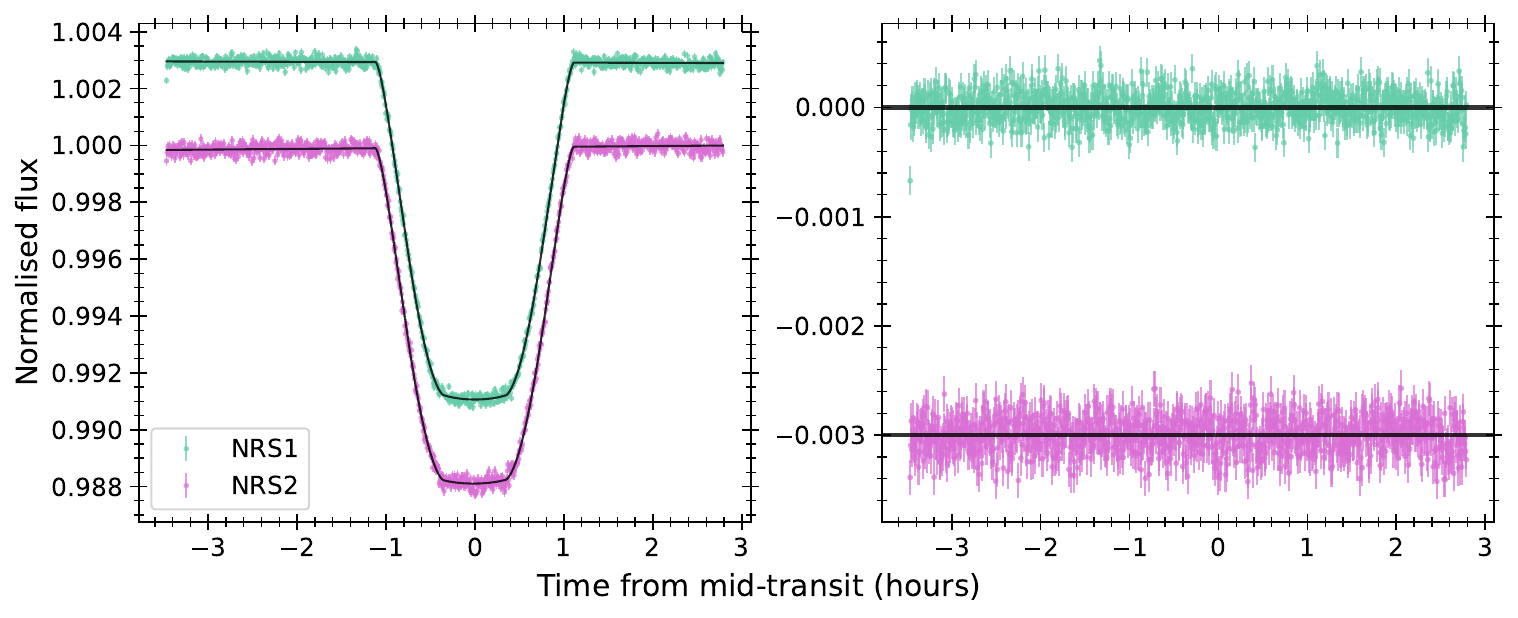}
    \caption{White light curves for HAT-P-30b for each detector, NRS1 (green) and NRS2 (magenta) from the Eureka data reduction (left) and the residuals from the light curve fits (right).}
    \label{fig:WL_lightcurves}
\end{figure*}

\begin{table*}
    \caption{The resulting system parameters from our fits to the individual JWST NIRSpec/G395H NRS1 and NRS2 white light curves, the system parameters from the combined fit to both the NRS1 and NRS2 data (Eureka only), and the weighted mean of both individual NRS1 and NRS2 fits (Tiberius only). 
    }
    \label{tab:system_params}
    \centering
    \begin{tabular}{c|c|c|c|c|c} \hline
         Pipeline & Instrument & $T_0$ (BJD) & $R_\mathrm{p}/R_*$ & $a/R_*$ & $i$ (\degree) \\ \hline \hline
         \texttt{Eureka!} & NRS1 & $ 2460380.639378 \pm 0.000030 $ & $ 0.110156 \pm 0.000090 $ & $ 6.819 \pm 0.021 $ & $ 82.746 \pm 0.031 $ \\
         \texttt{Eureka!} & NRS2 & $ 2460380.639344 \pm 0.000034 $ & $ 0.10991 \pm 0.00011 $ & $ 6.799 \pm 0.018 $ & $ 82.717 \pm 0.026  $\\ 
         \texttt{Eureka!} & Combined fit & $ 2460380.639363 \pm 0.000018   $ & $ 0.109920 \pm 0.000067 $ & $ 6.791 \pm 0.009  $ & $ 82.699 \pm 0.013 $  \\ \hline
         \texttt{Tiberius} & NRS1 & $ 2460380.639369 \pm 0.000023 $ & $ 0.110265 \pm 0.000070 $ & $ 6.798 \pm 0.012 $ & $ 82.716 \pm 0.017 $  \\
         \texttt{Tiberius} & NRS2 & $ 2460380.639335 \pm 0.000029 $ & $ 0.109843 \pm 0.000088 $ & $ 6.782 \pm 0.017 $ & $ 82.690 \pm 0.021 $  \\
         \texttt{Tiberius} & Weighted mean & $ 2460380.639356 \pm 0.000018 $ & $ 0.110101 \pm 0.000055 $ & $ 6.792 \pm 0.009 $ & $ 82.706 \pm 0.013 $ \\ \hline
    \end{tabular}
\end{table*}

\section{Observations}
\label{sec:observations}

We observed one transit of HAT-P-30b with JWST on 10 March 2024, 22:20:47 UT as part of the GO 3838 programme, using the NIRSpec instrument \citep{Jakobsen2022} in Bright Object Time Series (BOTS) mode, with the G395H grating, F290LP filter, and SUB2048 subarray. This setup provides spectroscopy from 2.8 -- 5.2 $\mu$m at a resolution of $R\sim2700$, with a gap in coverage from 3.72 -- 3.82 $\mu$m due to a physical gap between the NRS1 and NRS2 detectors. We use 26 groups per integration, with 925 integrations over 6.26 hours, including 2.25 hours during transit. We use a nearby faint star (2MASSJ08154583+0550218) for target acquistion in WATA mode with the SUB32 subarray and CLEAR filter.



\section{Data reduction}
\label{sec:data_reduction}

We perform two independent data reductions, as prescribed by the BOWIE-ALIGN data analysis strategy found in \citet{kirkBOWIEALIGNJWSTComparative2024}. We use two pipelines, \eureka \citep{Bell2022Eureka} and \tiberius \citep{Kirk2017,Kirk2021}, for the data reduction, broadly following the methodology for other BOWIE-ALIGN targets WASP-15\,b \citep{kirk2025bowie} TrES-4\,b \citep{meech2025bowie}, and KELT-7\,b \citep{ahrer2025bowie}, with exact details of the spectral extraction and light curve fitting detailed below.

\subsection{\eureka}
We conduct a data reduction of \myplanet using the open-source \texttt{python} package \eureka \citep{Bell2022Eureka}, which has been used extensively for analysing JWST exoplanet transmission spectra \citep[e.g.,][]{Ahrer2023,  MoranStevenson2023,alderson2024jwst,wallack2024jwst,xue2024jwst,teske2025jwst}.

\subsubsection{Light curve extraction}
We follow previous BOWIE-ALIGN analyses when conducting our \eureka analysis. We start with the uncalibrated files and run Stage\,1 and 2 of \eureka which is wrapped around the \texttt{jwst} pipeline (version 1.12.2, CRDS context pmap: 1253), with the following common modifications: we apply a correction factor to the \texttt{jwst} superbias (group 1, smooth, window length of 30 pixels), we increase the \texttt{jump\_rejection\_threshold} to 10.0$\sigma$ \citep[e.g., see][]{Alderson2023}, we run a group-level column-by-column background subtraction (with an outlier rejection threshold of $3 \times$ median), and we skip the \texttt{photom\_step}. Note that in order to compute the bias scale factor and run the 1/f background subtraction we mask the trace using 10 pixels around the central trace. 

In Stage\,3, we extract the stellar spectrum for each integration. We apply a constant column-by-column background subtraction (excluding the 8/10 pixel area on both sides from the central trace pixel for NRS1/NRS2, respectively) using a 5-sigma threshold for outlier rejection along both the time and spatial axes. We extract the stellar spectrum using optimal spectral extraction \citep{horne1986optimal} with an aperture half-width of 4 pixels. We further manually mask columns where the count is $>15\sigma$ from a rolling mean of $20$\,pixels across the frame.

The extracted stellar spectra are binned in Stage\,4 following the common BOWIE-ALIGN grid, at R=100 and R=400. We mask outliers $>5\sigma$ on the rolling median (25 pixels) of the binned light curves.

\subsubsection{Light curve fitting}
We follow the commonly used procedure that the broadband white light curves for NRS1 and NRS2 are fitted first, using the transit depth $R_\mathrm{p}/R_*$, the scaled stellar radius $a/R_*$, the mid-transit time $T_0$, and inclination $i$ as free parameters. We further fit for a baseline offset and a linear slope. The orbital period and eccentricity of \myplanet are fixed, to 2.8106 days \citep{Ivshina2022} and 0, respectively, and we fix the limb-darkening using the quadratic limb-darkening law and parameters $u_1, u_2$. The limb-darkening values are computed for each bin in \eureka's Stage\,4 using \texttt{ExoTiC-LD} \citep{grant2024exotic} and the 3D Stagger grid \citep{Magic2015}, based on the stellar parameters from Table\,\ref{tab:planetary-parameters}.

The best-fit parameters are derived using the \texttt{Python} Markov Chain Monte Carlo (MCMC) sampler \texttt{emcee} \citep{foreman2013emcee}, using 50 walkers and 3000 steps (of which 1000 were discarded as burn-in steps). The retrieved parameters for \myplanet's orbit based on the white light curve from NRS1 and NRS2 are in Table\,\ref{tab:system_params}, with the light curve fits presented in Figure\, \ref{fig:WL_lightcurves}. For our spectroscopic NRS1 and NRS2 fits, we then fixed the orbital parameters ($a/R_*$, $i$, $T_0$) to the best-fit values by the NRS1 or NRS2 white light curves, respectively. 
We fix the limb-darkening parameters to those calculated for each spectroscopic bin.
Therefore, each spectroscopic light curve fit included three free parameters: the transit depth $R_\mathrm{p}/R_*$ and the parameters describing the baseline offset and linear slope term of the light curve. The resulting transmission spectra at $R=100$ and $R=400$ are displayed in Fig.\,\ref{fig:transmission_spectra}.

We also perform a simultaneous fit to the white light curves from both detectors from which we derive the combined fit system parameters presented in Table \ref{tab:system_params}. Using these system parameters to generate a transmission spectrum in the same manner causes negligible differences in the spectrum when compared to that generated with the individual system parameters for each detector, except for a minor vertical offset (with an average of 32\,ppm for NRS1, 26\,ppm for NRS2) with no slope or change of shape.


\subsection{\tiberius}


We also conduct a data reduction of \myplanet using {\tt Tiberius} \citep{Kirk2017,Kirk2021}, an open-source package for data reduction and light-curve fitting originally developed for ground-based exoplanet atmosphere transmission spectroscopy, now used extensively for JWST observations \citep[e.g.,][]{Rustamkulov2023,MoranStevenson2023,Kirk2024}. 

\subsubsection{Light curve extraction}

For our \tiberius reduction, we followed an identical process as used in our BOWIE-ALIGN analyses of WASP-15b \citep{kirk2025bowie} and TrES-4b \citep{meech2025bowie}, and use the same pipeline version (\texttt{Tiberius}: v1.0.4, \texttt{jwst}: v1.8.2) and calibration reference files. We refer the reader to \cite{kirk2025bowie} for a detailed explanation of this process. In short, we perform stage 1 using a modified version of the \texttt{jwst} pipeline, skipping the \texttt{jump\_step} and adding our own 1/f correction. Stage 2 is performed by \tiberius after our own custom bad pixel flagging and correction procedure \citep[detailed in][]{kirk2025bowie}. We perform standard aperture photometry with an aperture full width of 8 pixels and perform an additional background subtraction at this stage, using a linear polynomial fitted to each pixel column after masking the 22 pixels centered on the stellar trace. 

To construct our light curves, we use the same two binning schemes as in our other BOWIE-ALIGN papers, namely a lower resolution $R=100$ wavelength grid and a higher resolution $R=400$ wavelength grid. Our NRS1 light curves span a wavelength range of 2.75--3.72\,\micron\ with NRS2 spanning 3.82--5.18\,\micron.

\subsubsection{Light curve fitting}

For our light curve fits, we again follow an identical procedure to that detailed in \cite{kirk2025bowie}. We fit our light curves with a quadratically limb-darkened analytic transit light curve (implemented through \texttt{batman}, \citealt{batman}) multiplied by a linear-in-time polynomial. For our white light curve fits, the free parameters are the planet-to-star radius ratio ($R_\mathrm{p}/R_*$), the planet's inclination ($i$), the semi-major axis to stellar radius ratio ($a/R_*$), the time of mid-transit ($T_0$), and the two parameters of the linear polynomial ($c_1$, $c_2$). We held the period fixed to 2.81060126\,d \citep{Ivshina2022} and the eccentricity to 0\footnote{We also tested fixing eccentricity to 0.035 according to the results of \cite{Johnson2011} but found that this led to a negligible difference in the resulting transmission spectra with a median difference of 1.5\,ppm.}. We fixed the limb darkening coefficients to the values found using the Stagger grid of 3D stellar atmosphere models \citep{Magic2015} as computed by \texttt{ExoTiC-LD} \citep{grant2024exotic} using the stellar parameters given in Table \ref{tab:planetary-parameters}. 

\begin{figure}
    \centering
    \includegraphics[width=0.99\linewidth]{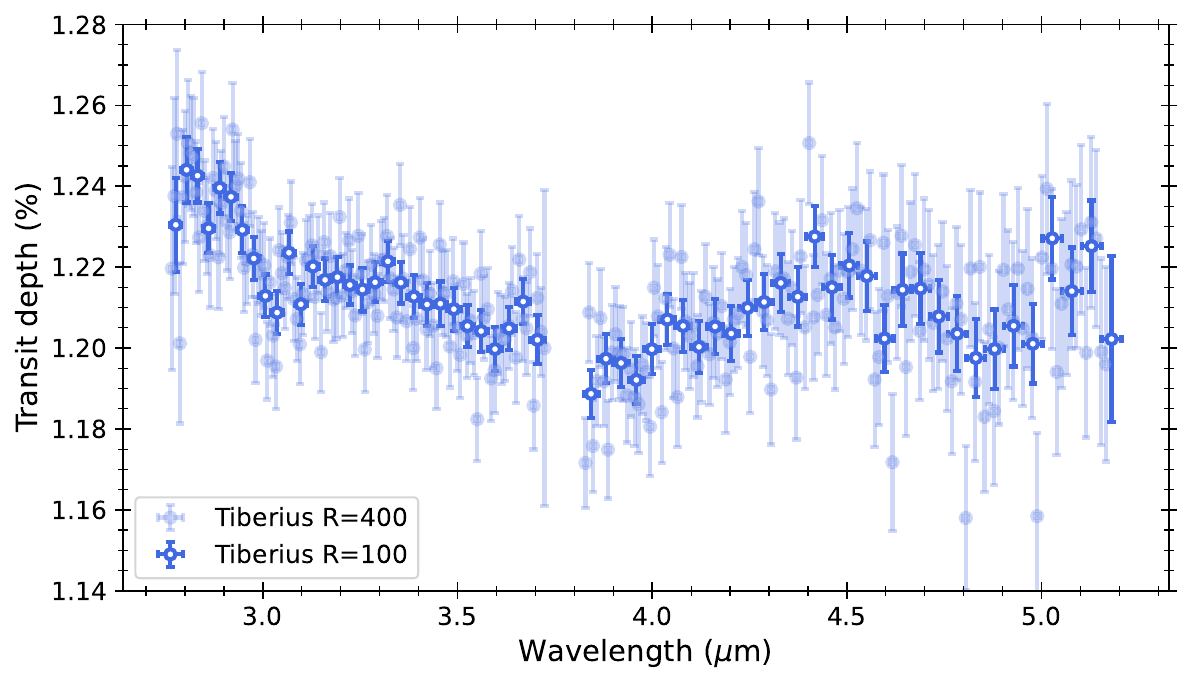}
    \includegraphics[width=0.99\linewidth]{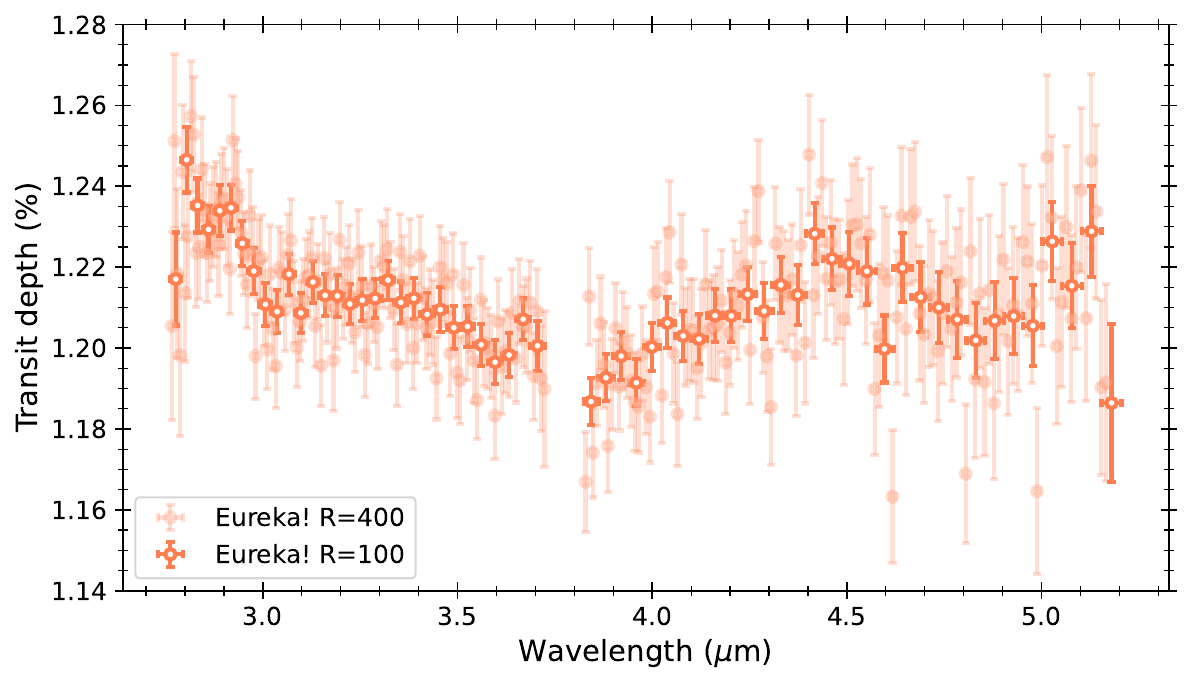}

    \caption{Transmission spectra of \myplanet from \tiberius (top) and \eureka (bottom) data reductions, binned to spectral resolutions of $R=$400 and $R=$100.}
    \label{fig:transmission_spectra}
\end{figure}

Our best-fitting system parameters are derived from the white light curves and are given in Table \ref{tab:system_params}. These are the result of optimization using a Levenberg-Marquardt algorithm. Following the fitting of the white light curves, we fit our spectroscopic light curves with the same model setup but with the system parameters fixed to the weighted mean values as given in Table \ref{tab:system_params}. The result of this was the transmission spectra shown in Fig. \ref{fig:transmission_spectra}.

\begin{figure}
    \centering
    \includegraphics[width=0.99\linewidth]{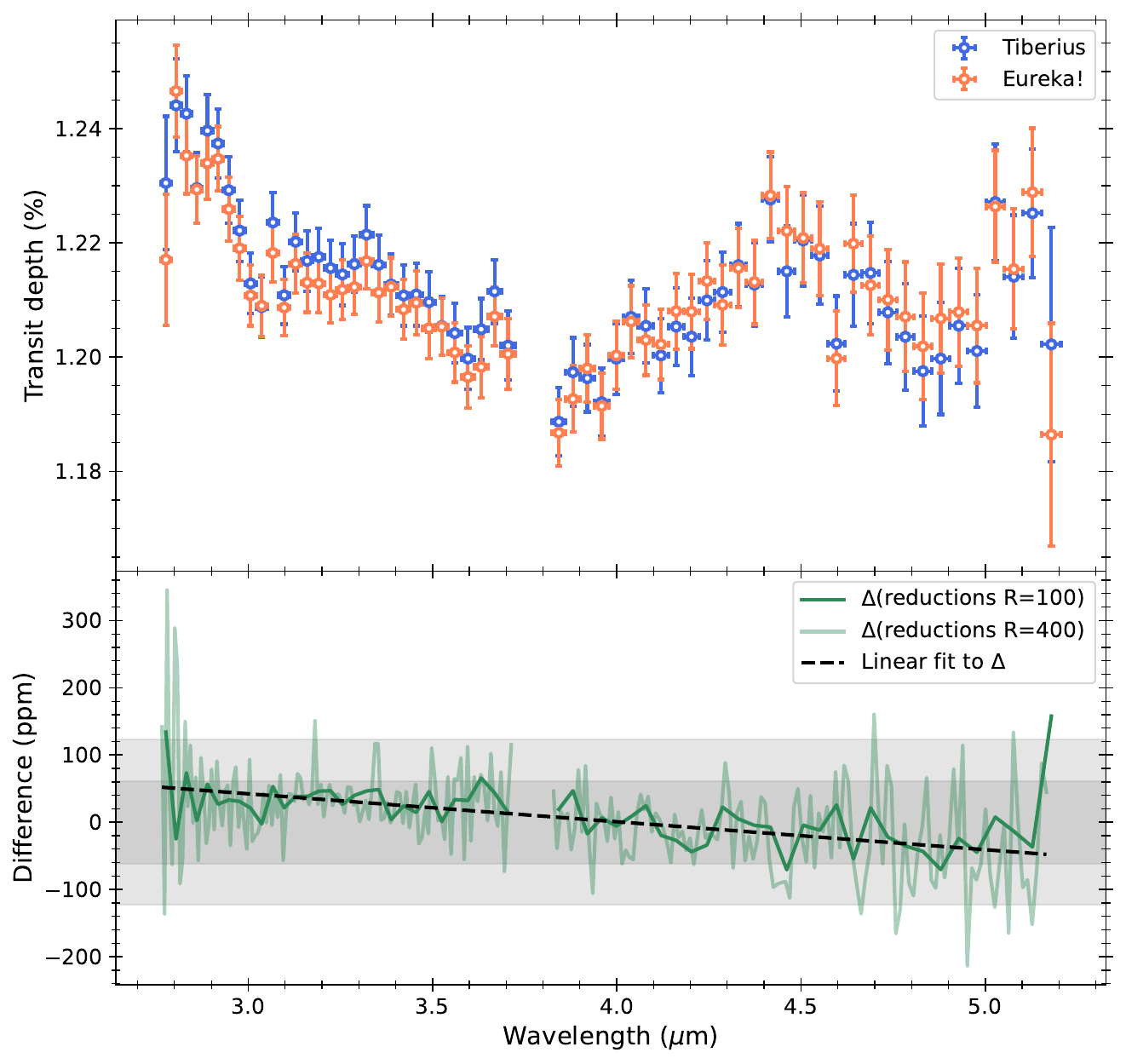}
    \caption{Comparison between transmission spectra obtained by our \tiberius and \eureka reductions at $R=$100, including the difference between the $R=100$ and $R=400$ spectra in green in the bottom panel. The shaded grey regions are the 1$\times$ and 2$\times$ the median transit depth uncertainty of the \eureka $R=100$ reduction. We find that the difference between reductions depends linearly on wavelength, with a mean offset of $+$35 ppm and $-$21 ppm in NRS1 and NRS2 respectively.}
    \label{fig:spectra_comparison}
\end{figure}

\subsection{The transmission spectrum of HAT-P-30b}

We present the transmission spectra obtained from each of our reduction pipelines at spectral resolutions of $R=100$ and $R=400$ in Fig. \ref{fig:transmission_spectra}. While these reductions agree well in overall feature shape and depths (with a 7 ppm mean offset), we do find that the difference between the reductions depends linearly on wavelength, as depicted in Fig. \ref{fig:spectra_comparison}. We perform a linear fit on the differences, demonstrating a difference in slope of 100 ppm across the whole spectrum. The light curve fitting, while performed independently for the two reductions, arrived at the same systematic models and used the same limb-darkening coefficients, only differing in the use of MCMC versus Levenberg-Marquardt, which has been demonstrated to not impact the results. While our derived \eureka transmission spectrum uses a different approach to system parameters (using the system parameters from each detector individually) than the \tiberius spectrum (using the weighted mean), we verify this is not the cause of the difference by testing a \eureka transmission spectrum using system parameters from a combined white light curve fit. This results in a near-identical transmission spectrum with no slope and minimal detector offset ($\sim5$\,ppm). Due to the high impact parameter of \myplanet, the limb-darkening is fixed in both of our reductions.

A detailed investigation into differences in reduction method was performed for the previous BOWIE-ALIGN target WASP-15b \citep{kirk2025bowie}, finding that offset and slopes between the \tiberius and \eureka reductions occur at the spectral extraction stage, as opposed to stage 1 extraction, system parameters, or limb-darkening coefficients. Understanding the difference between reductions is an ongoing effort for the field \citep[e.g.][]{carter2024benchmark}. 

\section{Atmospheric \new{Modelling}}
\label{sec:retrievals}

\new{To interpret the \eureka and \tiberius transmission spectra, we use a grid of simple 1D equilibrium forward models for an initial comparison to the data. We follow this with an atmospheric retrieval analysis to provide a detailed interpretation,} independently implemented using the publicly available packages \prt and \bear. We detail the modelling setup and results of each package below.

\begin{figure}
    \centering
    \includegraphics[width=0.99\linewidth]{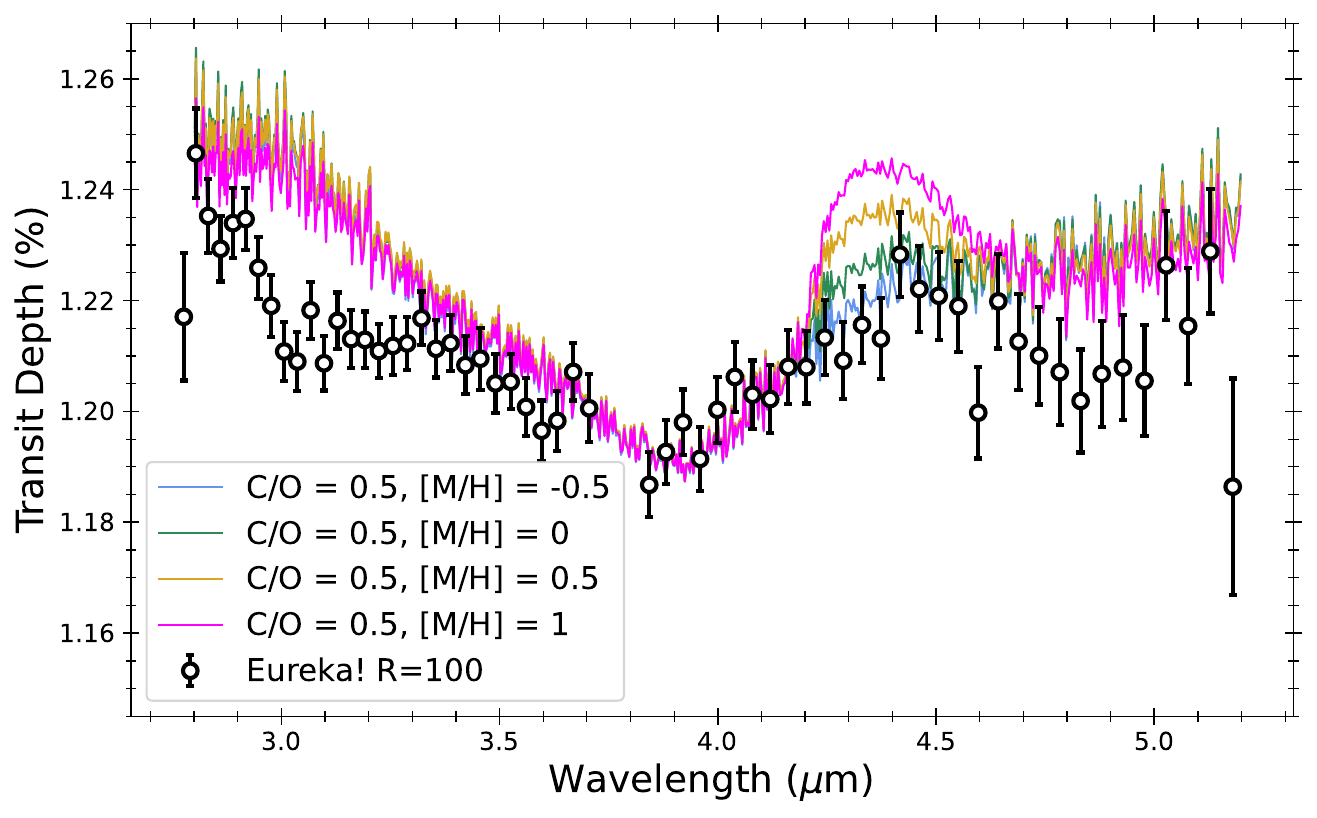}
    \includegraphics[width=0.99\linewidth]{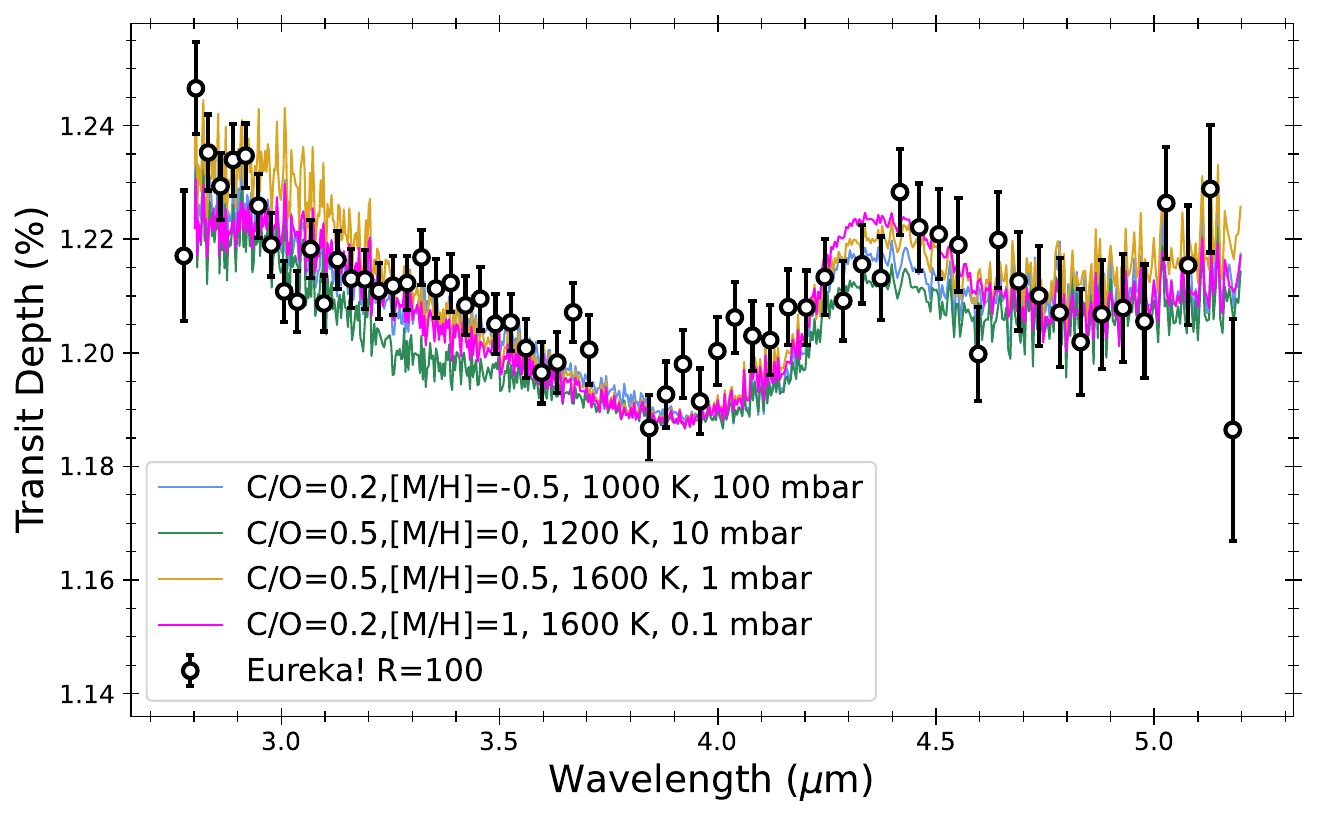}
    \caption{\new{1D isothermal chemical equilibrium forward models of \myplanet compared to the \eureka $R=100$ transmission spectrum. The left panel depicts cloud-free models with the isotherm temperature set to 1630 K, the equilibrium temperature of \myplanet. The right panel depicts a selection of models at different metallicities with comparable fits to the data using different C/O ratios, isotherm temperatures, and opaque grey cloud-top pressures, all noted in the legend. These models demonstrate that the spectral features of \myplanet are muted compared to those predicted by cloud-free models. The degeneracies between composition, temperature, and clouds highlighted in the right panel demonstrate the need for a full exploration of the parameter space using atmospheric retrievals to interpret the spectrum of \myplanet.}}
    \label{fig:forward-models}
\end{figure}

\subsection{\new{petitRADTRANS forward models}}
\label{subsec:forward}

\new{We simulate the atmosphere of \myplanet with the \prt\footnote{\url{https://petitradtrans.readthedocs.io/en/latest/}} \citep[v3.1.3,][]{molliere2019petitradtrans} package, assuming a 1D isothermal atmosphere and chemical equilibrium. In all cases we assume a \ch{H2}/He-dominated atmosphere, and include $R=1000$ correlated-$k$ line opacities from CO \citep{rothman2010hitemp}, \ch{H2O} \citep{polyansky2018exomol}, \ch{CO2} \citep{yurchenko2020exomol}, \ch{CH4} \citep{yurchenko2017hybrid}, \ch{H2S} \citep{azzam2016exomol}, \ch{SO2} \citep{underwood2016exomol}, SO \citep{brady2024exomol}, \ch{HCN} \citep{barber2014exomol}, and \ch{NH3} \citep{coles2019exomol}, as well as collisionally-induced absorption from \ch{H2-H2} and \ch{H2-He} \citep{borysow1988collison,borysow2001high,borysow2002collision}, and Rayleigh scattering from \ch{H2} and He \citep{dalgarno1962rayleigh,chan1965refractive}. We model the atmosphere using 100 log-spaced pressure layers from $10^{-6}$ to $10^2$\,bar.}

\new{We initially constructed a simple grid of cloud-free simulations with the temperature set to the equilibrium temperature, varying the C/O ratio from 0.2--1.0, and the log10 metallicity relative to solar ([M/H]) from -1 to 2. These simple models demonstrate visibly stronger spectral features than those present in the observed spectra, highlighting that the spectral features are muted to some degree, as shown in the first panel of Fig.\, \ref{fig:forward-models}.}

\begin{figure*}
    \centering
    \begin{minipage}[t]{0.75\textwidth}
        \centering
        \includegraphics[width=0.98\linewidth]{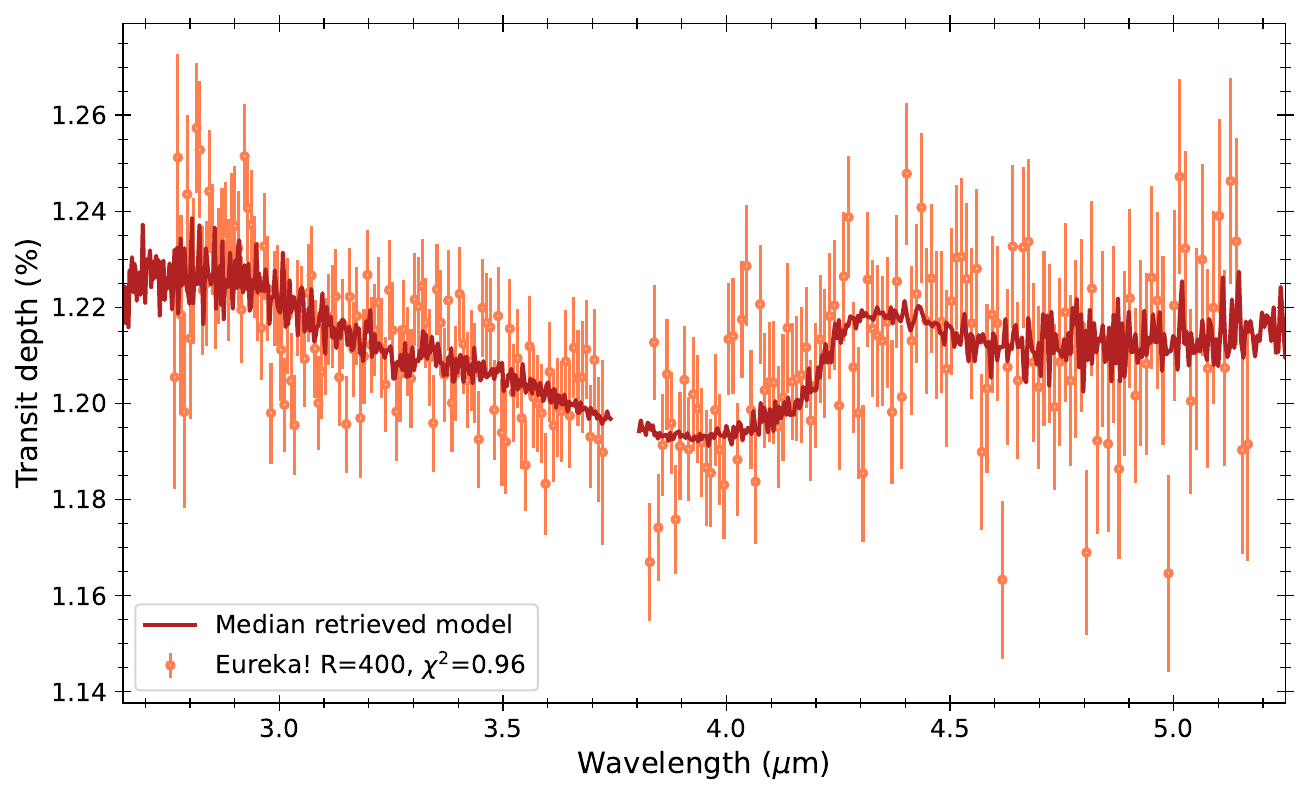}
    \end{minipage}
    \begin{minipage}[t]{0.2\textwidth}
        \vbox{ 
            \centering
            \includegraphics[width=0.85\linewidth]{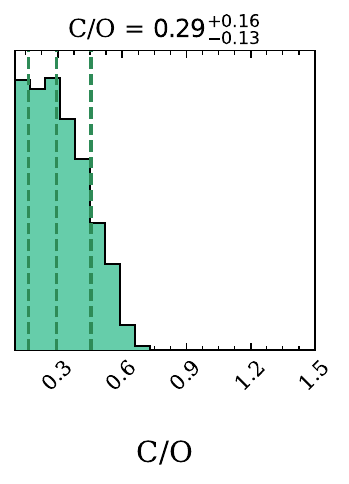} \\[1ex]
            \includegraphics[width=0.85\linewidth]{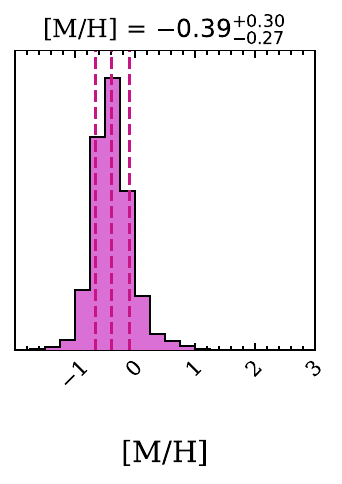}
        }
    \end{minipage}
    \caption{Median retrieved spectrum (red) of the \eureka $R=400$ data (orange) from the \prt equilibrium chemistry retrieval with no offset, with the posterior probability distributions of C/O ratio and [M/H].}
    \label{fig:retrieval_main}
\end{figure*}

\new{We then expanded this initial grid by including an opaque grey cloud deck at varying pressures (up to 0.1 mbar), and by reducing the isotherm temperature (down to 1000 K). By tuning the C/O ratio, isotherm temperature, and cloud-top pressure, we were able to create comparable, adequate fits to the data at a range of metallicities from $0.3-30\times$ solar, presented in the second panel of Fig.\, \ref{fig:forward-models}. This exemplifies the strong degeneracies in atmospheric models, particularly given the limited wavelength coverage of G395H data, with only subtle differences between these fits mainly visible at 4.4 $\mu$m and 3.3 $\mu$m. Our ability to constrain the atmosphere of \myplanet by eye with a course forward model grid alone is therefore limited, and we turn to the statistical power offered by atmospheric retrievals and nested sampling to fully explore the model parameter space.}

\subsection{petitRADTRANS \new{retrievals}}

We perform a variety of atmospheric retrievals on the \myplanet transmission spectrum with \prt \citep{Nasedkin2024}, using a similar setup to that used to analyse the spectrum of the BOWIE-ALIGN targets WASP-15\,b \citep{kirk2025bowie} and TrES-4\,b \citep{meech2025bowie}. \prt explores the multi-dimensional model transmission spectrum parameter space using Bayesian nested sampling \citep{Skilling2004} implemented through \texttt{MultiNest} \citep{Feroz2008} with \texttt{PyMultiNest} \citep{buchner2014x}. As with other BOWIE-ALIGN targets, we use three principle setups: equilibrium chemistry, free chemistry, and hybrid chemistry. 

\new{For all of our retrievals, we use the model setup properties and opacity sources described in Section \ref{subsec:forward}.} The stellar radius is fixed to the value of 1.215\,$R_{\odot}$ from \citet{Bonomo2017}. For our standard retrieval setups, we assume an isothermal pressure-temperature profile with a wide uniform temperature prior of 500--3000\,K. We also adopt a Gaussian prior for the gravity based on the mass and radius \citep{Bonomo2017}, and a wide uniform prior for the planetary radius of 0.8--2.2 $R_\mathrm{J}$, defined at a reference pressure of 1\,mbar. We parametrize the impact of aerosols by including a grey cloud deck, with a log-uniform prior on the cloud-top pressure from $10^{-6}\,$bar to 10$^2$ bar. 


We run our retrievals on the $R=100$ and $R=400$ transmission spectra from both the \eureka and \tiberius reductions. We permit for an offset between NRS1 and NRS2, with a uniform prior of $\pm$200\,ppm. We determine the detection significances of different species by computing the difference in Bayesian evidence $\ln{Z}$ between the retrieval with all species included, and the retrieval with the considered species omitted, converted into a frequentist significance value from the Bayes factor $B_m$ via the $p$-value using $p=\frac{1}{1+B_m}$ \citep{jeffreys1939theory}. We summarise all of our retrieval results, including evidence and posteriors, in Table\,\ref{tab:all_retrievals}.

\subsubsection{Equilibrium chemistry}


\begin{figure*}
    \centering
    \includegraphics[width=0.9\linewidth]{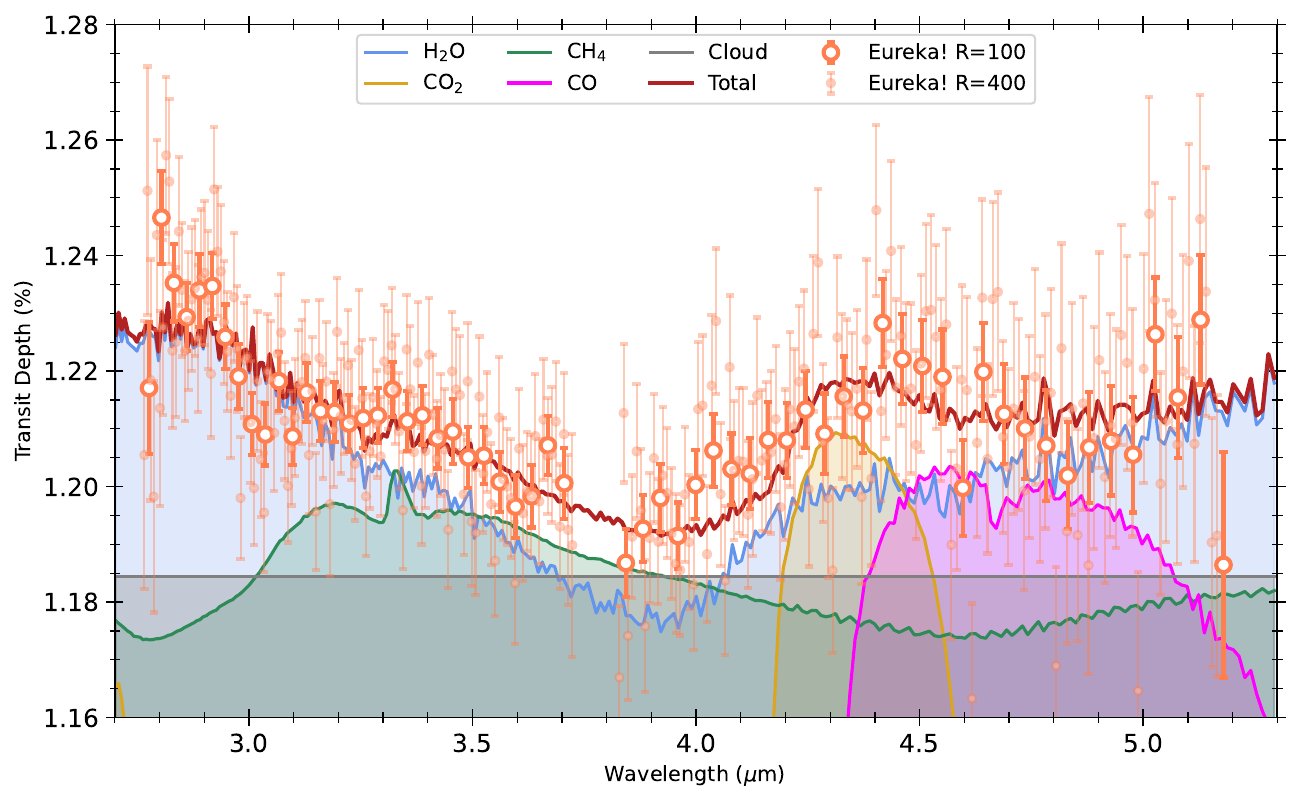}
    \caption{Median retrieved transmission spectrum of the \eureka $R=400$ data from the \prt equilibrium chemistry retrieval convolved to a spectral resolution of $R=400$, with the spectral contributions from \ch{H2O}, \ch{CO2}, \ch{CH4}, CO, and a grey cloud deck, and the combined absorption of all opacities (red).}
    \label{fig:eq-contributions}
\end{figure*}

In our equilibrium retrievals, we assume all the atmospheric species are in chemical equilibrium, with the abundances of \ch{CH4}, \ch{CO2}, \ch{CO}, \ch{H2O}, \ch{H2S}, HCN, and \ch{NH3} interpolated from a pre-computed grid parametrized with temperature, pressure, C/O ratio, and [M/H]. We use a wide uniform prior in both C/O ratio (0.1--1.5) and [M/H] (-2--3). In this parametrization, the C/H ratio is fixed by the [M/H] value \citep[using solar values from][]{Asplund2009}, while the O/H ratio is computed by the ratio of the C/H to the C/O.

We obtain consistent results across both reductions and resolutions, inferring a sub-Solar metallicity and a sub-Solar C/O ratio, with values of [M/H] = $-0.42^{+0.48}_{-0.38}$ and C/O = $0.28^{+0.17}_{-0.12}$ from the \eureka $R=400$ reduction, depicted in Fig. \ref{fig:retrieval_main}.
We also infer cool limb temperatures of $1010^{+110}_{-70}$\,K, as is commonly observed in transmission spectroscopy retrievals \citep{MacDonald2020,Welbanks2022}, and poor constraints on the cloud-top pressure (with a peak in the posterior at 10 mbar - see Fig. \ref{fig:prt_eq_cornerplot_R100}). We see evidence for \ch{H2O} and \ch{CO2} at 3.3 and 2.1 $\sigma$ significance respectively.


The \eureka reduction is consistent with no detector offset, while the \tiberius reduction has a median detector offset of $-56\pm35$\,ppm. Considering the metallicity posterior in Fig. \ref{fig:prt_eq_cornerplot_R100} highlights a degeneracy present in the \tiberius reduction, with a lower evidence secondary mode with high metallicity ($\sim60\times$solar) and a large $\sim110$\,ppm detector offset. This degeneracy is far weaker in the \eureka reduction, only representing a slight tail in the posterior metallicity distribution rather than a second peak.

Repeating the retrievals on the $R=400$ reductions with no offset, we confirm that no offset is necessary for the \eureka reduction, slightly increasing the Bayesian evidence with a Bayes factor of 4.5, and removing the high metallicity tail, resulting in a somewhat tighter posterior on the metallicity of $-0.39^{+0.30}_{-0.27}$. Not including an offset for the \tiberius reduction, on the other hand, is greatly disfavoured with a Bayes factor ($\frac{Z_1}{Z_2}$) of 1339, and gives discrepant results, with a lower C/O ratio, metallicity, and deeper clouds. We present this spectrum in Fig. \ref{fig:retrieval_main}, and the opacity contributions to this model in Fig. \ref{fig:eq-contributions}.

\begin{figure}
    \centering
    \includegraphics[width=0.95\linewidth]{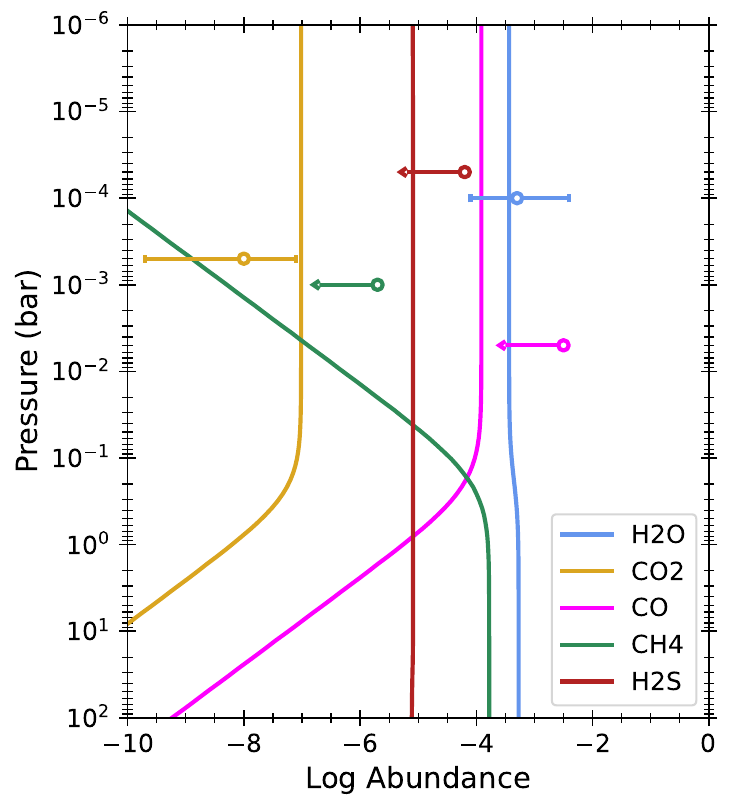}
    \caption{Vertical mixing ratios of \ch{H2O}, \ch{CO2}, \ch{CH4}, CO, and \ch{H2S} from the \eureka $R=400$ \prt equilibrium chemistry retrieval (lines), plotted with the median retrieved abundances with $1\sigma$ error bars for the detected species (\ch{H2O} and \ch{CO2}), and upper limits from the other species (\ch{CH4}, \ch{CO}, and \ch{H2S}) from the free chemistry retrieval (data points).}
    \label{fig:abundances}
\end{figure}


\subsubsection{Free chemistry}

For our free chemistry retrievals, the abundances of \ch{CH4}, \ch{CO2}, \ch{CO}, \ch{H2O}, \ch{H2S}, \ch{SO2}, and SO are free parameters, with a wide log-uniform prior in mass fraction from $10^{-12}$ to $10^{-0.5}$. We also tested retrievals including opacity from other species including \ch{NH3}, HCN, OCS, \ch{CS2}, and \ch{N2O}, but chose not to include them from our primary retrievals when the posteriors provided minimal constraints on their abundances. We obtain consistent results across all reductions and resolutions, retrieving a $\log(X_{\mathrm{H_2O}})$ abundance of $-3.3\pm0.8$, and a $\log(X_{\mathrm{CO_2}})$ abundance of $-8.1^{+0.9}_{-1.8}$ from the \eureka $R=400$ reduction. We also retrieve a very cold limb temperature of $650^{+120}_{-80}$ K, far cooler than the planet's equilibrium temperature of 1630 K. We place 2$\sigma$ upper limits on the $\log(X_{\mathrm{CH_4}})$ abundance of $-6.1$, $\log(X_{\mathrm{CO}})$ abundance of -3.6, and $\log(X_{\mathrm{SO2}})$ abundance of -5.3, respectively (see the posterior corner plot in Fig\,\ref{fig:prt_free_cornerplot_R100}). There is also a strong peak in the posterior of the SO abundance at $\sim10$\,ppm, with a feature at 4.5 $\mu$m, although repeating the retrievals without SO find its inclusion is slightly disfavoured by the Bayesian evidence, and it does not affect the other posteriors. The free retrieval abundance posteriors are consistent with the chemistry implied by the equilibrium  retrievals above 100 mbar, as demonstrated in Fig. \ref{fig:abundances}.

\subsubsection{Hybrid chemistry}

In our hybrid chemistry retrievals, we combine the equilibrium chemistry approach to set the abundances of the carbon- and nitrogen-bearing species \ch{CH4}, \ch{CO2}, \ch{CO}, \ch{H2O}, HCN, and \ch{NH3}, parameterized with C/O ratio and [M/H], and the free chemistry approach to set the abundances of sulfur species \ch{H2S}, \ch{SO2}, and SO to be free parameters. This allows us to model an atmosphere broadly in chemical equilibrium, but with the abundance of sulfur species modified by photochemistry or a variable S/O ratio.

Our hybrid chemistry results are broadly consistent with our equilibrium chemistry results, however, we do see stronger evidence for a high metallicity, high detector offset mode in both \eureka and \tiberius reductions now, becoming the higher evidence mode in \tiberius $R=400$. This appears to be caused by \ch{SO2}, which has a notable peak in the posterior at $\sim1$\,ppm abundance. The abundances of \ch{H2S} and SO are unconstrained in the hybrid retrievals. All the hybrid retrievals have poorer evidence than their equilibrium chemistry counterparts. Each sulfur species added further decreases the evidence (disfavoured with a Bayes factor of 3 with all three species for the \eureka reductions), except in the \tiberius $R=400$ reduction, where the inclusion of \ch{SO2} minimally improves the Bayesian evidence (Bayes factor of 1.3).

\subsection{BeAR \new{retrievals}}

We also perform retrievals using the open-source GPU-accelerated Bern Atmospheric Retrieval code (\bear\footnote{Formerly known at \texttt{Helios-r2}. \bear can be found at \url{https://github.com/newstrangeworlds/bear}}) \citep{kitzmann20}. \bear uses the \texttt{MultiNest} library \citep{Feroz2008} to perform the retrieval using Bayesian nested-sampling \citep{Skilling2004}, and using line-by-line opacity sampling. For our \bear retrievals, we use a similar setup as in \cite{kirk2025bowie}. We sample the opacities at a resolution of \SI{0.1}{\per\cm} in wavenumber (equivalent to $R\sim 20,000$--$30,000$), and include the following molecules and their associated \texttt{ExoMol} and \texttt{HITRAN} line-lists: \ch{H2O} \citep{polyansky2018exomol}, \ch{CH4} \citep{yurchenko14}, \ch{CO} \citep{li2015}, \ch{CO2} \citep{yurchenko2020exomol}, \ch{H2S} \citep{azzam2016exomol}, \ch{SO2} \citep{underwood2016exomol}, \ch{NH3} \citep{coles2019exomol}, and \ch{HCN} \citep{barber2014exomol}. All these opacities are computed using \texttt{Helios-k} \citep{grimm15} and are taken from the \texttt{DACE} database \citep{grimm21}. We also include opacity due to \ch{H2} Rayleigh scattering \citep{cox2000}, and collision-induced absorption from \ch{H2}-\ch{H2} \citep{abel2011} and \ch{H2}-\ch{He} \citep{abel2012}. The atmosphere is divided into 200 levels, equal in log-pressure space, ranging from 10 bar to $10^{-8}$ bar. 

For the retrievals, the stellar radius is again fixed to 1.215\,$R_{\odot}$ \citep{Bonomo2017}. The planet's gravity and radius at the 10 bar pressure level are free parameters in the retrieval. The gravity has a gaussian prior on $\log{g}$ with a mean of 2.96 and a standard deviation of 0.03 in cgs units, and the radius has a uniform prior of 1.25--1.65$R_{\rm J}$. The atmosphere is assumed to be isothermal, with a uniform prior on the temperature of 500--\SI{2500}{K}. A grey cloud deck is included, and the cloud-top pressure is a free parameter in the retrieval, with a log-uniform prior of $10^{-7}$--1.0 bar. We also include an offset between the NRS1 and NRS2 detectors, with a uniform prior from -100 to +100 ppm. 

We run two types of chemistry for our \bear retrievals. Firstly, we assume free chemistry. In this retrieval, each molecule's volume mixing ratio is a free parameter, with a log-uniform prior from $10^{-12}$ to 0.3. The mixing ratios are constant with altitude. The rest of the atmosphere is then filled with \ch{H2} and He, in a solar ratio of 0.17 \citep{Asplund2009}. Secondly, we assume equilibrium chemistry. For this, the free parameters are the [M/H], which has a log-uniform prior of 0.1--1000$\times$ solar, and the C/O ratio, which has a uniform prior of 0.1--2.0. [M/H] and C/O are related to the carbon and oxygen abundance in the same manner as in the \prt models. The chemistry is calculated using \texttt{FastChem} \citep{stock18,stock22}, which is already integrated into \bear. We apply both types of retrievals to the \eureka and \tiberius spectra, and $R=100$ and $R=400$. The results of our chemical equilibrium and free chemistry retrievals with \bear at $R=400$ are shown in Figures \ref{fig:BeAR_cornerplot_R400_chemeq} and \ref{fig:BeAR_cornerplot_R400}, respectively.

In the equilibrium chemistry case, we see a good agreement across the two data reductions. The only difference occurs in the detector offset, for which the retrieval on the \eureka reduction favours no offset, while the \tiberius reduction retrieval finds an offset of $-60^{+23}_{-21}$ ppm. For the Eureka $R=400$ retrieval, the retrieved [M/H] is $-0.52^{+0.31}_{-0.25}$, and the C/O ratio is $0.26^{+0.17}_{-0.11}$. The results are consistent with the \tiberius and $R=100$ retrievals (see full results in Table \ref{tab:all_retrievals}). 

In the free chemistry case, we again see a good agreement between \eureka and \tiberius reductions, with the exception of a bimodality in the \eureka $R=100$ case 
. The alternative solution in the \eureka case corresponds to an exceptionally high \ch{CO2} abundance, which is compensated by a higher temperature, larger planet radius, lower water abundance, and a substantial negative offset of NRS2. However, at $R=400$, the second solution in the \eureka case is not present, and the retrievals agree with both the $R=100$ and $R=400$ \tiberius solution.

Finally, we also tested the effects of adding stellar contamination to our free chemistry retrievals. For this, we use the PHOENIX grid of stellar models \citep{husser13}, and set the stellar values to those in Table \ref{tab:planetary-parameters}.
We retrieve the stellar effective temperature, using a Gaussian prior with a mean of 6304 K and a standard deviation of 88 K \citep{Johnson2011}.  We then include the effects of hot and cold spots on the star, and retrieve their temperatures and covering fractions. For the hot spots, the temperature is a retrieved as a positive $\Delta T$ with respect to the stellar effective temperature, using a uniform prior of 0--1000 K. For the cold spots, the temperature is retrieved as a negative $\Delta T$ with respect to the stellar effective temperature, using a uniform prior of 0--1500 K. For both, the covering fractions use uniform priors of 0--0.5. For both reductions and resolutions, we find that adding stellar contamination to the retrievals has little effect on the results, and a Bayesian evidence comparison favours the retrieval without contamination in all cases except using the \eureka reduction at $R=100$. In the latter case, however, the Bayes factor is negligible at only 1.65. Overall we conclude that this constitutes no evidence for stellar contamination in the spectrum of HAT-P-30 b.

\begin{figure*}
    \centering
    \includegraphics[width=0.95\linewidth]{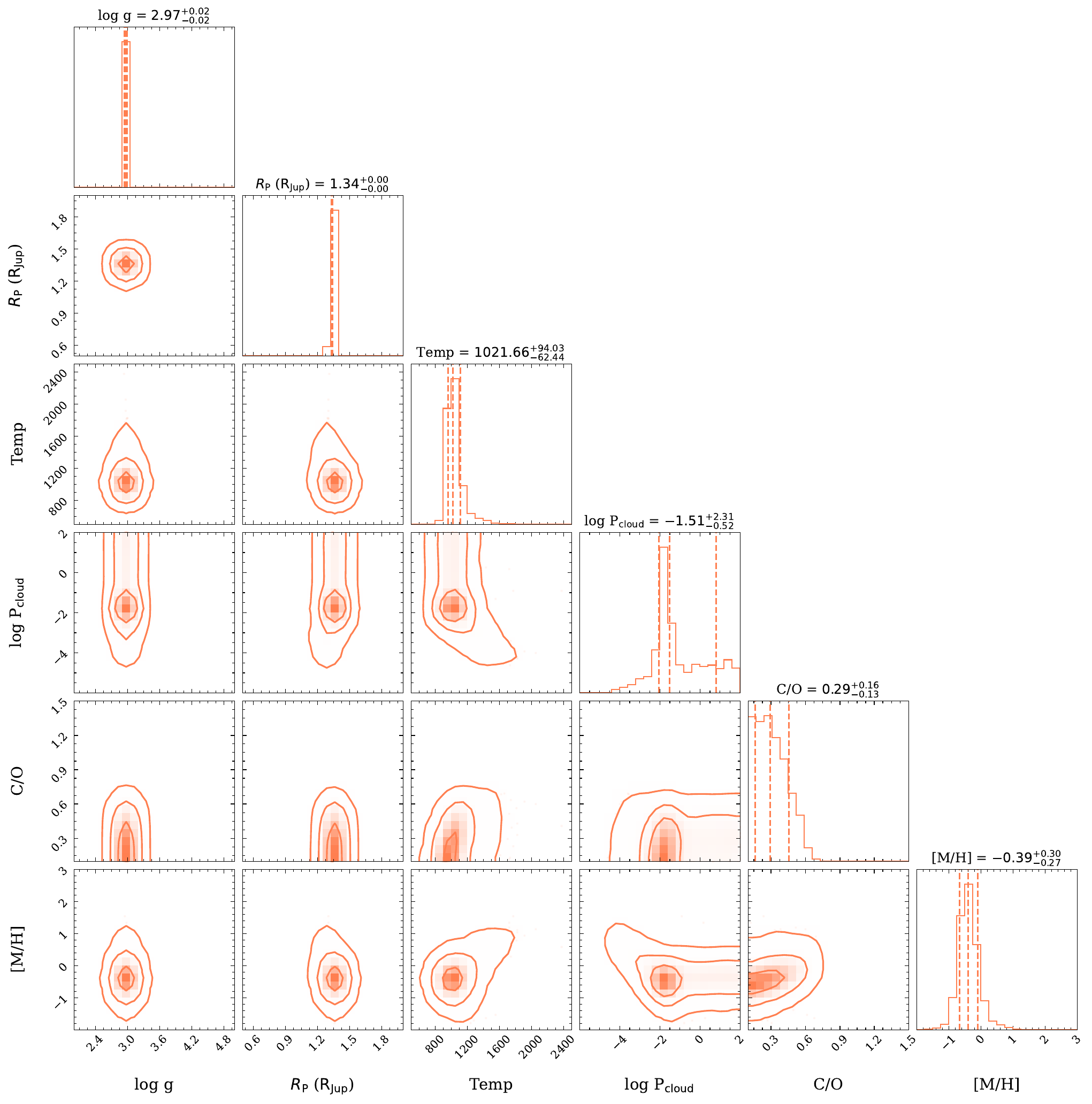}
    \caption{Corner plot showing the posterior probability distributions from the favoured retrieval setup: the \prt equilibrium chemistry retrieval on the $R=400$ \eureka transmission spectrum with no detector offset.}
    \label{fig:corner-main}
\end{figure*}

\subsection{Overview of retrieval results}

Using two independent retrieval analyses with \prt and \bear on two independent data reductions at two spectral resolutions, with a variety of retrieval setups, we are able to put together a consistent picture for the atmosphere of \myplanet. The complete set of results are presented in Table \ref{tab:all_retrievals}, with the posterior corner plots for the $R=400$ retrievals displayed in Appendix A. We find that retrievals on the \tiberius reductions require an offset between the detectors, while the \eureka reduction retrievals do not. We also generally find we get slightly tighter constraints using the $R=400$ retrievals, given the better ability to sample narrow spectral features. Therefore we favour the interpretations from the \eureka $R=400$ retrieval with no offset.
We present the median retrieved \prt transmission spectrum to the \eureka $R=400$ spectrum in Fig.\,\ref{fig:retrieval_main} with equilibrium chemistry and the posteriors for metallicity and C/O ratio, as well as the spectral contributions of each opacity source in Fig.\,\ref{fig:eq-contributions}, and the posterior corner plot in Fig.\,\ref{fig:corner-main}.

All the equilibrium retrievals have a consistent main result: sub-solar C/O ratio ($0.29^{+0.16}_{-0.13}$) and metallicity ($-0.39^{+0.30}_{-0.27}\times$solar), with exact numbers taken from the \prt \eureka $R=400$ equilibrium retrieval with no offest. The free chemistry retrievals are similarly concordant, and yield consistent abundances with those implied by the equilibrium chemistry, with the vertical abundance profiles depicted in Fig.\,\ref{fig:abundances}.

In a sub-sample of the retrievals, we do see a degree of bimodality, with an alternative explanation of a high-metallicity, \ch{CO2}-dominated spectrum. This secondary mode appears in the \bear free retrieval of the \eureka $R=100$ spectrum, the \prt equilibrium retrievals of both \tiberius spectra, and all the \prt hybrid retrievals. The mode is highly degenerate with detector offset, requiring an offset of $\sim60$ ppm in the \eureka data and $\sim110$ ppm in the \tiberius data. This mode represents a 50--100$\times$solar metallicity, the lower end of which does fit within the constraints of a fully-mixed atmosphere from the interior structure modelling (see ahead to Section \ref{sec:interior}). This mode is only favoured over the low-metallicity mode in a single retrieval, the \tiberius $R=400$ \prt hybrid retrieval. However, the evidence generally does not support the extra free parameters necessary for utilising a hybrid retrieval over equilibrium chemistry, and the $\sim110$ ppm offset required is much larger than the typical offset observed in other data sets \citep[e.g.,][]{Alderson2023,MoranStevenson2023}. We therefore highly emphasize the lower metallicity mode as our favoured interpretation.


We also note that all of our free chemistry retrievals have very cold limb temperatures of $\sim$650\,K. Cool limb temperatures are commonly observed in transmission spectroscopy of hot Jupiters, and can be attributed to the 1D modelling of the atmosphere \citep{MacDonald2020} (i.e. limb asymmetries) or the isothermal parametrization used \citep{Welbanks2022}. This case, however, seems particularly extreme, with a 1000\,K discrepancy from equilibrium temperature to observed isothermal temperature. We can see by comparing to the equilibrium chemistry retrievals, which are effectively forced into a higher temperature by the absence of \ch{CH4} in the spectrum (the dominant carbon-carrier at low temperatures), that we may in fact be probing the cloud-top pressure--scale height degeneracy. The particular noise profile of the observation may be affecting the feature width and pushing us towards a low temperature explanation for the muted scale height, while in reality this is more likely an impact of clouds. This is demonstrated when restricting the temperature prior to higher temperatures, as the cloud top pressure is pushed higher to compensate. This does not impact our interpretation of the chemistry, however, as the abundances are consistent with the higher temperature equilibrium chemistry retrieval counterparts. \new{A limb asymmetry analysis of the complete sample of BOWIE-ALIGN planets, including \myplanet, will be presented in future work.}

\new{We do note that the contribution from \ch{CH4} appears to be non-negligible (see Fig.\, 6), despite \ch{CH4} not being expected at the high temperatures of \myplanet. If the retrieved low temperature of \myplanet is a result of the temperature profile parametrization or limb asymmetry, the retrieval may then be trying to minimise the contribution from \ch{CH4} by suppressing the carbon content of the planet, reflected as a low C/O ratio and metallicity. To account for this, we run an equilibrium retrieval discarding \ch{CH4} as an opacity source. This results in a temperature posterior of $\sim650$\,K, matching the free retrievals, as the absence of strong \ch{CH4} features no longer sets a lower limit on the temperature. We account for this by repeating the retrieval without \ch{CH4} with a temperature lower bound of 1000 K. This results in a completely consistent best-fit spectrum to the equilibrium retrieval including \ch{CH4}, although the constraints on temperature, cloud-top pressure, and metallicity are all weaker. We therefore demonstrate that rather than the low temperature driving the inclusion of \ch{CH4} in the spectrum, it is in fact the other way around, with the enhanced opacity at $\sim3.3-3.5$ $\mu$m driving the retrieval to a lower temperature. We therefore do not believe that the low observed temperatures are suppressing the C/O ratio, but that our low temperature conclusion may instead be the result of enhanced transit depths at $\sim3.3-3.5$ $\mu$m from the particular random noise draw of this observation, the presence of disequilibrium \ch{CH4} or species with similar opacity, or a non-grey continuum opacity putting in a bluewards-increasing slope in that wavelength range. We conclude that the observed sub-solar C/O ratio is robust, while the metallicity, while consistent with sub-solar, may be less well constrained ($-0.38^{+0.83}_{-0.68}$) than our equilibrium models suggest, should the real temperature be higher and the opacity discovered to not be \ch{CH4}.} \newer{Fixing the temperature at the equilibrium temperature of 1630 K biases the metallicity posterior towards the higher end of that distribution ($0.43^{+0.22}_{-0.46}$) but is still consistent with the nominal sub-solar metallicities within $2\sigma$, while ruling out metallicities greater than 10xSolar. The C/O ratio is unchanged by the fixed temperature, while the need for clouds around 0.1-1 mbar to sufficiently mute the spectral features becomes evident. The Bayesian evidence does not support fixing the temperature (decrease in $\ln{Z}$ of 6.7), and we note true temperature profile could very well diverge from an isotherm fixed to the equilibrium temperature.}

For the purpose of studying the implications of our retrieved parameters for planet formation and relation to other planets in the BOWIE-ALIGN sample, we recommend the use of the posteriors from the \eureka $R=400$ \prt retrieval with no offset between the detectors (Fig\,\ref{fig:corner-main}), as it is unimpacted by the degeneracy with the higher-metallicity mode. We therefore proceed with a C/O ratio of $0.29^{+0.16}_{-0.13}$ and metallicity of $-0.39^{+0.30}_{-0.27}$.

\section{Interior Structure Models}
\label{sec:interior}
Using the Bayesian framework of \citet{Thorngren2019}, we estimate the bulk composition of HAT-P-30 b, accounting for uncertainties in the mass, radius, and age of the planet.  The evolution models underlying the statistical model are again from \citet{Thorngren2018}, and solve the equations of hydrostatic equilibrium, conservation of mass, and the equation of state in one dimension.  The planet's thermal state is evolved forward from a hot initial state using the atmosphere models of \citet{Fortney2007} to regulate heat flow out of the planet's interior.  The equilibrium temperature is well into the hot Jupiter range, so we account for the anomalous heating using the flux-heating relation of \citet{Thorngren2018}.  This implies a quite high intrinsic temperature of $653 \pm 45$.  

The retrieval finds a bulk metallicity of $Z=0.28 \pm 0.03$ (this is the statistical error from parameter uncertainties, not modeling uncertainty).  This is within the expected range for a planet of this mass \citep{Thorngren2016}, being only moderately above average.  If the planet were fully mixed, it would imply an atmospheric metallicity 2-sigma upper limit of $71.1\times$Solar.  The observed atmospheric metallicity could be less than this if the bulk metal is contained within a core or beneath a compositional gradient.  It could not be greater, however, as this would be unstable to convection.

The substantial enhancement of the bulk metallicity relative to that observed in its atmosphere is not an inconsistency -- it instead implies that the planet has a massive core.  To match the radius and atmospheric metallicity, the core must be at least $50 M_\oplus$.  This could be either with a sharp core boundary or as a fuzzy core featuring an extensive compositional gradient \citep[see e.g.][]{Helled2017} -- structure models such as these don't distinguish between those cases \citep{Bloot2023}.  A planet like this could have formed if most of the metal accretion occurred early on, with the final stages being the accretion of metal-poor gas.  Alternatively, if the planet formed in the ice line, there might be a substantial quantity of rock and iron mixed into its interior that has either settled out or cannot be seen in the atmosphere due to condensation.

\section{Discussion}
\label{sec:discussion}

\subsection{The atmosphere of \myplanet}
\label{subsec:interpretation}

Our transmission spectrum, obtained from two consistent independent reductions, describes the wavelength-dependent opacity in the atmosphere of \myplanet from 2.8 $\mu$m and 5.2 $\mu$m at a resolution of $R=400$. We have used parametric radiative transfer models, implemented using atmospheric retrievals from the \prt and \bear retrieval packages, with a variety of retrieval setups including equilibrium and free chemistry, to analyze this spectrum and make inferences about the atmospheric properties of \myplanet.
In 22 of the 23 retrievals performed, the favoured interpretation is one of a low-metallicity and C/O ratio atmosphere, representing sub-solar abundances of carbon and potentially also oxygen, derived from the equilbrium chemistry retreivals ($0.41^{+0.40}_{-0.19}$ and $0.77^{+0.84}_{-0.56}$ $\times$ their solar abundances respectively, computed from the C/H, which is fixed by the metallicity, and the C/O ratio). In this interpretation, we require some muting of spectral features from cooler-than-expected temperatures ($\sim 1000$ K in the equilibrium retrievals, $\sim 650$ K in the free retrievals due to the lack of chemical constraints) and clouds. No offset between detectors is favoured by the Bayesian evidence for the \eureka spectra, while a $\sim 55$ ppm offset is required for the \tiberius spectra, matching the observed linear offset between the reductions. We see evidence for both \ch{H2O} and \ch{CO2} at 3.3 and 2.1 $\sigma$ respectively.

Favoured in the Tiberius $R=400$ \prt hybrid chemistry retrieval, and present as a notable secondary mode in 4 other retrievals, is an alternative interpretation, with a highly super-solar metallicity of $\sim60\times$ solar. This is consistent with the upper limits of metallicity of a fully-mixed atmosphere suggested by our interior structure modelling. This mode requires freedom in both detector offset, needing a $\sim60$ ppm offset in the \eureka retrievals and $\sim110$ ppm in the \tiberius retrievals, and \ch{SO2} abundance, to appropriately fit the spectra with \ch{CO2} features.

We favour the low metallicity interpretation for the atmosphere of \myplanet for a number of reasons. Firstly, it is favoured in all of the standard equilibrium and free chemistry retrievals, and it is only not favoured with a single combination of reduction, resolution and retrieval pipeline, using the hybrid retrieval method - where the increase in complexity from the equilibrium method is not justified by the Bayesian evidence.
Secondly, the large detector offset of 110 ppm that gives rise to the higher metallicity solution is greater than the offsets that have previously been present in NIRSpec/G395H observations of hot Jupiters \citep[e.g.,][]{Alderson2023}.
Thirdly, the high metallicity modes only notably appears when the \ch{SO2} abundance is a free parameter, to appropriately scale the transit depth at 4 $\mu$m. While \ch{SO2} has been observed in exoplanet atmospheres \citep[e.g.,][]{Alderson2023} and is predicted by photochemical models \citep[e.g.,][]{Tsai2023,kirk2025bowie} in high metallicity hot Jupiters, including \ch{SO2} does not improve the Bayesian evidence in all free retrievals, and 3 of 4 hybrid retrievals. 

We therefore highly emphasize our low metallicity ($-0.39^{+0.30}_{-0.27}$) and C/O ratio ($0.29^{+0.16}_{-0.13}$)  interpretation as that most favoured by our data and physical models. We do however note the existence of this physically-plausible alternative interpretation, with a high metallicity ($1.77^{+0.31}_{-2.07}$) and potential \ch{SO2}, representing a fully-mixed atmosphere, although we highlight it is by-in-large not favoured by our data. Further observations could definitively resolve between these scenarios, including NIRISS SOSS observations taken in March 2025 as part of JWST programme 5924 (PI: Sing). \new{We also note that the apparent presense of \ch{CH4} is somewhat impacting our interpretation of the chemistry and atmospheric temperature, and while this does not affect our inferred C/O ratio, disregarding \ch{CH4} as a potential equilibrium opacity source does weaken the constraint on the metallicity to $-0.38^{+0.83}_{-0.68}$. We will investigate limb asymmetries of the BOWIE-ALIGN planets, a potential cause of the low observed temperature, in a future work,} \newer{which could further refine our metallicity and temperature constraints.}

\subsection{Implications for formation}

\label{subsec:formation}

The preferred atmospheric retrieval solutions of \myplanet show a sub-solar C/O ratio and sub-solar, sub-stellar metallicity with C/O $= 0.18$--0.38 and  [M/H] $= -0.83$ to $-0.05$, dervied from \prt equilibrium chemistry retrievals of the $R=400$ \eureka reduction.
The C/O ratio deviates from the solar value at the $\sim1.5\,\sigma$ level.
In formation models, supersolar metallicities and subsolar C/O are the expected outcome of high solid accretion of oxygen-rich dust and ice \citep[e.g.][]{Madhusudhan2014}.
However, we disfavour the high metallicity solutions with large detector offsets that appear in a subset of our retrievals (see Section \ref{subsec:interpretation}).

Constraints from the C/O ratio and metallicity alone lead to large degeneracies in the possible formation history of the planet. However, a sub-solar C/O ratio and a sub-solar metallicity together are harder to reproduce because the gas in protoplanetary discs typically has sub-solar metallicity but super-solar C/O \cite[see, e.g][]{bergin2024carbon}. For this reason, the retrieved atmosphere is inconsistent with the fiducial models of \citet{penzlinBOWIEALIGNHowFormation2024} by at least 1-$\sigma$. The closest match is for a planet that accreted most of its gas far from the star, where the gas is extremely metal-poor and all carbon and oxygen carriers are frozen out on the dust. Since the solid-phase abundance at this location has a solar C/O ratio, adding solids at this location produces an atmosphere with a solar C/O. If the planet accretes solids after migrating further in, this is no longer the case -- the composition of solids can reach a sub-solar C/O ratio due to, e.g., the sublimation of CO and \ch{CH4}. Thus, by accreting gas far out but being enriched by a modest amount of solids as it migrates, a planet can reach sub-solar C/O ratios and metallicities in the outer disc in our models \citep{penzlinBOWIEALIGNHowFormation2024}.
Despite this, the fiducial simulations do not reach C/O ratios below 0.35 for sub-solar metallicity and, hence, exceed the best fit values of \myplanet. This mismatch is, however, sensitive to the composition of the disc -- the depletion of CO into \ch{CH4} or \ch{C2H6} instead \citep[e.g.][]{bosman2019} can lower the C/O ratio in the solids further and lead to a better agreement between models and observation.

As noted by \citet{meech2025bowie} when discussing TrES-4b (which has similar metallicity and C/O ratio to \myplanet: $\log{( Z / Z_\odot )}=-0.41$ to $-0.04$ and $\mathrm{C/O}=0.30$--0.42), this is not the only possible channel of formation of sub-solar C/O and metallicities. Two further mechanisms allow a planet to reach sub-solar C/O and metallicity while migrating through the disc to a close-in orbit.
In the first case, the abundances are dominated by the gas accreted. This requires the gas runaway phase of planet growth to happen just inside the water ice line (where the gas-phase C/O ratio can be sub-solar if the dust grains contain a significant amount of refractory carbon) and, therefore, accretes its bulk mass through water vapour-enriched gas and would predict that \myplanet's metallicity is at most only slightly sub-stellar.  
The second case relies on accretion in an environment where the gas is depleted in metallicity by trapping volatile species in a dust trap further out in the disc. 

TrES-4b differs from \myplanet due to its orbital alignment: TrES-4b has an aligned orbit while \myplanet is misaligned. If this difference in orbital alignment reflects an essential role of high-eccentricity migration in \myplanet but not in TrES-4b, then TrES-4b will likely have migrated further through the disc to a close-in orbit and have probably accreted more gas and planetesimals in regions close to the star. This puts an additional caveat on the gas accretion-dominated scenario, in that the water snow line would have to be far enough out that \myplanet formed inside the water snow line while remaining far enough out that high-eccentricity migration remains possible ($\gtrsim 0.6$\,AU, \citealt{2016Munoz}). As it stands, the similarity in composition between TrES-4b and \myplanet may, or may not, point to similar formation histories despite their different migration history, making it difficult to draw strong conclusions by comparing them. However, with the full sample of eight planets in the BOWIE-ALIGN survey, we can make statistically more meaningful comparisons between the populations, hopefully gaining new insights into the formation of gas giants.

\vspace{-0.5cm}

\section{Conclusions}
\label{sec:conclusions}

We present the JWST NIRSpec/G395H transmission of \myplanet, the fourth planet observed as part of the BOWIE-ALIGN programme, and one of four misaligned planets in the eight planet BOWIE-ALIGN sample. We use two independent data reductions and a host of atmospheric retrievals from two different pipelines to demonstrate a consistent interpretation for the planet's atmosphere. We find that \myplanet has a sub-solar metallicity of $0.41^{+0.40}_{-0.19}\times$solar and a sub-solar C/O ratio of $0.29^{+0.16}_{-0.13}$.
We detect \ch{H2O} ($3.1\sigma$) and see evidence for \ch{CO2} ($2.1\sigma$), with consistent abundances between free and equilibrium chemistry retrievals, and evidence for muting of spectral features due to clouds and/or cool limb temperatures.
We do not find evidence for stellar contamination or sulfur chemistry. We see no evidence for a detector offset in the \eureka reduction, while we see a $\sim55$ ppm detector offset between the detectors in the \tiberius reduction, becoming consistent with \eureka with the free offset and therefore matching the observed linear offset between the reductions. We see evidence for a second possible atmospheric composition in a sub-set of the retrievals corresponding to a higher metallicity of $\sim60\times$solar, consistent with the upper limit of a fully-mixed atmosphere from our internal structure modelling. This alternative mode is favoured in only one of 23 performed retrievals, and is absent in most of the performed retrievals, requiring a larger offset ($\sim110$ ppm for \tiberius, $\sim65$ ppm for \eureka) and \ch{SO2}-abundance as a free parameter to adequately fit the observed transmission spectrum. We therefore strongly favour the low metallicity interpretation (discussed in more detail in Section \ref{subsec:interpretation}), \new{ although we note that the metallicity constraints may be weaker depending on the cause of the low observed temperature.}

The retrieved abundances suggest a composition similar to another BOWIE-ALIGN target, TrES-4\,b \citep{meech2025bowie}, which has a metallicity of $0.58^{+0.59}_{-0.32}\times$solar and a C/O of $0.32^{+0.11}_{-0.08/}$, comparing \prt equilibrium chemistry retrievals with identical setups. Obtaining a low metallicity and low C/O ratio deviates from the expected outcome of formation models, being inconsistent with \citet{penzlinBOWIEALIGNHowFormation2024} by $>1\sigma$. Several scenarios could explain this discrepancy, including accretion in the outer disc together with some CO depletion relative to the fiducial models, a composition dominated by gas accreted inside the water ice line, or accretion in a metallicity-depleted environment due to the trapping of volatile species further out in the disc. While \myplanet is misaligned in its orbit, TrES-4\,b is aligned, suggesting that TrES-4\,b may have migrated through disc migration rather than high-eccentricity migration. Given the constraints these migration mechanisms may place on the planet's evolution relative to the ice lines, we could very well be probing different explanations that give rise to similar atmospheric compositions. We therefore await the full BOWIE-ALIGN sample of eight planets to fully explore the formation of these two populations of hot Jupiters and draw statistically meaningful inferences. We do note that \myplanet has a significant difference in metallicity and C/O ratio compared to its fellow misaligned hot Jupiter WASP-15\,b, demonstrating significant diversity can exist in planets with presumed similar formation histories.

\section*{Acknowledgements}

This work is based on observations made with the NASA/ESA/CSA JWST. The data were obtained from the Mikulski Archive for Space Telescopes at the Space Telescope Science Institute, which is operated by the Association of Universities for Research in Astronomy, Inc., under NASA contract NAS 5-03127 for JWST. These observations are associated with program \#3838. This work was inspired by collaboration through the UK-led BOWIE+ collaboration. Support for program JWST-GO-3838 was provided by NASA
through a grant from the Space Telescope Science Institute, which is operated by the
Association of Universities for Research in Astronomy, Inc., under NASA contract NAS
5-03127. 

We thank the reviewer for improving the manuscript with their helpful comments. A.B.C acknowledges studentship support from the UK Science and Technology Facilities Council (STFC). 
C.E.F acknowledges financial support from the European Research Council (ERC) under the European Union’s Horizon 2020 research and innovation program under grant agreement no. 805445. 
J.K acknowledges financial support from Imperial College London through an Imperial College Research Fellowship grant. 
A.P acknowledges funding from the European Union under the European Union's Horizon Europe Research and Innovation Programme 101124282 (EARLYBIRD). Views and opinions expressed are, however, those of the authors only and do not necessarily reflect those of the European Union or the European Research Council. Neither the European Union nor the granting authority can be held responsible for them.
PJW acknowledges support from the UK Science and Technology Facilities Council (STFC) through consolidated grant ST/X001121/1.
R.A.B thanks the Royal Society for their support through a University Research Fellowship.
N.J.M., D.E.S. and M.Z. acknowledge support from a UKRI Future Leaders Fellowship [Grant MR/T040866/1], a Science and Technology Facilities Funding Council Small Award [Grant ST/T000082/1], and the Leverhulme Trust through a research project grant [RPG-2020-82].

\section*{Data Availability}

The raw data are available on the Mikulski Archive for Space Telescopes at the Space Telescope Science Institute, under program number \#3838. The data products associated with this manuscript will also be made available on Zenodo.
 



\bibliographystyle{mnras}
\bibliography{alignment} 

@ARTICLE{husser13,
       author = {{Husser}, T. -O. and {Wende-von Berg}, S. and {Dreizler}, S. and {Homeier}, D. and {Reiners}, A. and {Barman}, T. and {Hauschildt}, P.~H.},
        title = "{A new extensive library of PHOENIX stellar atmospheres and synthetic spectra}",
      journal = {\aap},
         year = 2013,
        month = may,
       volume = {553},
          eid = {A6},
        pages = {A6},
          doi = {10.1051/0004-6361/201219058},
archivePrefix = {arXiv},
       eprint = {1303.5632},
 primaryClass = {astro-ph.SR},
       adsurl = {https://ui.adsabs.harvard.edu/abs/2013A&A...553A...6H}
}

@ARTICLE{stock22,
       author = {{Stock}, Joachim W. and {Kitzmann}, Daniel and {Patzer}, A. Beate C.},
        title = "{FASTCHEM 2 : an improved computer program to determine the gas-phase chemical equilibrium composition for arbitrary element distributions}",
      journal = {\mnras},
         year = 2022,
        month = dec,
       volume = {517},
       number = {3},
        pages = {4070-4080},
          doi = {10.1093/mnras/stac2623},
archivePrefix = {arXiv},
       eprint = {2206.08247},
 primaryClass = {astro-ph.EP},
       adsurl = {https://ui.adsabs.harvard.edu/abs/2022MNRAS.517.4070S}
}

@ARTICLE{stock18,
       author = {{Stock}, Joachim W. and {Kitzmann}, Daniel and {Patzer}, A. Beate C. and {Sedlmayr}, Erwin},
        title = "{FastChem: A computer program for efficient complex chemical equilibrium calculations in the neutral/ionized gas phase with applications to stellar and planetary atmospheres}",
      journal = {\mnras},
         year = 2018,
        month = sep,
       volume = {479},
       number = {1},
        pages = {865-874},
          doi = {10.1093/mnras/sty1531},
archivePrefix = {arXiv},
       eprint = {1804.05010},
 primaryClass = {astro-ph.EP},
       adsurl = {https://ui.adsabs.harvard.edu/abs/2018MNRAS.479..865S}
}

@ARTICLE{grimm15,
       author = {{Grimm}, Simon L. and {Heng}, Kevin},
        title = "{HELIOS-K: An Ultrafast, Open-source Opacity Calculator for Radiative Transfer}",
      journal = {\apj},
         year = 2015,
        month = aug,
       volume = {808},
       number = {2},
          eid = {182},
        pages = {182},
          doi = {10.1088/0004-637X/808/2/182},
archivePrefix = {arXiv},
       eprint = {1503.03806},
 primaryClass = {astro-ph.EP},
       adsurl = {https://ui.adsabs.harvard.edu/abs/2015ApJ...808..182G}
}

@ARTICLE{grimm21,
       author = {{Grimm}, Simon L. and {Malik}, Matej and {Kitzmann}, Daniel and {Guzm{\'a}n-Mesa}, Andrea and {Hoeijmakers}, H. Jens and {Fisher}, Chloe and {Mendon{\c{c}}a}, Jo{\~a}o M. and {Yurchenko}, Sergey N. and {Tennyson}, Jonathan and {Alesina}, Fabien and {Buchschacher}, Nicolas and {Burnier}, Julien and {Segransan}, Damien and {Kurucz}, Robert L. and {Heng}, Kevin},
        title = "{HELIOS-K 2.0 Opacity Calculator and Open-source Opacity Database for Exoplanetary Atmospheres}",
      journal = {\apjs},
         year = 2021,
        month = mar,
       volume = {253},
       number = {1},
          eid = {30},
        pages = {30},
          doi = {10.3847/1538-4365/abd773},
archivePrefix = {arXiv},
       eprint = {2101.02005},
 primaryClass = {astro-ph.EP},
       adsurl = {https://ui.adsabs.harvard.edu/abs/2021ApJS..253...30G}
}

@BOOK{cox2000,
       author = {{Cox}, Arthur N.},
        title = "{Allen's astrophysical quantities}",
         year = 2000,
       adsurl = {https://ui.adsabs.harvard.edu/abs/2000asqu.book.....C}
}

@article{abel2011,
author = {Abel, Martin and Frommhold, Lothar and Li, Xiaoping and Hunt, Katharine L. C.},
title = {Collision-Induced Absorption by H2 Pairs: From Hundreds to Thousands of Kelvin},
journal = {The Journal of Physical Chemistry A},
volume = {115},
number = {25},
pages = {6805-6812},
year = {2011},
doi = {10.1021/jp109441f},
    note ={PMID: 21207941}
}

@ARTICLE{abel2012,
       author = {{Abel}, Martin and {Frommhold}, Lothar and {Li}, Xiaoping and
         {Hunt}, Katharine L.~C.},
        title = "{Infrared absorption by collisional H$_{2}$-He complexes at temperatures up to 9000 K and frequencies from 0 to 20 000 cm$^{-1}$}",
      journal = {\jcp},
         year = "2012",
        month = "Jan",
       volume = {136},
       number = {4},
        pages = {044319-044319},
          doi = {10.1063/1.3676405},
       adsurl = {https://ui.adsabs.harvard.edu/abs/2012JChPh.136d4319A}
}

@ARTICLE{li2015,
       author = {{Li}, Gang and {Gordon}, Iouli E. and {Rothman}, Laurence S. and {Tan}, Yan and {Hu}, Shui-Ming and {Kassi}, Samir and {Campargue}, Alain and {Medvedev}, Emile S.},
        title = "{Rovibrational Line Lists for Nine Isotopologues of the CO Molecule in the X $^{1}${\ensuremath{\Sigma}}$^{+}$ Ground Electronic State}",
      journal = {\apjs},
         year = 2015,
        month = jan,
       volume = {216},
       number = {1},
          eid = {15},
        pages = {15},
          doi = {10.1088/0067-0049/216/1/15},
       adsurl = {https://ui.adsabs.harvard.edu/abs/2015ApJS..216...15L}
}

@ARTICLE{yurchenko14,
       author = {{Yurchenko}, Sergei N. and {Tennyson}, Jonathan},
        title = "{ExoMol line lists - IV. The rotation-vibration spectrum of methane up to 1500 K}",
      journal = {\mnras},
         year = 2014,
        month = may,
       volume = {440},
       number = {2},
        pages = {1649-1661},
          doi = {10.1093/mnras/stu326},
archivePrefix = {arXiv},
       eprint = {1401.4852},
 primaryClass = {astro-ph.EP},
       adsurl = {https://ui.adsabs.harvard.edu/abs/2014MNRAS.440.1649Y}
}

@ARTICLE{kitzmann20,
       author = {{Kitzmann}, Daniel and {Heng}, Kevin and {Oreshenko}, Maria and {Grimm}, Simon L. and {Apai}, D{\'a}niel and {Bowler}, Brendan P. and {Burgasser}, Adam J. and {Marley}, Mark S.},
        title = "{Helios-r2: A New Bayesian, Open-source Retrieval Model for Brown Dwarfs and Exoplanet Atmospheres}",
      journal = {\apj},
         year = 2020,
        month = feb,
       volume = {890},
       number = {2},
          eid = {174},
        pages = {174},
          doi = {10.3847/1538-4357/ab6d71},
archivePrefix = {arXiv},
       eprint = {1910.01070},
 primaryClass = {astro-ph.EP},
       adsurl = {https://ui.adsabs.harvard.edu/abs/2020ApJ...890..174K}
}

@article{MoranStevenson2023,
    title = {{High Tide or Riptide on the Cosmic Shoreline? A Water-Rich Atmosphere or Stellar Contamination for the Warm Super-Earth GJ{\~{}}486b from JWST Observations}},
    year = {2023},
    journal = {The Astrophysical Journal Letters},
    author = {Moran, Sarah E. and Stevenson, Kevin B. and Sing, David K. and MacDonald, Ryan J. and Kirk, James and Lustig-Yaeger, Jacob and Peacock, Sarah and Mayorga, L. C. and Bennett, Katherine A. and L{\'{o}}pez-Morales, Mercedes and May, E. M. and Rustamkulov, Zafar and Valenti, Jeff A. and Adams Redai, Jéa I. and Alam, Munazza K. and Batalha, Natasha E. and Fu, Guangwei and Gonzalez-Quiles, Junellie and Highland, Alicia N. and Kruse, Ethan and Lothringer, Joshua D. and Ortiz Ceballos, Kevin N. and Sotzen, Kristin S. and Wakeford, Hannah R. and Moran, Sarah E. and Stevenson, Kevin B. and Sing, David K. and MacDonald, Ryan J. and Kirk, James and Lustig-Yaeger, Jacob and Peacock, Sarah and Mayorga, L. C. and Bennett, Katherine A. and L{\'{o}}pez-Morales, Mercedes and May, E. M. and Rustamkulov, Zafar and Valenti, Jeff A. and Adams Redai, Jéa I. and Alam, Munazza K. and Batalha, Natasha E. and Fu, Guangwei and Gonzalez-Quiles, Junellie and Highland, Alicia N. and Kruse, Ethan and Lothringer, Joshua D. and Ortiz Ceballos, Kevin N. and Sotzen, Kristin S. and Wakeford, Hannah R.},
    number = {1},
    month = {5},
    pages = {L11},
    volume = {948},
    url = {https://ui.adsabs.harvard.edu/abs/2023arXiv230500868M/abstract},
    doi = {10.48550/ARXIV.2305.00868},
    arxivId = {arXiv:2305.00868}
}

@article{Bell2022Eureka,
    title = {{Eureka!: An End-to-End Pipeline for JWST Time-Series Observations}},
    year = {2022},
    journal = {JOSS},
    author = {Bell, Taylor J and Ahrer, Eva-Maria and Brande, Jonathan and Carter, Aarynn L and Feinstein, Adina D and Guzman Caloca, Giannina and Mansfield, Megan and Zieba, Sebastian and Piaulet, Caroline and Benneke, Björn and Filippazzo, Joseph and May, Erin M and Roy, Pierre-Alexis and Kreidberg, Laura and Stevenson, Kevin B},
    number = {79},
    month = {7},
    pages = {4503},
    volume = {7},
    arxivId = {2207.03585}
}

@ARTICLE{Madhusudhan2014,
       author = {{Madhusudhan}, Nikku and {Amin}, Mustafa A. and {Kennedy}, Grant M.},
        title = "{Toward Chemical Constraints on Hot Jupiter Migration}",
      journal = {\apjl},
         year = 2014,
        month = oct,
       volume = {794},
       number = {1},
          eid = {L12},
        pages = {L12},
          doi = {10.1088/2041-8205/794/1/L12},
archivePrefix = {arXiv},
       eprint = {1408.3668},
 primaryClass = {astro-ph.EP},
       adsurl = {https://ui.adsabs.harvard.edu/abs/2014ApJ...794L..12M}
}

@ARTICLE{Booth2017,
       author = {{Booth}, Richard A. and {Clarke}, Cathie J. and {Madhusudhan}, Nikku and {Ilee}, John D.},
        title = "{Chemical enrichment of giant planets and discs due to pebble drift}",
      journal = {\mnras},
         year = 2017,
        month = aug,
       volume = {469},
       number = {4},
        pages = {3994-4011},
          doi = {10.1093/mnras/stx1103},
archivePrefix = {arXiv},
       eprint = {1705.03305},
 primaryClass = {astro-ph.EP},
       adsurl = {https://ui.adsabs.harvard.edu/abs/2017MNRAS.469.3994B}
}

@ARTICLE{Espinoza2017,
       author = {{Espinoza}, N{\'e}stor and {Fortney}, Jonathan J. and {Miguel}, Yamila and {Thorngren}, Daniel and {Murray-Clay}, Ruth},
        title = "{Metal Enrichment Leads to Low Atmospheric C/O Ratios in Transiting Giant Exoplanets}",
      journal = {\apjl},
         year = 2017,
        month = mar,
       volume = {838},
       number = {1},
          eid = {L9},
        pages = {L9},
          doi = {10.3847/2041-8213/aa65ca},
archivePrefix = {arXiv},
       eprint = {1611.08616},
 primaryClass = {astro-ph.EP},
       adsurl = {https://ui.adsabs.harvard.edu/abs/2017ApJ...838L...9E}
}

@ARTICLE{Oberg2011,
       author = {{{\"O}berg}, Karin I. and {Murray-Clay}, Ruth and {Bergin}, Edwin A.},
        title = "{The Effects of Snowlines on C/O in Planetary Atmospheres}",
      journal = {\apjl},
         year = 2011,
        month = dec,
       volume = {743},
       number = {1},
          eid = {L16},
        pages = {L16},
          doi = {10.1088/2041-8205/743/1/L16},
archivePrefix = {arXiv},
       eprint = {1110.5567},
 primaryClass = {astro-ph.GA},
       adsurl = {https://ui.adsabs.harvard.edu/abs/2011ApJ...743L..16O}
}

@ARTICLE{Rasio1996,
       author = {{Rasio}, Frederic A. and {Ford}, Eric B.},
        title = "{Dynamical instabilities and the formation of extrasolar planetary systems}",
      journal = {Science},
         year = 1996,
        month = nov,
       volume = {274},
        pages = {954-956},
          doi = {10.1126/science.274.5289.954},
       adsurl = {https://ui.adsabs.harvard.edu/abs/1996Sci...274..954R}
}

@ARTICLE{Wu2003,
       author = {{Wu}, Y. and {Murray}, N.},
        title = "{Planet Migration and Binary Companions: The Case of HD 80606b}",
      journal = {\apj},
         year = 2003,
        month = may,
       volume = {589},
       number = {1},
        pages = {605-614},
          doi = {10.1086/374598},
archivePrefix = {arXiv},
       eprint = {astro-ph/0303010},
 primaryClass = {astro-ph},
       adsurl = {https://ui.adsabs.harvard.edu/abs/2003ApJ...589..605W}
}

@ARTICLE{Owen2020,
       author = {{Owen}, James E.},
        title = "{Snow lines can be thermally unstable}",
      journal = {\mnras},
         year = 2020,
        month = jul,
       volume = {495},
       number = {3},
        pages = {3160-3174},
          doi = {10.1093/mnras/staa1309},
archivePrefix = {arXiv},
       eprint = {2005.03665},
 primaryClass = {astro-ph.EP},
       adsurl = {https://ui.adsabs.harvard.edu/abs/2020MNRAS.495.3160O}
}

@ARTICLE{Alderson2023,
       author = {{Alderson}, Lili and {Wakeford}, Hannah R. and {Alam}, Munazza K. and {Batalha}, Natasha E. and {Lothringer}, Joshua D. and {Adams Redai}, Jea and {Barat}, Saugata and {Brande}, Jonathan and {Damiano}, Mario and {Daylan}, Tansu and {Espinoza}, N{\'e}stor and {Flagg}, Laura and {Goyal}, Jayesh M. and {Grant}, David and {Hu}, Renyu and {Inglis}, Julie and {Lee}, Elspeth K.~H. and {Mikal-Evans}, Thomas and {Ramos-Rosado}, Lakeisha and {Roy}, Pierre-Alexis and {Wallack}, Nicole L. and {Batalha}, Natalie M. and {Bean}, Jacob L. and {Benneke}, Bj{\"o}rn and {Berta-Thompson}, Zachory K. and {Carter}, Aarynn L. and {Changeat}, Quentin and {Col{\'o}n}, Knicole D. and {Crossfield}, Ian J.~M. and {D{\'e}sert}, Jean-Michel and {Foreman-Mackey}, Daniel and {Gibson}, Neale P. and {Kreidberg}, Laura and {Line}, Michael R. and {L{\'o}pez-Morales}, Mercedes and {Molaverdikhani}, Karan and {Moran}, Sarah E. and {Morello}, Giuseppe and {Moses}, Julianne I. and {Mukherjee}, Sagnick and {Schlawin}, Everett and {Sing}, David K. and {Stevenson}, Kevin B. and {Taylor}, Jake and {Aggarwal}, Keshav and {Ahrer}, Eva-Maria and {Allen}, Natalie H. and {Barstow}, Joanna K. and {Bell}, Taylor J. and {Blecic}, Jasmina and {Casewell}, Sarah L. and {Chubb}, Katy L. and {Crouzet}, Nicolas and {Cubillos}, Patricio E. and {Decin}, Leen and {Feinstein}, Adina D. and {Fortney}, Joanthan J. and {Harrington}, Joseph and {Heng}, Kevin and {Iro}, Nicolas and {Kempton}, Eliza M. -R. and {Kirk}, James and {Knutson}, Heather A. and {Krick}, Jessica and {Leconte}, J{\'e}r{\'e}my and {Lendl}, Monika and {MacDonald}, Ryan J. and {Mancini}, Luigi and {Mansfield}, Megan and {May}, Erin M. and {Mayne}, Nathan J. and {Miguel}, Yamila and {Nikolov}, Nikolay K. and {Ohno}, Kazumasa and {Palle}, Enric and {Parmentier}, Vivien and {Petit dit de la Roche}, Dominique J.~M. and {Piaulet}, Caroline and {Powell}, Diana and {Rackham}, Benjamin V. and {Redfield}, Seth and {Rogers}, Laura K. and {Rustamkulov}, Zafar and {Tan}, Xianyu and {Tremblin}, P. and {Tsai}, Shang-Min and {Turner}, Jake D. and {de Val-Borro}, Miguel and {Venot}, Olivia and {Welbanks}, Luis and {Wheatley}, Peter J. and {Zhang}, Xi},
        title = "{Early Release Science of the exoplanet WASP-39b with JWST NIRSpec G395H}",
      journal = {\nat},
         year = 2023,
        month = feb,
       volume = {614},
       number = {7949},
        pages = {664-669},
          doi = {10.1038/s41586-022-05591-3},
archivePrefix = {arXiv},
       eprint = {2211.10488},
 primaryClass = {astro-ph.EP},
       adsurl = {https://ui.adsabs.harvard.edu/abs/2023Natur.614..664A}
}

@ARTICLE{Rustamkulov2023,
       author = {{Rustamkulov}, Z. and {Sing}, D.~K. and {Mukherjee}, S. and {May}, E.~M. and {Kirk}, J. and {Schlawin}, E. and {Line}, M.~R. and {Piaulet}, C. and {Carter}, A.~L. and {Batalha}, N.~E. and {Goyal}, J.~M. and {L{\'o}pez-Morales}, M. and {Lothringer}, J.~D. and {MacDonald}, R.~J. and {Moran}, S.~E. and {Stevenson}, K.~B. and {Wakeford}, H.~R. and {Espinoza}, N. and {Bean}, J.~L. and {Batalha}, N.~M. and {Benneke}, B. and {Berta-Thompson}, Z.~K. and {Crossfield}, I.~J.~M. and {Gao}, P. and {Kreidberg}, L. and {Powell}, D.~K. and {Cubillos}, P.~E. and {Gibson}, N.~P. and {Leconte}, J. and {Molaverdikhani}, K. and {Nikolov}, N.~K. and {Parmentier}, V. and {Roy}, P. and {Taylor}, J. and {Turner}, J.~D. and {Wheatley}, P.~J. and {Aggarwal}, K. and {Ahrer}, E. and {Alam}, M.~K. and {Alderson}, L. and {Allen}, N.~H. and {Banerjee}, A. and {Barat}, S. and {Barrado}, D. and {Barstow}, J.~K. and {Bell}, T.~J. and {Blecic}, J. and {Brande}, J. and {Casewell}, S. and {Changeat}, Q. and {Chubb}, K.~L. and {Crouzet}, N. and {Daylan}, T. and {Decin}, L. and {D{\'e}sert}, J. and {Mikal-Evans}, T. and {Feinstein}, A.~D. and {Flagg}, L. and {Fortney}, J.~J. and {Harrington}, J. and {Heng}, K. and {Hong}, Y. and {Hu}, R. and {Iro}, N. and {Kataria}, T. and {Kempton}, E.~M. -R. and {Krick}, J. and {Lendl}, M. and {Lillo-Box}, J. and {Louca}, A. and {Lustig-Yaeger}, J. and {Mancini}, L. and {Mansfield}, M. and {Mayne}, N.~J. and {Miguel}, Y. and {Morello}, G. and {Ohno}, K. and {Palle}, E. and {Petit dit de la Roche}, D.~J.~M. and {Rackham}, B.~V. and {Radica}, M. and {Ramos-Rosado}, L. and {Redfield}, S. and {Rogers}, L.~K. and {Shkolnik}, E.~L. and {Southworth}, J. and {Teske}, J. and {Tremblin}, P. and {Tucker}, G.~S. and {Venot}, O. and {Waalkes}, W.~C. and {Welbanks}, L. and {Zhang}, X. and {Zieba}, S.},
        title = "{Early Release Science of the exoplanet WASP-39b with JWST NIRSpec PRISM}",
      journal = {\nat},
         year = 2023,
        month = feb,
       volume = {614},
       number = {7949},
        pages = {659-663},
          doi = {10.1038/s41586-022-05677-y},
archivePrefix = {arXiv},
       eprint = {2211.10487},
 primaryClass = {astro-ph.EP},
       adsurl = {https://ui.adsabs.harvard.edu/abs/2023Natur.614..659R}
}

@ARTICLE{Ahrer2023,
       author = {{Ahrer}, Eva-Maria and {Stevenson}, Kevin B. and {Mansfield}, Megan and {Moran}, Sarah E. and {Brande}, Jonathan and {Morello}, Giuseppe and {Murray}, Catriona A. and {Nikolov}, Nikolay K. and {Petit dit de la Roche}, Dominique J.~M. and {Schlawin}, Everett and {Wheatley}, Peter J. and {Zieba}, Sebastian and {Batalha}, Natasha E. and {Damiano}, Mario and {Goyal}, Jayesh M. and {Lendl}, Monika and {Lothringer}, Joshua D. and {Mukherjee}, Sagnick and {Ohno}, Kazumasa and {Batalha}, Natalie M. and {Battley}, Matthew P. and {Bean}, Jacob L. and {Beatty}, Thomas G. and {Benneke}, Bj{\"o}rn and {Berta-Thompson}, Zachory K. and {Carter}, Aarynn L. and {Cubillos}, Patricio E. and {Daylan}, Tansu and {Espinoza}, N{\'e}stor and {Gao}, Peter and {Gibson}, Neale P. and {Gill}, Samuel and {Harrington}, Joseph and {Hu}, Renyu and {Kreidberg}, Laura and {Lewis}, Nikole K. and {Line}, Michael R. and {L{\'o}pez-Morales}, Mercedes and {Parmentier}, Vivien and {Powell}, Diana K. and {Sing}, David K. and {Tsai}, Shang-Min and {Wakeford}, Hannah R. and {Welbanks}, Luis and {Alam}, Munazza K. and {Alderson}, Lili and {Allen}, Natalie H. and {Anderson}, David R. and {Barstow}, Joanna K. and {Bayliss}, Daniel and {Bell}, Taylor J. and {Blecic}, Jasmina and {Bryant}, Edward M. and {Burleigh}, Matthew R. and {Carone}, Ludmila and {Casewell}, S.~L. and {Changeat}, Quentin and {Chubb}, Katy L. and {Crossfield}, Ian J.~M. and {Crouzet}, Nicolas and {Decin}, Leen and {D{\'e}sert}, Jean-Michel and {Feinstein}, Adina D. and {Flagg}, Laura and {Fortney}, Jonathan J. and {Gizis}, John E. and {Heng}, Kevin and {Iro}, Nicolas and {Kempton}, Eliza M. -R. and {Kendrew}, Sarah and {Kirk}, James and {Knutson}, Heather A. and {Komacek}, Thaddeus D. and {Lagage}, Pierre-Olivier and {Leconte}, J{\'e}r{\'e}my and {Lustig-Yaeger}, Jacob and {MacDonald}, Ryan J. and {Mancini}, Luigi and {May}, E.~M. and {Mayne}, N.~J. and {Miguel}, Yamila and {Mikal-Evans}, Thomas and {Molaverdikhani}, Karan and {Palle}, Enric and {Piaulet}, Caroline and {Rackham}, Benjamin V. and {Redfield}, Seth and {Rogers}, Laura K. and {Roy}, Pierre-Alexis and {Rustamkulov}, Zafar and {Shkolnik}, Evgenya L. and {Sotzen}, Kristin S. and {Taylor}, Jake and {Tremblin}, P. and {Tucker}, Gregory S. and {Turner}, Jake D. and {de Val-Borro}, Miguel and {Venot}, Olivia and {Zhang}, Xi},
        title = "{Early Release Science of the exoplanet WASP-39b with JWST NIRCam}",
      journal = {\nat},
         year = 2023,
        month = feb,
       volume = {614},
       number = {7949},
        pages = {653-658},
          doi = {10.1038/s41586-022-05590-4},
archivePrefix = {arXiv},
       eprint = {2211.10489},
 primaryClass = {astro-ph.EP},
       adsurl = {https://ui.adsabs.harvard.edu/abs/2023Natur.614..653A}
}

@ARTICLE{JWST2023,
       author = {{JWST Transiting Exoplanet Community Early Release Science Team} and {Ahrer}, Eva-Maria and {Alderson}, Lili and {Batalha}, Natalie M. and {Batalha}, Natasha E. and {Bean}, Jacob L. and {Beatty}, Thomas G. and {Bell}, Taylor J. and {Benneke}, Bj{\"o}rn and {Berta-Thompson}, Zachory K. and {Carter}, Aarynn L. and {Crossfield}, Ian J.~M. and {Espinoza}, N{\'e}stor and {Feinstein}, Adina D. and {Fortney}, Jonathan J. and {Gibson}, Neale P. and {Goyal}, Jayesh M. and {Kempton}, Eliza M. -R. and {Kirk}, James and {Kreidberg}, Laura and {L{\'o}pez-Morales}, Mercedes and {Line}, Michael R. and {Lothringer}, Joshua D. and {Moran}, Sarah E. and {Mukherjee}, Sagnick and {Ohno}, Kazumasa and {Parmentier}, Vivien and {Piaulet}, Caroline and {Rustamkulov}, Zafar and {Schlawin}, Everett and {Sing}, David K. and {Stevenson}, Kevin B. and {Wakeford}, Hannah R. and {Allen}, Natalie H. and {Birkmann}, Stephan M. and {Brande}, Jonathan and {Crouzet}, Nicolas and {Cubillos}, Patricio E. and {Damiano}, Mario and {D{\'e}sert}, Jean-Michel and {Gao}, Peter and {Harrington}, Joseph and {Hu}, Renyu and {Kendrew}, Sarah and {Knutson}, Heather A. and {Lagage}, Pierre-Olivier and {Leconte}, J{\'e}r{\'e}my and {Lendl}, Monika and {MacDonald}, Ryan J. and {May}, E.~M. and {Miguel}, Yamila and {Molaverdikhani}, Karan and {Moses}, Julianne I. and {Murray}, Catriona Anne and {Nehring}, Molly and {Nikolov}, Nikolay K. and {Petit dit de la Roche}, D.~J.~M. and {Radica}, Michael and {Roy}, Pierre-Alexis and {Stassun}, Keivan G. and {Taylor}, Jake and {Waalkes}, William C. and {Wachiraphan}, Patcharapol and {Welbanks}, Luis and {Wheatley}, Peter J. and {Aggarwal}, Keshav and {Alam}, Munazza K. and {Banerjee}, Agnibha and {Barstow}, Joanna K. and {Blecic}, Jasmina and {Casewell}, S.~L. and {Changeat}, Quentin and {Chubb}, K.~L. and {Col{\'o}n}, Knicole D. and {Coulombe}, Louis-Philippe and {Daylan}, Tansu and {de Val-Borro}, Miguel and {Decin}, Leen and {Dos Santos}, Leonardo A. and {Flagg}, Laura and {France}, Kevin and {Fu}, Guangwei and {Garc{\'\i}a Mu{\~n}oz}, A. and {Gizis}, John E. and {Glidden}, Ana and {Grant}, David and {Heng}, Kevin and {Henning}, Thomas and {Hong}, Yu-Cian and {Inglis}, Julie and {Iro}, Nicolas and {Kataria}, Tiffany and {Komacek}, Thaddeus D. and {Krick}, Jessica E. and {Lee}, Elspeth K.~H. and {Lewis}, Nikole K. and {Lillo-Box}, Jorge and {Lustig-Yaeger}, Jacob and {Mancini}, Luigi and {Mandell}, Avi M. and {Mansfield}, Megan and {Marley}, Mark S. and {Mikal-Evans}, Thomas and {Morello}, Giuseppe and {Nixon}, Matthew C. and {Ortiz Ceballos}, Kevin and {Piette}, Anjali A.~A. and {Powell}, Diana and {Rackham}, Benjamin V. and {Ramos-Rosado}, Lakeisha and {Rauscher}, Emily and {Redfield}, Seth and {Rogers}, Laura K. and {Roman}, Michael T. and {Roudier}, Gael M. and {Scarsdale}, Nicholas and {Shkolnik}, Evgenya L. and {Southworth}, John and {Spake}, Jessica J. and {Steinrueck}, Maria E. and {Tan}, Xianyu and {Teske}, Johanna K. and {Tremblin}, Pascal and {Tsai}, Shang-Min and {Tucker}, Gregory S. and {Turner}, Jake D. and {Valenti}, Jeff A. and {Venot}, Olivia and {Waldmann}, Ingo P. and {Wallack}, Nicole L. and {Zhang}, Xi and {Zieba}, Sebastian},
        title = "{Identification of carbon dioxide in an exoplanet atmosphere}",
      journal = {\nat},
         year = 2023,
        month = feb,
       volume = {614},
       number = {7949},
        pages = {649-652},
          doi = {10.1038/s41586-022-05269-w},
archivePrefix = {arXiv},
       eprint = {2208.11692},
 primaryClass = {astro-ph.EP},
       adsurl = {https://ui.adsabs.harvard.edu/abs/2023Natur.614..649J}
}

@ARTICLE{Law2021,
       author = {{Law}, Charles J. and {Loomis}, Ryan A. and {Teague}, Richard and {{\"O}berg}, Karin I. and {Czekala}, Ian and {Andrews}, Sean M. and {Huang}, Jane and {Aikawa}, Yuri and {Alarc{\'o}n}, Felipe and {Bae}, Jaehan and {Bergin}, Edwin A. and {Bergner}, Jennifer B. and {Boehler}, Yann and {Booth}, Alice S. and {Bosman}, Arthur D. and {Calahan}, Jenny K. and {Cataldi}, Gianni and {Cleeves}, L. Ilsedore and {Furuya}, Kenji and {Guzm{\'a}n}, Viviana V. and {Ilee}, John D. and {Le Gal}, Romane and {Liu}, Yao and {Long}, Feng and {M{\'e}nard}, Fran{\c{c}}ois and {Nomura}, Hideko and {Qi}, Chunhua and {Schwarz}, Kamber R. and {Sierra}, Anibal and {Tsukagoshi}, Takashi and {Yamato}, Yoshihide and {van't Hoff}, Merel L.~R. and {Walsh}, Catherine and {Wilner}, David J. and {Zhang}, Ke},
        title = "{Molecules with ALMA at Planet-forming Scales (MAPS). III. Characteristics of Radial Chemical Substructures}",
      journal = {\apjs},
         year = 2021,
        month = nov,
       volume = {257},
       number = {1},
          eid = {3},
        pages = {3},
          doi = {10.3847/1538-4365/ac1434},
archivePrefix = {arXiv},
       eprint = {2109.06210},
 primaryClass = {astro-ph.EP},
       adsurl = {https://ui.adsabs.harvard.edu/abs/2021ApJS..257....3L}
}

@ARTICLE{Schneider2021,
       author = {{Schneider}, Aaron David and {Bitsch}, Bertram},
        title = "{How drifting and evaporating pebbles shape giant planets. I. Heavy element content and atmospheric C/O}",
      journal = {\aap},
         year = 2021,
        month = oct,
       volume = {654},
          eid = {A71},
        pages = {A71},
          doi = {10.1051/0004-6361/202039640},
archivePrefix = {arXiv},
       eprint = {2105.13267},
 primaryClass = {astro-ph.EP},
       adsurl = {https://ui.adsabs.harvard.edu/abs/2021A&A...654A..71S}
}

@ARTICLE{Ivshina2022,
       author = {{Ivshina}, Ekaterina S. and {Winn}, Joshua N.},
        title = "{TESS Transit Timing of Hundreds of Hot Jupiters}",
      journal = {\apjs},
         year = 2022,
        month = apr,
       volume = {259},
       number = {2},
          eid = {62},
        pages = {62},
          doi = {10.3847/1538-4365/ac545b10.48550/arXiv.2202.03401},
archivePrefix = {arXiv},
       eprint = {2202.03401},
 primaryClass = {astro-ph.EP},
       adsurl = {https://ui.adsabs.harvard.edu/abs/2022ApJS..259...62I}
}

@ARTICLE{Morbidelli2016,
       author = {{Morbidelli}, A. and {Bitsch}, B. and {Crida}, A. and {Gounelle}, M. and {Guillot}, T. and {Jacobson}, S. and {Johansen}, A. and {Lambrechts}, M. and {Lega}, E.},
        title = "{Fossilized condensation lines in the Solar System protoplanetary disk}",
      journal = {\icarus},
         year = 2016,
        month = mar,
       volume = {267},
        pages = {368-376},
          doi = {10.1016/j.icarus.2015.11.027},
archivePrefix = {arXiv},
       eprint = {1511.06556},
 primaryClass = {astro-ph.EP},
       adsurl = {https://ui.adsabs.harvard.edu/abs/2016Icar..267..368M}
}

@ARTICLE{Johnson2011,
       author = {{Johnson}, John Asher and {Winn}, J.~N. and {Bakos}, G. {\'A}. and {Hartman}, J.~D. and {Morton}, T.~D. and {Torres}, G. and {Kov{\'a}cs}, G{\'e}za and {Latham}, D.~W. and {Noyes}, R.~W. and {Sato}, B. and {Esquerdo}, G.~A. and {Fischer}, D.~A. and {Marcy}, G.~W. and {Howard}, A.~W. and {Buchhave}, L.~A. and {F{\H{u}}r{\'e}sz}, G. and {Quinn}, S.~N. and {B{\'e}ky}, B. and {Sasselov}, D.~D. and {Stefanik}, R.~P. and {L{\'a}z{\'a}r}, J. and {Papp}, I. and {S{\'a}ri}, P.},
        title = "{HAT-P-30b: A Transiting Hot Jupiter on a Highly Oblique Orbit}",
      journal = {\apj},
         year = 2011,
        month = jul,
       volume = {735},
       number = {1},
          eid = {24},
        pages = {24},
          doi = {10.1088/0004-637X/735/1/24},
archivePrefix = {arXiv},
       eprint = {1103.3825},
 primaryClass = {astro-ph.EP},
       adsurl = {https://ui.adsabs.harvard.edu/abs/2011ApJ...735...24J}
}

@ARTICLE{Kraft1967,
       author = {{Kraft}, Robert P.},
        title = "{Studies of Stellar Rotation. V. The Dependence of Rotation on Age among Solar-Type Stars}",
      journal = {\apj},
         year = 1967,
        month = nov,
       volume = {150},
        pages = {551},
          doi = {10.1086/149359},
       adsurl = {https://ui.adsabs.harvard.edu/abs/1967ApJ...150..551K}
}

@ARTICLE{2016Munoz,
       author = {{Mu{\~n}oz}, Diego J. and {Lai}, Dong and {Liu}, Bin},
        title = "{The formation efficiency of close-in planets via Lidov-Kozai migration: analytic calculations}",
      journal = {\mnras},
         year = 2016,
        month = jul,
       volume = {460},
       number = {1},
        pages = {1086-1093},
          doi = {10.1093/mnras/stw983},
archivePrefix = {arXiv},
       eprint = {1601.05814},
 primaryClass = {astro-ph.EP},
       adsurl = {https://ui.adsabs.harvard.edu/abs/2016MNRAS.460.1086M}
}

@ARTICLE{Blazek2022,
       author = {{Bla{\v{z}}ek}, Martin and {Kab{\'a}th}, Petr and {Piette}, Anjali A.~A. and {Madhusudhan}, Nikku and {Skarka}, Marek and {{\v{S}}ubjak}, J{\'a}n and {Anderson}, David R. and {Boffin}, Henri M.~J. and {C{\'a}ceres}, Claudio C. and {Gibson}, Neale P. and {Hoyer}, Sergio and {Ivanov}, Valentin D. and {Rojo}, Patricio M.},
        title = "{Constraints on TESS albedos for five hot Jupiters}",
      journal = {\mnras},
         year = 2022,
        month = jul,
       volume = {513},
       number = {3},
        pages = {3444-3457},
          doi = {10.1093/mnras/stac992},
archivePrefix = {arXiv},
       eprint = {2204.03327},
 primaryClass = {astro-ph.EP},
       adsurl = {https://ui.adsabs.harvard.edu/abs/2022MNRAS.513.3444B}
}

@ARTICLE{Bonomo2017,
       author = {{Bonomo}, A.~S. and {Desidera}, S. and {Benatti}, S. and {Borsa}, F. and {Crespi}, S. and {Damasso}, M. and {Lanza}, A.~F. and {Sozzetti}, A. and {Lodato}, G. and {Marzari}, F. and {Boccato}, C. and {Claudi}, R.~U. and {Cosentino}, R. and {Covino}, E. and {Gratton}, R. and {Maggio}, A. and {Micela}, G. and {Molinari}, E. and {Pagano}, I. and {Piotto}, G. and {Poretti}, E. and {Smareglia}, R. and {Affer}, L. and {Biazzo}, K. and {Bignamini}, A. and {Esposito}, M. and {Giacobbe}, P. and {H{\'e}brard}, G. and {Malavolta}, L. and {Maldonado}, J. and {Mancini}, L. and {Martinez Fiorenzano}, A. and {Masiero}, S. and {Nascimbeni}, V. and {Pedani}, M. and {Rainer}, M. and {Scandariato}, G.},
        title = "{The GAPS Programme with HARPS-N at TNG . XIV. Investigating giant planet migration history via improved eccentricity and mass determination for 231 transiting planets}",
      journal = {\aap},
         year = 2017,
        month = jun,
       volume = {602},
          eid = {A107},
        pages = {A107},
          doi = {10.1051/0004-6361/201629882},
archivePrefix = {arXiv},
       eprint = {1704.00373},
 primaryClass = {astro-ph.EP},
       adsurl = {https://ui.adsabs.harvard.edu/abs/2017A&A...602A.107B}
}

@ARTICLE{Jakobsen2022,
       author = {{Jakobsen}, P. and {Ferruit}, P. and {Alves de Oliveira}, C. and {Arribas}, S. and {Bagnasco}, G. and {Barho}, R. and {Beck}, T.~L. and {Birkmann}, S. and {B{\"o}ker}, T. and {Bunker}, A.~J. and {Charlot}, S. and {de Jong}, P. and {de Marchi}, G. and {Ehrenwinkler}, R. and {Falcolini}, M. and {Fels}, R. and {Franx}, M. and {Franz}, D. and {Funke}, M. and {Giardino}, G. and {Gnata}, X. and {Holota}, W. and {Honnen}, K. and {Jensen}, P.~L. and {Jentsch}, M. and {Johnson}, T. and {Jollet}, D. and {Karl}, H. and {Kling}, G. and {K{\"o}hler}, J. and {Kolm}, M. -G. and {Kumari}, N. and {Lander}, M.~E. and {Lemke}, R. and {L{\'o}pez-Caniego}, M. and {L{\"u}tzgendorf}, N. and {Maiolino}, R. and {Manjavacas}, E. and {Marston}, A. and {Maschmann}, M. and {Maurer}, R. and {Messerschmidt}, B. and {Moseley}, S.~H. and {Mosner}, P. and {Mott}, D.~B. and {Muzerolle}, J. and {Pirzkal}, N. and {Pittet}, J. -F. and {Plitzke}, A. and {Posselt}, W. and {Rapp}, B. and {Rauscher}, B.~J. and {Rawle}, T. and {Rix}, H. -W. and {R{\"o}del}, A. and {Rumler}, P. and {Sabbi}, E. and {Salvignol}, J. -C. and {Schmid}, T. and {Sirianni}, M. and {Smith}, C. and {Strada}, P. and {te Plate}, M. and {Valenti}, J. and {Wettemann}, T. and {Wiehe}, T. and {Wiesmayer}, M. and {Willott}, C.~J. and {Wright}, R. and {Zeidler}, P. and {Zincke}, C.},
        title = "{The Near-Infrared Spectrograph (NIRSpec) on the James Webb Space Telescope. I. Overview of the instrument and its capabilities}",
      journal = {\aap},
         year = 2022,
        month = may,
       volume = {661},
          eid = {A80},
        pages = {A80},
          doi = {10.1051/0004-6361/202142663},
archivePrefix = {arXiv},
       eprint = {2202.03305},
 primaryClass = {astro-ph.IM},
       adsurl = {https://ui.adsabs.harvard.edu/abs/2022A&A...661A..80J}
}

@ARTICLE{Kirk2017,
       author = {{Kirk}, J. and {Wheatley}, P.~J. and {Louden}, T. and {Doyle}, A.~P. and {Skillen}, I. and {McCormac}, J. and {Irwin}, P.~G.~J. and {Karjalainen}, R.},
        title = "{Rayleigh scattering in the transmission spectrum of HAT-P-18b}",
      journal = {\mnras},
         year = 2017,
        month = jul,
       volume = {468},
       number = {4},
        pages = {3907-3916},
          doi = {10.1093/mnras/stx752},
archivePrefix = {arXiv},
       eprint = {1611.06916},
 primaryClass = {astro-ph.EP},
       adsurl = {https://ui.adsabs.harvard.edu/abs/2017MNRAS.468.3907K}
}

@ARTICLE{Kirk2021,
         author = {{Kirk}, James and {Rackham}, Benjamin V. and {MacDonald}, Ryan J. and {L{\'o}pez-Morales}, Mercedes and {Espinoza}, N{\'e}stor and {Lendl}, Monika and {Wilson}, Jamie and {Osip}, David J. and {Wheatley}, Peter J. and {Skillen}, Ian and {Apai}, D{\'a}niel and {Bixel}, Alex and {Gibson}, Neale P. and {Jord{\'a}n}, Andr{\'e}s and {Lewis}, Nikole K. and {Louden}, Tom and {McGruder}, Chima D. and {Nikolov}, Nikolay and {Rodler}, Florian and {Weaver}, Ian C.},
          title = "{ACCESS and LRG-BEASTS: A Precise New Optical Transmission Spectrum of the Ultrahot Jupiter WASP-103b}",
        journal = {\aj},
           year = 2021,
          month = jul,
         volume = {162},
         number = {1},
            eid = {34},
          pages = {34},
            doi = {10.3847/1538-3881/abfcd2},
  archivePrefix = {arXiv},
         eprint = {2105.00012},
   primaryClass = {astro-ph.EP},
         adsurl = {https://ui.adsabs.harvard.edu/abs/2021AJ....162...34K}
  }

@ARTICLE{Kirk2024,
       author = {{Kirk}, James and {Stevenson}, Kevin B. and {Fu}, Guangwei and {Lustig-Yaeger}, Jacob and {Moran}, Sarah E. and {Peacock}, Sarah and {Alam}, Munazza K. and {Batalha}, Natasha E. and {Bennett}, Katherine A. and {Gonzalez-Quiles}, Junellie and {L{\'o}pez-Morales}, Mercedes and {Lothringer}, Joshua D. and {MacDonald}, Ryan J. and {May}, E.~M. and {Mayorga}, L.~C. and {Rustamkulov}, Zafar and {Sing}, David K. and {Sotzen}, Kristin S. and {Valenti}, Jeff A. and {Wakeford}, Hannah R.},
        title = "{JWST/NIRCam Transmission Spectroscopy of the Nearby Sub-Earth GJ 341b}",
      journal = {\aj},
         year = 2024,
        month = mar,
       volume = {167},
       number = {3},
          eid = {90},
        pages = {90},
          doi = {10.3847/1538-3881/ad19df},
archivePrefix = {arXiv},
       eprint = {2401.06043},
 primaryClass = {astro-ph.EP},
       adsurl = {https://ui.adsabs.harvard.edu/abs/2024AJ....167...90K}
}

@article{Magic2015,
  adsurl = {http://adsabs.harvard.edu/abs/2015A%26A...573A..90M},
  archiveprefix = {arXiv},
  author = {{Magic}, Z. and {Chiavassa}, A. and {Collet}, R. and {Asplund}, M.},
  doi = {10.1051/0004-6361/201423804},
  eid = {A90},
  eprint = {1403.3487},
  journal = {\aap},
  month = jan,
  pages = {A90},
  primaryclass = {astro-ph.SR},
  title = {{The Stagger-grid: A grid of 3D stellar atmosphere models. IV. Limb darkening coefficients}},
  volume = 573,
  year = 2015}

@ARTICLE{batman,
       author = {{Kreidberg}, Laura},
        title = "{batman: BAsic Transit Model cAlculatioN in Python}",
      journal = {\pasp},
         year = 2015,
        month = nov,
       volume = {127},
       number = {957},
        pages = {1161},
          doi = {10.1086/683602},
archivePrefix = {arXiv},
       eprint = {1507.08285},
 primaryClass = {astro-ph.EP},
       adsurl = {https://ui.adsabs.harvard.edu/abs/2015PASP..127.1161K}
}

@ARTICLE{Tsai2023,
       author = {{Tsai}, Shang-Min and {Lee}, Elspeth K.~H. and {Powell}, Diana and {Gao}, Peter and {Zhang}, Xi and {Moses}, Julianne and {H{\'e}brard}, Eric and {Venot}, Olivia and {Parmentier}, Vivien and {Jordan}, Sean and {Hu}, Renyu and {Alam}, Munazza K. and {Alderson}, Lili and {Batalha}, Natalie M. and {Bean}, Jacob L. and {Benneke}, Bj{\"o}rn and {Bierson}, Carver J. and {Brady}, Ryan P. and {Carone}, Ludmila and {Carter}, Aarynn L. and {Chubb}, Katy L. and {Inglis}, Julie and {Leconte}, J{\'e}r{\'e}my and {Line}, Michael and {L{\'o}pez-Morales}, Mercedes and {Miguel}, Yamila and {Molaverdikhani}, Karan and {Rustamkulov}, Zafar and {Sing}, David K. and {Stevenson}, Kevin B. and {Wakeford}, Hannah R. and {Yang}, Jeehyun and {Aggarwal}, Keshav and {Baeyens}, Robin and {Barat}, Saugata and {de Val-Borro}, Miguel and {Daylan}, Tansu and {Fortney}, Jonathan J. and {France}, Kevin and {Goyal}, Jayesh M. and {Grant}, David and {Kirk}, James and {Kreidberg}, Laura and {Louca}, Amy and {Moran}, Sarah E. and {Mukherjee}, Sagnick and {Nasedkin}, Evert and {Ohno}, Kazumasa and {Rackham}, Benjamin V. and {Redfield}, Seth and {Taylor}, Jake and {Tremblin}, Pascal and {Visscher}, Channon and {Wallack}, Nicole L. and {Welbanks}, Luis and {Youngblood}, Allison and {Ahrer}, Eva-Maria and {Batalha}, Natasha E. and {Behr}, Patrick and {Berta-Thompson}, Zachory K. and {Blecic}, Jasmina and {Casewell}, S.~L. and {Crossfield}, Ian J.~M. and {Crouzet}, Nicolas and {Cubillos}, Patricio E. and {Decin}, Leen and {D{\'e}sert}, Jean-Michel and {Feinstein}, Adina D. and {Gibson}, Neale P. and {Harrington}, Joseph and {Heng}, Kevin and {Henning}, Thomas and {Kempton}, Eliza M. -R. and {Krick}, Jessica and {Lagage}, Pierre-Olivier and {Lendl}, Monika and {Lothringer}, Joshua D. and {Mansfield}, Megan and {Mayne}, N.~J. and {Mikal-Evans}, Thomas and {Palle}, Enric and {Schlawin}, Everett and {Shorttle}, Oliver and {Wheatley}, Peter J. and {Yurchenko}, Sergei N.},
        title = "{Photochemically produced SO$_{2}$ in the atmosphere of WASP-39b}",
      journal = {\nat},
         year = 2023,
        month = may,
       volume = {617},
       number = {7961},
        pages = {483-487},
          doi = {10.1038/s41586-023-05902-2},
archivePrefix = {arXiv},
       eprint = {2211.10490},
 primaryClass = {astro-ph.EP},
       adsurl = {https://ui.adsabs.harvard.edu/abs/2023Natur.617..483T}
}

@article{molliere2019petitradtrans,
  title={petitRADTRANS-A Python radiative transfer package for exoplanet characterization and retrieval},
  author={Molli{\`e}re, P and Wardenier, JP and Van Boekel, R and Henning, Th and Molaverdikhani, K and Snellen, IAG},
  journal={Astronomy \& Astrophysics},
  volume={627},
  pages={A67},
  year={2019},
  publisher={EDP Sciences}
}

@article{Nasedkin2024, doi = {10.21105/joss.05875}, url = {https://doi.org/10.21105/joss.05875}, year = {2024}, publisher = {The Open Journal}, volume = {9}, number = {96}, pages = {5875}, author = {Evert Nasedkin and Paul Mollière and Doriann Blain}, title = {Atmospheric Retrievals with petitRADTRANS}, journal = {Journal of Open Source Software} }

@ARTICLE{Helling2016,
       author = {{Helling}, Ch. and {Lee}, E. and {Dobbs-Dixon}, I. and {Mayne}, N. and {Amundsen}, D.~S. and {Khaimova}, J. and {Unger}, A.~A. and {Manners}, J. and {Acreman}, D. and {Smith}, C.},
        title = "{The mineral clouds on HD 209458b and HD 189733b}",
      journal = {\mnras},
         year = 2016,
        month = jul,
       volume = {460},
       number = {1},
        pages = {855-883},
          doi = {10.1093/mnras/stw662},
archivePrefix = {arXiv},
       eprint = {1603.04022},
 primaryClass = {astro-ph.EP},
       adsurl = {https://ui.adsabs.harvard.edu/abs/2016MNRAS.460..855H}
}

@article{Muller2024,
archivePrefix = {arXiv},
arxivId = {2403.16273},
author = {M{\"{u}}ller, Simon and Helled, Ravit},
doi = {10.3847/1538-4357/ad3738},
eprint = {2403.16273},
issn = {0004-637X},
journal = {The Astrophysical Journal},
number = {1},
pages = {7},
publisher = {IOP Publishing},
title = {{Can Jupiter's Atmospheric Metallicity Be Different from the Deep Interior?}},
url = {http://dx.doi.org/10.3847/1538-4357/ad3738},
volume = {967},
year = {2024}
}

@article{Asplund2009,
archivePrefix = {arXiv},
arxivId = {0909.0948},
author = {Asplund, Martin and Grevesse, Nicolas and Sauval, A. Jacques and Scott, Pat},
doi = {10.1146/annurev.astro.46.060407.145222},
eprint = {0909.0948},
issn = {00664146},
journal = {Annual Review of Astronomy and Astrophysics},
pages = {481--522},
title = {{The Chemical Composition of the Sun}},
volume = {47},
year = {2009}
}

@article{kirk2025bowie,
  title={BOWIE-ALIGN: JWST reveals hints of planetesimal accretion and complex sulphur chemistry in the atmosphere of the misaligned hot Jupiter WASP-15b},
  author={Kirk, James and Ahrer, Eva-Maria and Claringbold, Alastair B and Zamyatina, Maria and Fisher, Chloe and McCormack, Mason and Panwar, Vatsal and Powell, Diana and Taylor, Jake and Thorngren, Daniel P and others},
  journal={Monthly Notices of the Royal Astronomical Society},
  pages={staf208},
  year={2025},
  publisher={Oxford University Press}
}

@article{carter2024benchmark,
  title={A benchmark JWST near-infrared spectrum for the exoplanet WASP-39 b},
  author={Carter, AL and May, EM and Espinoza, N and Welbanks, L and Ahrer, E and Alderson, L and Brahm, R and Feinstein, AD and Grant, D and Line, M and others},
  journal={Nature Astronomy},
  volume={8},
  number={8},
  pages={1008--1019},
  year={2024},
  publisher={Nature Publishing Group UK London}
}

@article{meech2025bowie,
  title={BOWIE-ALIGN: Sub-stellar metallicity and carbon depletion in the aligned TrES-4b with JWST NIRSpec transmission spectroscopy},
  author={Meech, Annabella and Claringbold, Alastair B and Ahrer, Eva-Maria and Kirk, James and L{\'o}pez-Morales, Mercedes and Taylor, Jake and Booth, Richard A and Penzlin, Anna BT and Alderson, Lili and Christie, Duncan A and others},
  journal={Monthly Notices of the Royal Astronomical Society},
  pages={staf530},
  year={2025},
  publisher={Oxford University Press}
}

@article{rothman2010hitemp,
  title={HITEMP, the high-temperature molecular spectroscopic database},
  author={Rothman, Laurence S and Gordon, IE and Barber, RJ and Dothe, H and Gamache, Robert R and Goldman, A and Perevalov, VI and Tashkun, SA and Tennyson, J},
  journal={Journal of Quantitative Spectroscopy and Radiative Transfer},
  volume={111},
  number={15},
  pages={2139--2150},
  year={2010},
  publisher={Elsevier}
}

@article{polyansky2018exomol,
  title={ExoMol molecular line lists XXX: a complete high-accuracy line list for water},
  author={Polyansky, Oleg L and Kyuberis, Aleksandra A and Zobov, Nikolai F and Tennyson, Jonathan and Yurchenko, Sergei N and Lodi, Lorenzo},
  journal={Monthly Notices of the Royal Astronomical Society},
  volume={480},
  number={2},
  pages={2597--2608},
  year={2018},
  publisher={Oxford University Press}
}

@article{yurchenko2017hybrid,
  title={A hybrid line list for CH4 and hot methane continuum},
  author={Yurchenko, Sergei N and Amundsen, David S and Tennyson, Jonathan and Waldmann, Ingo P},
  journal={Astronomy \& Astrophysics},
  volume={605},
  pages={A95},
  year={2017},
  publisher={EDP Sciences}
}

@article{azzam2016exomol,
  title={ExoMol molecular line lists--XVI. The rotation--vibration spectrum of hot H2S},
  author={Azzam, Ala'a AA and Tennyson, Jonathan and Yurchenko, Sergei N and Naumenko, Olga V},
  journal={Monthly Notices of the Royal Astronomical Society},
  volume={460},
  number={4},
  pages={4063--4074},
  year={2016},
  publisher={Oxford University Press}
}

@article{barber2014exomol,
  title={ExoMol line lists--III. An improved hot rotation-vibration line list for HCN and HNC},
  author={Barber, RJ and Strange, JK and Hill, C and Polyansky, OL and Mellau, G Ch and Yurchenko, SN and Tennyson, Jonathan},
  journal={Monthly Notices of the Royal Astronomical Society},
  volume={437},
  number={2},
  pages={1828--1835},
  year={2014},
  publisher={Oxford University Press}
}

@article{yurchenko2020exomol,
  title={ExoMol line lists--XXXIX. Ro-vibrational molecular line list for CO2},
  author={Yurchenko, SN and Mellor, Thomas M and Freedman, Richard S and Tennyson, J},
  journal={Monthly Notices of the Royal Astronomical Society},
  volume={496},
  number={4},
  pages={5282--5291},
  year={2020},
  publisher={Oxford University Press}
}

@article{underwood2016exomol,
  title={ExoMol molecular line lists--XIV. The rotation--vibration spectrum of hot SO2},
  author={Underwood, Daniel S and Tennyson, Jonathan and Yurchenko, Sergei N and Huang, Xinchuan and Schwenke, David W and Lee, Timothy J and Clausen, S{\o}nnik and Fateev, Alexander},
  journal={Monthly Notices of the Royal Astronomical Society},
  volume={459},
  number={4},
  pages={3890--3899},
  year={2016},
  publisher={Oxford University Press}
}

@article{brady2024exomol,
  title={ExoMol line lists--LVI. The SO line list, MARVEL analysis of experimental transition data and refinement of the spectroscopic model},
  author={Brady, Ryan P and Yurchenko, Sergei N and Tennyson, Jonathan and Kim, Gap-Sue},
  journal={Monthly Notices of the Royal Astronomical Society},
  volume={527},
  number={3},
  pages={6675--6690},
  year={2024},
  publisher={Oxford University Press}
}

@article{coles2019exomol,
  title={ExoMol molecular line lists--XXXV. A rotation-vibration line list for hot ammonia},
  author={Coles, Phillip A and Yurchenko, Sergei N and Tennyson, Jonathan},
  journal={Monthly Notices of the Royal Astronomical Society},
  volume={490},
  number={4},
  pages={4638--4647},
  year={2019},
  publisher={Oxford University Press}
}

@article{Skilling2004,
  title={Nested sampling},
  author={Skilling, John},
  journal={Bayesian inference and maximum entropy methods in science and engineering},
  volume={735},
  pages={395--405},
  year={2004}
}

@ARTICLE{Feroz2008,
       author = {{Feroz}, F. and {Hobson}, M.~P.},
        title = "{Multimodal nested sampling: an efficient and robust alternative to Markov Chain Monte Carlo methods for astronomical data analyses}",
      journal = {\mnras},
         year = 2008,
        month = feb,
       volume = {384},
       number = {2},
        pages = {449-463},
          doi = {10.1111/j.1365-2966.2007.12353.x},
archivePrefix = {arXiv},
       eprint = {0704.3704},
 primaryClass = {astro-ph},
       adsurl = {https://ui.adsabs.harvard.edu/abs/2008MNRAS.384..449F}
}

@article{buchner2014x,
  title={X-ray spectral modelling of the AGN obscuring region in the CDFS: Bayesian model selection and catalogue},
  author={Buchner, J and Georgakakis, A and Nandra, K and Hsu, L and Rangel, C and Brightman, M and Merloni, A and Salvato, M and Donley, J and Kocevski, D},
  journal={Astronomy \& Astrophysics},
  volume={564},
  pages={A125},
  year={2014},
  publisher={EDP Sciences}
}

@ARTICLE{Feinstein2023,
       author = {{Feinstein}, Adina D. and {Radica}, Michael and {Welbanks}, Luis and {Murray}, Catriona Anne and {Ohno}, Kazumasa and {Coulombe}, Louis-Philippe and {Espinoza}, N{\'e}stor and {Bean}, Jacob L. and {Teske}, Johanna K. and {Benneke}, Bj{\"o}rn and {Line}, Michael R. and {Rustamkulov}, Zafar and {Saba}, Arianna and {Tsiaras}, Angelos and {Barstow}, Joanna K. and {Fortney}, Jonathan J. and {Gao}, Peter and {Knutson}, Heather A. and {MacDonald}, Ryan J. and {Mikal-Evans}, Thomas and {Rackham}, Benjamin V. and {Taylor}, Jake and {Parmentier}, Vivien and {Batalha}, Natalie M. and {Berta-Thompson}, Zachory K. and {Carter}, Aarynn L. and {Changeat}, Quentin and {dos Santos}, Leonardo A. and {Gibson}, Neale P. and {Goyal}, Jayesh M. and {Kreidberg}, Laura and {L{\'o}pez-Morales}, Mercedes and {Lothringer}, Joshua D. and {Miguel}, Yamila and {Molaverdikhani}, Karan and {Moran}, Sarah E. and {Morello}, Giuseppe and {Mukherjee}, Sagnick and {Sing}, David K. and {Stevenson}, Kevin B. and {Wakeford}, Hannah R. and {Ahrer}, Eva-Maria and {Alam}, Munazza K. and {Alderson}, Lili and {Allen}, Natalie H. and {Batalha}, Natasha E. and {Bell}, Taylor J. and {Blecic}, Jasmina and {Brande}, Jonathan and {Caceres}, Claudio and {Casewell}, S.~L. and {Chubb}, Katy L. and {Crossfield}, Ian J.~M. and {Crouzet}, Nicolas and {Cubillos}, Patricio E. and {Decin}, Leen and {D{\'e}sert}, Jean-Michel and {Harrington}, Joseph and {Heng}, Kevin and {Henning}, Thomas and {Iro}, Nicolas and {Kempton}, Eliza M. -R. and {Kendrew}, Sarah and {Kirk}, James and {Krick}, Jessica and {Lagage}, Pierre-Olivier and {Lendl}, Monika and {Mancini}, Luigi and {Mansfield}, Megan and {May}, E.~M. and {Mayne}, N.~J. and {Nikolov}, Nikolay K. and {Palle}, Enric and {Petit dit de la Roche}, Dominique J.~M. and {Piaulet}, Caroline and {Powell}, Diana and {Redfield}, Seth and {Rogers}, Laura K. and {Roman}, Michael T. and {Roy}, Pierre-Alexis and {Nixon}, Matthew C. and {Schlawin}, Everett and {Tan}, Xianyu and {Tremblin}, P. and {Turner}, Jake D. and {Venot}, Olivia and {Waalkes}, William C. and {Wheatley}, Peter J. and {Zhang}, Xi},
        title = "{Early Release Science of the exoplanet WASP-39b with JWST NIRISS}",
      journal = {\nat},
         year = 2023,
        month = feb,
       volume = {614},
       number = {7949},
        pages = {670-675},
          doi = {10.1038/s41586-022-05674-1},
archivePrefix = {arXiv},
       eprint = {2211.10493},
 primaryClass = {astro-ph.EP},
       adsurl = {https://ui.adsabs.harvard.edu/abs/2023Natur.614..670F}
}

@article{molliere2022interpreting,
  title={Interpreting the atmospheric composition of exoplanets: sensitivity to planet formation assumptions},
  author={Molli{\`e}re, Paul and Molyarova, Tamara and Bitsch, Bertram and Henning, Thomas and Schneider, Aaron and Kreidberg, Laura and Eistrup, Christian and Burn, Remo and Nasedkin, Evert and Semenov, Dmitry and others},
  journal={The Astrophysical Journal},
  volume={934},
  number={1},
  pages={74},
  year={2022},
  publisher={IOP Publishing}
}

@article{ligterink2024mind,
  title={Mind the trap-Non-negligible effect of volatile trapping in ice on C/O ratios in protoplanetary disks and exoplanetary atmospheres},
  author={Ligterink, Niels FW and Kipfer, KA and Gavino, S},
  journal={Astronomy \& Astrophysics},
  volume={687},
  pages={A224},
  year={2024},
  publisher={EDP Sciences}
}

@article{law2021molecules,
  title={Molecules with ALMA at planet-forming scales (MAPS). III. Characteristics of radial chemical substructures},
  author={Law, Charles J and Loomis, Ryan A and Teague, Richard and {\"O}berg, Karin I and Czekala, Ian and Andrews, Sean M and Huang, Jane and Aikawa, Yuri and Alarc{\'o}n, Felipe and Bae, Jaehan and others},
  journal={The Astrophysical Journal Supplement Series},
  volume={257},
  number={1},
  pages={3},
  year={2021},
  publisher={IOP Publishing}
}

@article{penzlinBOWIEALIGNHowFormation2024,
  title = {{{BOWIE-ALIGN}}: How Formation and Migration Histories of Giant Planets Impact Atmospheric Compositions},
  shorttitle = {{{BOWIE-ALIGN}}},
  author = {Penzlin, Anna B. T. and Booth, Richard A. and Kirk, James and Owen, James E. and Ahrer, E. and Christie, Duncan A. and Claringbold, Alastair B. and {Esparza-Borges}, Emma and {L{\'o}pez-Morales}, M. and Mayne, N. J. and McCormack, Mason and Meech, Annabella and Panwar, Vatsal and Powell, Diana and Sergeev, Denis E. and Taylor, Jake and Wheatley, Peter J. and Zamyatina, Maria},
  year = {2024},
  month = nov,
  journal = {Monthly Notices of the Royal Astronomical Society},
  volume = {535},
  pages = {171--186},
  publisher = {OUP},
  issn = {0035-8711},
  doi = {10.1093/mnras/stae2362},
  urldate = {2025-01-13}
}

@article{kirkBOWIEALIGNJWSTComparative2024,
  title = {{{BOWIE-ALIGN}}: {{A JWST}} Comparative Survey of Aligned versus Misaligned Hot {{Jupiters}} to Test the Dependence of Atmospheric Composition on Migration History},
  shorttitle = {{{BOWIE-ALIGN}}},
  author = {Kirk, James and Ahrer, Eva-Maria and Penzlin, Anna B. T. and Owen, James E. and Booth, Richard A. and Alderson, Lili and Christie, Duncan A. and Claringbold, Alastair B. and {Esparza-Borges}, Emma and Fisher, Chloe E. and {L{\'o}pez-Morales}, Mercedes and Mayne, N. J. and McCormack, Mason and Meech, Annabella and Panwar, Vatsal and Powell, Diana and Sergeev, Denis E. and Taylor, Jake and Tsai, Shang-Min and Valentine, Daniel and Wakeford, Hannah R. and Wheatley, Peter J. and Zamyatina, Maria},
  year = {2024},
  month = jan,
  journal = {RAS Techniques and Instruments},
  volume = {3},
  pages = {691--704},
  doi = {10.1093/rasti/rzae043},
  urldate = {2025-01-13}
}

@article{beyerKraftBreakSharply2024,
  title = {The {{Kraft Break Sharply Divides Low Mass}} and {{Intermediate Mass Stars}}},
  author = {Beyer, Alexa and White, Russel},
  year = {2024},
  month = feb,
  journal = {American Astronomical Society Meeting Abstracts},
  volume = {243},
  pages = {205.08},
  urldate = {2024-12-24},
  langid = {english}
}

@article{albrechtOBLIQUITIESHOTJUPITER2012,
  title = {{{OBLIQUITIES OF HOT JUPITER HOST STARS}}: {{EVIDENCE FOR TIDAL INTERACTIONS AND PRIMORDIAL MISALIGNMENTS}}*},
  shorttitle = {{{OBLIQUITIES OF HOT JUPITER HOST STARS}}},
  author = {Albrecht, Simon and Winn, Joshua N. and Johnson, John A. and Howard, Andrew W. and Marcy, Geoffrey W. and Butler, R. Paul and Arriagada, Pamela and Crane, Jeffrey D. and Shectman, Stephen A. and Thompson, Ian B. and Hirano, Teruyuki and Bakos, Gaspar and Hartman, Joel D.},
  year = {2012},
  month = aug,
  journal = {The Astrophysical Journal},
  volume = {757},
  number = {1},
  pages = {18},
  publisher = {The American Astronomical Society},
  issn = {0004-637X},
  doi = {10.1088/0004-637X/757/1/18},
  urldate = {2024-12-24},
  langid = {english}
}

@article{winnHotStarsHot2010,
  title = {Hot {{Stars}} with {{Hot Jupiters Have High Obliquities}}},
  author = {Winn, Joshua N. and Fabrycky, Daniel and Albrecht, Simon and Johnson, John Asher},
  year = {2010},
  month = aug,
  journal = {The Astrophysical Journal},
  volume = {718},
  number = {2},
  pages = {L145-L149},
  issn = {0004-637X},
  doi = {10.1088/2041-8205/718/2/L145},
  urldate = {2024-12-24},
  langid = {english}
}

@ARTICLE{Welbanks2022,
       author = {{Welbanks}, Luis and {Madhusudhan}, Nikku},
        title = "{On Atmospheric Retrievals of Exoplanets with Inhomogeneous Terminators}",
      journal = {\apj},
         year = 2022,
        month = jul,
       volume = {933},
       number = {1},
          eid = {79},
        pages = {79},
          doi = {10.3847/1538-4357/ac6df1},
archivePrefix = {arXiv},
       eprint = {2112.09125},
 primaryClass = {astro-ph.EP},
       adsurl = {https://ui.adsabs.harvard.edu/abs/2022ApJ...933...79W}
}

@ARTICLE{MacDonald2020,
       author = {{MacDonald}, Ryan J. and {Goyal}, Jayesh M. and {Lewis}, Nikole K.},
        title = "{Why Is it So Cold in Here? Explaining the Cold Temperatures Retrieved from Transmission Spectra of Exoplanet Atmospheres}",
      journal = {\apjl},
         year = 2020,
        month = apr,
       volume = {893},
       number = {2},
          eid = {L43},
        pages = {L43},
          doi = {10.3847/2041-8213/ab8238},
archivePrefix = {arXiv},
       eprint = {2003.11548},
 primaryClass = {astro-ph.EP},
       adsurl = {https://ui.adsabs.harvard.edu/abs/2020ApJ...893L..43M}
}

@article{horne1986optimal,
  title={An optimal extraction algorithm for CCD spectroscopy.},
  author={Horne, Keith},
  journal={Publications of the Astronomical Society of the Pacific},
  volume={98},
  number={604},
  pages={609},
  year={1986},
  publisher={IOP Publishing}
}

@article{foreman2013emcee,
  title={emcee: the MCMC hammer},
  author={Foreman-Mackey, Daniel and Hogg, David W and Lang, Dustin and Goodman, Jonathan},
  journal={Publications of the Astronomical Society of the Pacific},
  volume={125},
  number={925},
  pages={306},
  year={2013},
  publisher={IOP Publishing}
}

@article{grant2024exotic,
  title={ExoTiC-LD: thirty seconds to stellar limb-darkening coefficients},
  author={Grant, David and Wakeford, Hannah R},
  journal={arXiv preprint arXiv:2408.10341},
  year={2024}
}

@article{cegla2023exploring,
  title={Exploring the stellar surface phenomena of WASP-52 and HAT-P-30 with ESPRESSO},
  author={Cegla, HM and Roguet-Kern, N and Lendl, M and Akinsanmi, B and McCormac, J and Oshagh, M and Wheatley, PJ and Chen, G and Allart, R and Mortier, A and others},
  journal={Astronomy \& Astrophysics},
  volume={674},
  pages={A174},
  year={2023},
  publisher={EDP Sciences}
}

@article{bergin2024carbon,
  title={The Carbon Isotopic Ratio and Planet Formation},
  author={Bergin, Edwin A and Bosman, Arthur and Teague, Richard and Calahan, Jenny and Willacy, Karen and Cleeves, L Ilsedore and Schwarz, Kamber and Zhang, Ke and Bruderer, Simon},
  journal={The Astrophysical Journal},
  volume={965},
  number={2},
  pages={147},
  year={2024},
  publisher={IOP Publishing}
}

@article{bosman2019,
  title={Probing planet formation and disk substructures in the inner disk of Herbig Ae stars with CO rovibrational emission},
  author={Bosman, Arthur D and Banzatti, Andrea and Bruderer, Simon and Tielens, Alexander GGM and Blake, Geoffrey A and van Dishoeck, Ewine F},
  journal={Astronomy \& Astrophysics},
  volume={631},
  pages={A133},
  year={2019},
  publisher={EDP Sciences}
}

@article{alderson2024jwst,
  title={JWST COMPASS: NIRSpec/G395H Transmission Observations of the Super-Earth TOI-836b},
  author={Alderson, Lili and Batalha, Natasha E and Wakeford, Hannah R and Wallack, Nicole L and Aguichine, Artyom and Teske, Johanna and Redai, Jea Adams and Alam, Munazza K and Batalha, Natalie M and Gao, Peter and others},
  journal={The Astronomical Journal},
  volume={167},
  number={5},
  pages={216},
  year={2024},
  publisher={IOP Publishing}
}

@article{wallack2024jwst,
  title={JWST COMPASS: A NIRSpec/G395H Transmission Spectrum of the Sub-Neptune TOI-836c},
  author={Wallack, Nicole L and Batalha, Natasha E and Alderson, Lili and Scarsdale, Nicholas and Redai, Jea I Adams and Aguichine, Artyom and Alam, Munazza K and Gao, Peter and Wolfgang, Angie and Batalha, Natalie M and others},
  journal={The Astronomical Journal},
  volume={168},
  number={2},
  pages={77},
  year={2024},
  publisher={IOP Publishing}
}

@article{teske2025jwst,
  title={JWST COMPASS: NIRSpec/G395H Transmission Observations of TOI-776 c, a 2 R⊕ M Dwarf Planet},
  author={Teske, Johanna and Batalha, Natasha E and Wallack, Nicole L and Kirk, James and Wogan, Nicholas F and Gordon, Tyler A and Alam, Munazza K and Aguichine, Artyom and Wolfgang, Angie and Wakeford, Hannah R and others},
  journal={The Astronomical Journal},
  volume={169},
  number={5},
  pages={249},
  year={2025},
  publisher={IOP Publishing}
}

@article{xue2024jwst,
  title={JWST Transmission Spectroscopy of HD 209458b: A Supersolar Metallicity, a Very Low C/O, and No Evidence of CH4, HCN, or C2H2},
  author={Xue, Qiao and Bean, Jacob L and Zhang, Michael and Welbanks, Luis and Lunine, Jonathan and August, Prune},
  journal={The Astrophysical Journal Letters},
  volume={963},
  number={1},
  pages={L5},
  year={2024},
  publisher={IOP Publishing}
}

@article{jeffreys1939theory,
  title={Theory of Probability},
  author={Jeffreys, Harold},
  journal={Theory of Probability},
  year={1939}
}

@ARTICLE{Thorngren2016,
       author = {{Thorngren}, Daniel P. and {Fortney}, Jonathan J. and {Murray-Clay}, Ruth A. and {Lopez}, Eric D.},
        title = "{The Mass-Metallicity Relation for Giant Planets}",
      journal = {\apj},
         year = 2016,
        month = nov,
       volume = {831},
       number = {1},
          eid = {64},
        pages = {64},
          doi = {10.3847/0004-637X/831/1/64},
archivePrefix = {arXiv},
       eprint = {1511.07854},
 primaryClass = {astro-ph.EP},
       adsurl = {https://ui.adsabs.harvard.edu/abs/2016ApJ...831...64T}
}

@ARTICLE{Thorngren2018,
       author = {{Thorngren}, Daniel P. and {Fortney}, Jonathan J.},
        title = "{Bayesian Analysis of Hot-Jupiter Radius Anomalies: Evidence for Ohmic Dissipation?}",
      journal = {\aj},
         year = 2018,
        month = may,
       volume = {155},
       number = {5},
          eid = {214},
        pages = {214},
          doi = {10.3847/1538-3881/aaba13},
archivePrefix = {arXiv},
       eprint = {1709.04539},
 primaryClass = {astro-ph.EP},
       adsurl = {https://ui.adsabs.harvard.edu/abs/2018AJ....155..214T}
}

@ARTICLE{Thorngren2019,
       author = {{Thorngren}, Daniel and {Fortney}, Jonathan J.},
        title = "{Connecting Giant Planet Atmosphere and Interior Modeling: Constraints on Atmospheric Metal Enrichment}",
      journal = {\apjl},
         year = 2019,
        month = apr,
       volume = {874},
       number = {2},
          eid = {L31},
        pages = {L31},
          doi = {10.3847/2041-8213/ab1137},
archivePrefix = {arXiv},
       eprint = {1811.11859},
 primaryClass = {astro-ph.EP},
       adsurl = {https://ui.adsabs.harvard.edu/abs/2019ApJ...874L..31T}
}

@article{Helled2017,
  title={The fuzziness of giant planets’ cores},
  author={Helled, Ravit and Stevenson, David},
  journal={The Astrophysical Journal Letters},
  volume={840},
  number={1},
  pages={L4},
  year={2017},
  publisher={IOP Publishing}
}

@article{Bloot2023,
  title={Exoplanet interior retrievals: core masses and metallicities from atmospheric abundances},
  author={Bloot, Sanne and Miguel, Yamila and Bazot, Micha{\"e}l and Howard, Saburo},
  journal={Monthly Notices of the Royal Astronomical Society},
  volume={523},
  number={4},
  pages={6282--6292},
  year={2023},
  publisher={Oxford University Press}
}

@article{Fortney2007,
  title={Planetary radii across five orders of magnitude in mass and stellar insolation: application to transits},
  author={Fortney, Jonathan J and Marley, Mark S and Barnes, Jason W},
  journal={The Astrophysical Journal},
  volume={659},
  number={2},
  pages={1661},
  year={2007},
  publisher={IOP Publishing}
}

@article{borysow2001high,
  title={High-temperature (1000--7000 K) collision-induced absorption of H2 pairs computed from the first principles, with application to cool and dense stellar atmospheres},
  author={Borysow, Aleksandra and J{\o}rgensen, Uffe G and Fu, Yi},
  journal={Journal of Quantitative Spectroscopy and Radiative Transfer},
  volume={68},
  number={3},
  pages={235--255},
  year={2001},
  publisher={Elsevier}
}

@article{borysow2002collision,
  title={Collision-induced absorption coefficients of H2 pairs at temperatures from 60 K to 1000 K},
  author={Borysow, A},
  journal={Astronomy \& Astrophysics},
  volume={390},
  number={2},
  pages={779--782},
  year={2002},
  publisher={EDP Sciences}
}

@article{borysow1988collison,
  title={Collison-induced rototranslational absorption spectra of H2-He pairs at temperatures from 40 to 3000 K},
  author={Borysow, Jacek and Frommhold, Lothar and Birnbaum, George},
  journal={Astrophysical Journal, Part 1 (ISSN 0004-637X), vol. 326, March 1, 1988, p. 509-515. NASA-supported research.},
  volume={326},
  pages={509--515},
  year={1988}
}

@article{dalgarno1962rayleigh,
  title={Rayleigh Scattering by Molecular Hydrogen.},
  author={Dalgarno, A and Williams, DA},
  journal={Astrophysical Journal, vol. 136, p. 690-692},
  volume={136},
  pages={690--692},
  year={1962}
}

@article{chan1965refractive,
  title={The refractive index of helium},
  author={Chan, YM and Dalgarno, A},
  journal={Proceedings of the Physical Society},
  volume={85},
  number={2},
  pages={227},
  year={1965},
  publisher={IOP Publishing}
}

@article{ahrer2025bowie,
  title={BOWIE-ALIGN: weak spectral features in KELT-7b’s JWST NIRSpec/G395H transmission spectrum imply a high cloud deck or a low-metallicity atmosphere},
  author={Ahrer, Eva-Maria and Fairman, Charlotte and Kirk, James and Wakeford, Hannah R and Barstow, Joanna K and Penzlin, Anna BT and Alderson, Lili and Booth, Richard A and Christie, Duncan A and Claringbold, Alastair B and others},
  journal={Monthly Notices of the Royal Astronomical Society},
  volume={543},
  number={3},
  pages={2442--2462},
  year={2025},
  publisher={Oxford University Press}
}




 \appendix

 \section{Full atmospheric retrieval results}

We provide additional information from the atmospheric retrieval analysis described in Section \ref{sec:retrievals}. Table \ref{tab:all_retrievals} summarizes the results from all 23 retrievals performed, while the corner plot of posterior probability distributions for the $R=400$ equilibrium results from \prt and \bear are presented in Fig. \ref{fig:prt_eq_cornerplot_R100} and Fig. \ref{fig:BeAR_cornerplot_R400_chemeq} respectively, and the $R=400$ free chemistry results are presented in Fig. \ref{fig:prt_free_cornerplot_R100} and Fig. \ref{fig:BeAR_cornerplot_R400}.

{

\renewcommand{\arraystretch}{1.3}
\begin{landscape}

\begin{table}
    
    \centering
    
    \caption{Retrieval results from \prt and \bear, as described in Section \ref{sec:retrievals}. For posteriors with  reasonably well-formed peaks in their posteriors we provide the median and $1\sigma$ confidence intervals, for unconstrained parameters, we provide the $2\sigma$ upper limits. Parameters which weren't fitted in the retrieval model are marked with `--'. }
    
    \label{tab:all_retrievals}
    \begin{adjustbox}{width=1.3\textwidth}
             
        \begin{tabular}{l c c c c c c c c c c c c c c} \toprule
             Input spectrum & $\ln{Z}$ & $R_\mathrm{P}$ (\Rjup) & $\log g$ (cgs) & $T_{\mathrm{iso}}$ (K) & $\log P_{\mathrm{cloud}}$ (bar) & C/O & [M/H] & \ch{H2O} & \ch{CO2} & \ch{CO} & \ch{CH4} & \ch{SO2} & \ch{SO} & Offset (ppm)\\
             \midrule\texttt{petitRADTRANS} & & & & & & & & & & & & & &  \\
             \textit{Equilibrium Chemistry:} & & & & & & & & & & & & & &  \\
             \eureka $R=100$ & $476.5\pm0.3$ & $1.337\pm0.003$ & $2.96\pm0.02$ & $1025{\substack{+80\\-59}}$ & $-1.20{\substack{+2.17\\-0.83}}$ & $0.29{\substack{+0.18\\-0.13}}$ & $-0.47{\substack{+0.42\\-0.38}}$ & - & - & - & - & - & - & $8\pm31$ \\
             \eureka $R=400$ & $1797.8\pm0.1$ & $1.336\pm0.004$ & $2.96\pm0.02$ & $1026{\substack{+110\\-68}}$ & $-1.41{\substack{+2.26\\-0.97}}$ & $0.28{\substack{+0.17\\-0.12}}$ & $-0.42{\substack{+0.48\\-0.38}}$ & - & - & - & - & - & - & $3\pm30$\\
             \tiberius $R=100$ & $474.5\pm0.1$ & $1.338\pm0.005$ & $2.96\pm0.02$ & $1014{\substack{+106\\-73}}$ & $-1.63{\substack{+2.31\\-0.80}}$ & $0.30{\substack{+0.19\\-0.14}}$ & $-0.34{\substack{+0.53\\-0.44}}$ & - & - & - & - & - & - & $-46\pm36$\\
             \tiberius $R=400$ & $1803.9\pm0.1$ & $1.337\pm0.006$ & $2.96\pm0.02$ & $1009{\substack{+107\\-69}}$ & $-1.69{\substack{+2.33\\-0.66}}$ & $0.30{\substack{+0.19\\-0.13}}$ & $-0.29{\substack{+0.63\\-0.40}}$ & - & - & - & - & - & - & $-56\pm35$\\
             \textit{Eq. Chemistry, no offset:} & & & & & & & & & & & & & &  \\
             \eureka $R=400$$^\dagger$ & $1799.3\pm0.1$ & $1.336\pm0.003$ & $2.97\pm0.02$ & $1021{\substack{+94\\-62}}$ & $-1.51{\substack{+2.31\\-0.52}}$ & $0.29{\substack{+0.16\\-0.13}}$ & $-0.39{\substack{+0.30\\-0.27}}$ & - & - & - & - & - & - & - \\
             \tiberius $R=400$ & $1796.7\pm0.1$ & $1.340\pm0.003$ & $2.96\pm0.02$ &  $992{\substack{+59\\-48}}$ & $0.07{\substack{+1.34\\-1.25}}$  & $0.18{\substack{+0.11\\-0.06}}$  & $-0.83{\substack{+0.26\\-0.32}}$ & - & - & - & - & - & -  & - \\
             \textit{Free Chemistry:} & & & & & & & & & & & & & &  \\
             \eureka $R=100$ & $477.0\pm0.4$ & $1.341\pm0.004$ & $2.96\pm0.02$ & $668{\substack{+110\\-86}}$ & $0.61{\substack{+0.92\\-0.99}}$ & - & - & $-3.2\pm0.9$ & $-8.9{\substack{+1.3\\-2.6}}$ & $<-3.0$ & $<-5.9$ & $<-5.1$ & $-5.0{\substack{+1.0\\-2.9}}$ & $-5\pm24$\\
             \eureka $R=400$ & $1798.0\pm0.1$ & $1.342\pm0.004$ & $2.96\pm0.02$ & $666{\substack{+115\\-81}}$ & $0.50{\substack{+0.95\\-0.96}}$ & - & - & $-3.3\pm0.8$ & $-8.1{\substack{+0.9\\-1.8}}$ & $<-3.6$ & $<-6.1$ & $<-5.3$ & $-5.4{\substack{+1.2\\-2.5}}$ & $-11\pm24$ \\
             \tiberius $R=100$ & $474.6\pm0.1$ & $1.345\pm0.005$ & $2.96\pm0.02$ & $672{\substack{+126\\-90}}$ & $0.43{\substack{+0.97\\-1.02}}$ & - & - & $-3.5{\substack{+0.9\\-0.8}}$ & $-8.7{\substack{+1.2\\-2.4}}$ & $<-3.1$ & $<-4.1$ & $<-5.9$ & $-5.2{\substack{+1.0\\-2.9}}$ & $-55\pm23$ \\
             \tiberius $R=400$ & $1805.6\pm0.1$ & $1.344\pm0.004$ & $2.96\pm0.02$ & $650{\substack{+117\\-77}}$ &  $0.52{\substack{+0.96\\-1.03}}$ & - & - & $-3.4\pm0.8$ & $-7.9{\substack{+0.9\\-1.4}}$  & $<-2.7$ & $<-5.9$ & $<-5.5$ & $<-2.6$ & $-66\pm24$ \\
             \textit{Free Chemistry, no SO:} & & & & & & & & & & & & & &  \\
             \eureka $R=400$ & $1797.6\pm0.1$ & $1.342\pm0.005$ & $2.96\pm0.02$ & $686{\substack{+123\\-86}}$ & $0.56{\substack{+0.89\\-0.97}}$ & - & - & $-3.3{\substack{+0.9\\-0.8}}$  & $-8.0{\substack{+0.9\\-1.7}}$ & $<-2.5$ & $<-5.7$ & $<-5.6$ & - & $-3\pm24$\\
             \textit{Hybrid Chemistry:} & & & & & & & & & & & & & & \\
             \eureka $R=100$ & $475.4\pm0.1$ & $1.336\pm0.006$ & $2.96\pm0.02$ & $1008{\substack{+74\\-68}}$ & $-0.25{\substack{+1.49\\-1.46}}$ & $0.29{\substack{+0.16\\-0.12}}$ & $-0.36{\substack{+2.09\\-0.41}}$ & - & - & - & - & $-6.4{\substack{+0.6\\-3.6}}$ & $<-3.3$ & $-10\pm44$ \\
             \eureka $R=400$ & $1796.9\pm0.1$ & $1.333\pm0.007$ & $2.96\pm0.02$ & $1019{\substack{+124\\-78}}$ & $-1.70{\substack{+1.78\\-0.68}}$ & $0.31{\substack{+0.18\\-0.13}}$ & $-0.05{\substack{+1.96\\-0.52}}$ & - & - & - & - & $<-3.3$ & $<-2.9$ & $-32\pm40$ \\
             \tiberius $R=100$ & $474.1\pm0.2$ & $1.336\pm0.006$ & $2.96\pm0.02$ & $981{\substack{+93\\-76}}$ & $-0.91{\substack{+1.92\\-1.43}}$ & $0.28{\substack{+0.16\\-0.13}}$ & $-0.16{\substack{+2.07\\-0.49}}$ & - & - & - & - & $<-3.8$ & $<-1.8$ & $-70\pm47$ \\
             \tiberius $R=400$ & $1803.5\pm0.1$ & $1.329\pm0.006$ & $2.96\pm0.02$ & $989{\substack{+123\\-95}}$ & $-1.66{\substack{+2.32\\-0.71}}$ & $0.34{\substack{+0.16\\-0.14}}$ & $1.77{\substack{+0.31\\-2.07}}$ & - & - & - & - & $<-3.8$ & $<-1.6$ & $-116\pm39$ \\
             \midrule\texttt{BeAR} & & & & & & & & & & & & & & \\
             \textit{Equilibrium Chemistry:} & & & & & & & & & & & & & & \\
             \eureka $R=100$ & $-354.8\pm0.1$ & $1.256\pm0.003$ & $2.96\pm0.02$ & $998{\substack{+58 \\ -51}}$ & $-1.75{\substack{+0.41 \\ -0.24}}$ & $0.3{\substack{+0.19 \\ -0.13}}$ & $-0.5{\substack{+0.32 \\ -0.26}}$ & - & - & - & - & - & - & $-6\pm26$ \\
             \eureka $R=400$ & $-1522.3\pm0.1$ & $1.257\pm0.003$ & $2.96\pm0.03$ & $985{\substack{+58 \\ -46}}$ & $-1.76{\substack{+0.44 \\ -0.26}}$ & $0.26{\substack{+0.17 \\ -0.11}}$ & $-0.52{\substack{+0.31 \\ -0.25}}$ & - & - & - & - & - & - & $-4{\substack{+23 \\ -24}}$ \\
             \tiberius $R=100$ & $-356.4\pm0.1$ & $1.258{\substack{+0.003 \\ -0.004}}$ & $2.96{\substack{+0.03 \\ -0.02}}$ & $974{\substack{+62 \\ -48}}$ & $-1.84{\substack{+0.36 \\ -0.26}}$ & $0.26{\substack{+0.19 \\ -0.11}}$ & $-0.5{\substack{+0.3 \\ -0.27}}$ & - & - & - & - & - & - & $-49{\substack{+27 \\ -23}}$ \\
             \tiberius $R=400$ & $-1524.7\pm0.1$ & $1.259\pm0.003$ & $2.96\pm0.02$ & $969{\substack{+56 \\ -47}}$ & $-1.9{\substack{+0.27 \\ -0.24}}$ & $0.25{\substack{+0.18 \\ -0.11}}$ & $-0.46{\substack{+0.29 \\ -0.27}}$ & - & - & - & - & - & - & $-60{\substack{+23 \\ -21}}$ \\
             \textit{Free Chemistry:} & & & & & & & & & & & & & & \\
             \eureka $R=100$ & $-354.8\pm0.1$ & $1.278{\substack{+0.013 \\ -0.005}}$ & $2.96\pm0.03$ & $688{\substack{+245 \\ -85}}$ & $-0.63{\substack{+0.4 \\ -0.55}}$ & - & - & $-2.9{\substack{+1.0 \\ -1.7}}$ & $-7.6{\substack{+6.9 \\ -2.0}}$ & $<-2.3$ & $<-5.7$ & $<-5.5$ & - & $-3{\substack{+25 \\ -88}}$ \\
             \eureka $R=400$ & $-1521.2\pm0.1$ & $1.277\pm0.004$ & $2.96\pm0.03$ & $674{\substack{+93 \\ -76}}$ & $-0.66{\substack{+0.43 \\ -0.57}}$ & - & - & $-2.7\pm1.0$ & $-7.5{\substack{+1.2 \\ -1.4}}$ & $<-3.1$ & $<-6.0$ & $<-5.9$ & - & $-1{\substack{+23 \\ -24}}$ \\
             \tiberius $R=100$ & $-355.5\pm0.1$ & $1.279{\substack{+0.004 \\ -0.005}}$ & $2.96\pm0.03$ & $659{\substack{+105 \\ -77}}$ & $-0.7{\substack{+0.46 \\ -0.56}}$ & - & - & $-2.9{\substack{+0.9 \\ -1.0}}$ & $-8.2{\substack{+1.3 \\ -1.9}}$ & $<-3.8$ & $<-6.1$ & $<-6.1$ & - & $-41{\substack{+22 \\ -24}}$ \\
             \tiberius $R=400$ & $-1522.6\pm0.1$ & $1.28{\substack{+0.003 \\ -0.004}}$ & $2.96\pm0.03$ & $644{\substack{+92 \\ -70}}$ & $-0.66{\substack{+0.43 \\ -0.57}}$ & - & - & $-2.9\pm1.0$ & $-7.4{\substack{+1.0 \\ -1.2}}$ & $<-3.8$ & $<-6.2$ & $<-6.0$ & - & $-58{\substack{+22 \\ -20}}$ \\
              \bottomrule
              \multicolumn{12}{l}{$^\dagger$ Favoured interpretation.}
              
        \end{tabular}
    \end{adjustbox}
\end{table}
\end{landscape}
}
\onecolumn
\begin{figure*}
    \centering
    \settototalheight{\dimen0}{\includegraphics[width=\textwidth]{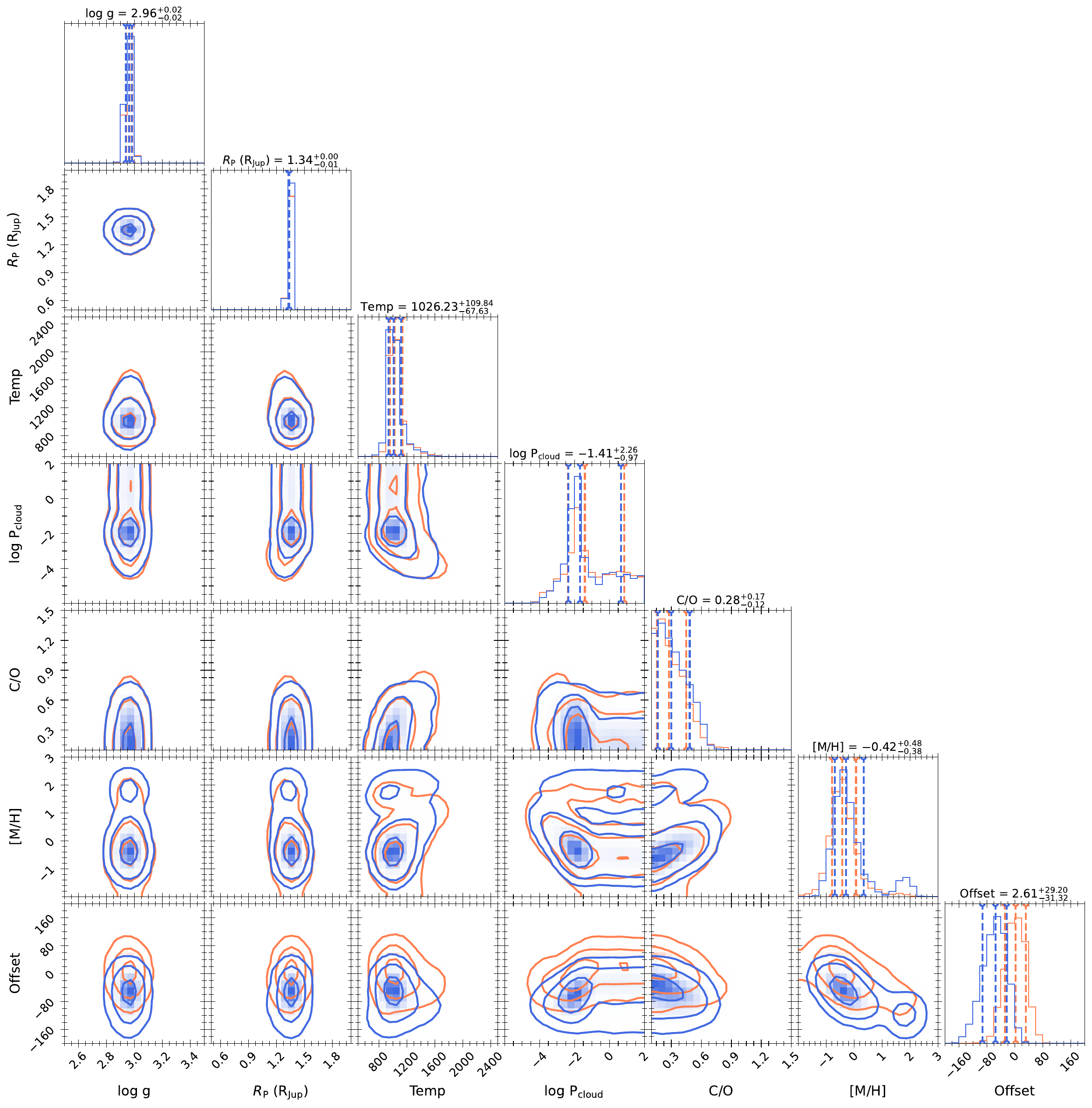}}
    \includegraphics[width=\textwidth]{figures/prt_eq_R400_corner.pdf}%
    \llap{\raisebox{\dimen0-4cm}{%
    \includegraphics[height=4cm]{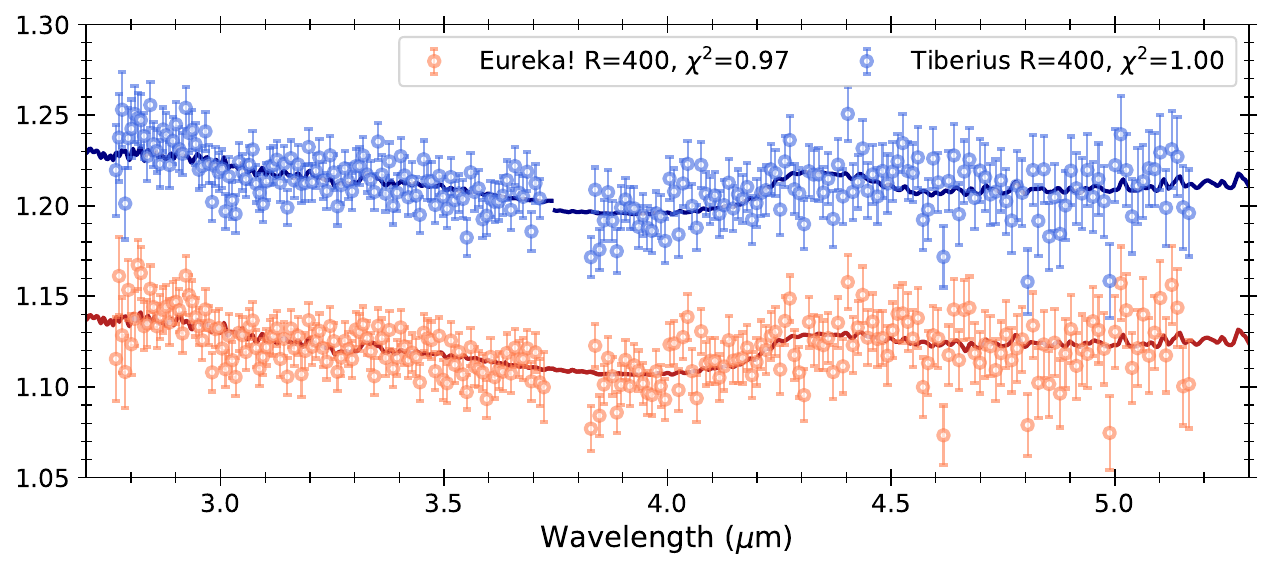}%
    }}
    \caption{Corner plot showing the posterior probability distributions from the \prt equilibrium chemistry retrievals on the $R=400$ transmission spectra from \tiberius (blue) and \eureka (orange, from which the numbers are derived). The top right panel displays the best-fit models from each retrieval, offset by 500 ppm, with the reduced $\chi^2$ of the fit indicated in the legend.}
    \label{fig:prt_eq_cornerplot_R100}
\end{figure*}

\begin{figure*}
    \centering
    \settototalheight{\dimen0}{\includegraphics[width=\textwidth]{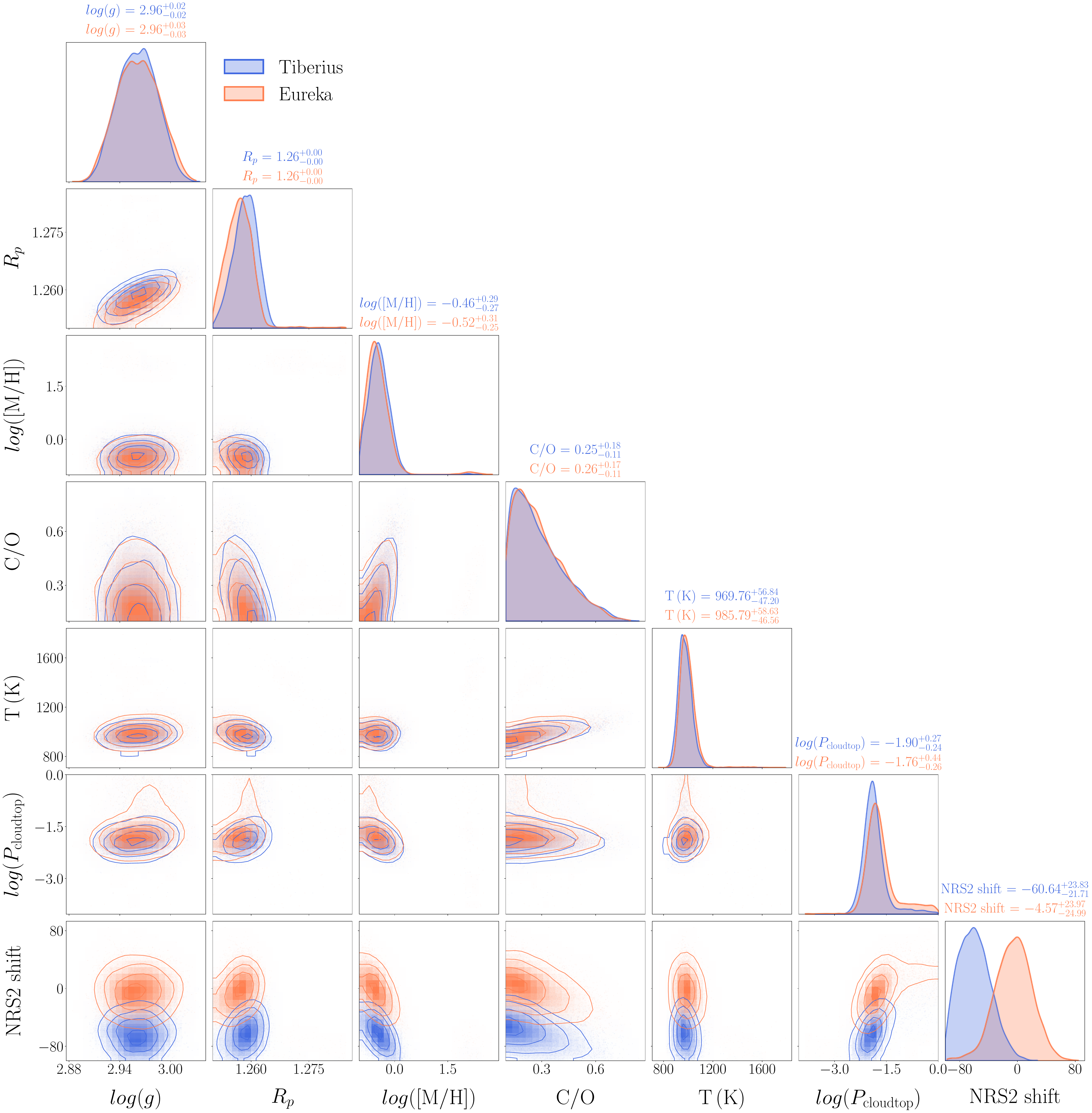}}
    \includegraphics[width=\textwidth]{figures/BeAR_retrieval_comparison_R400_chemeq.pdf}%
    \llap{\raisebox{\dimen0-6cm}{%
    \includegraphics[height=6cm]{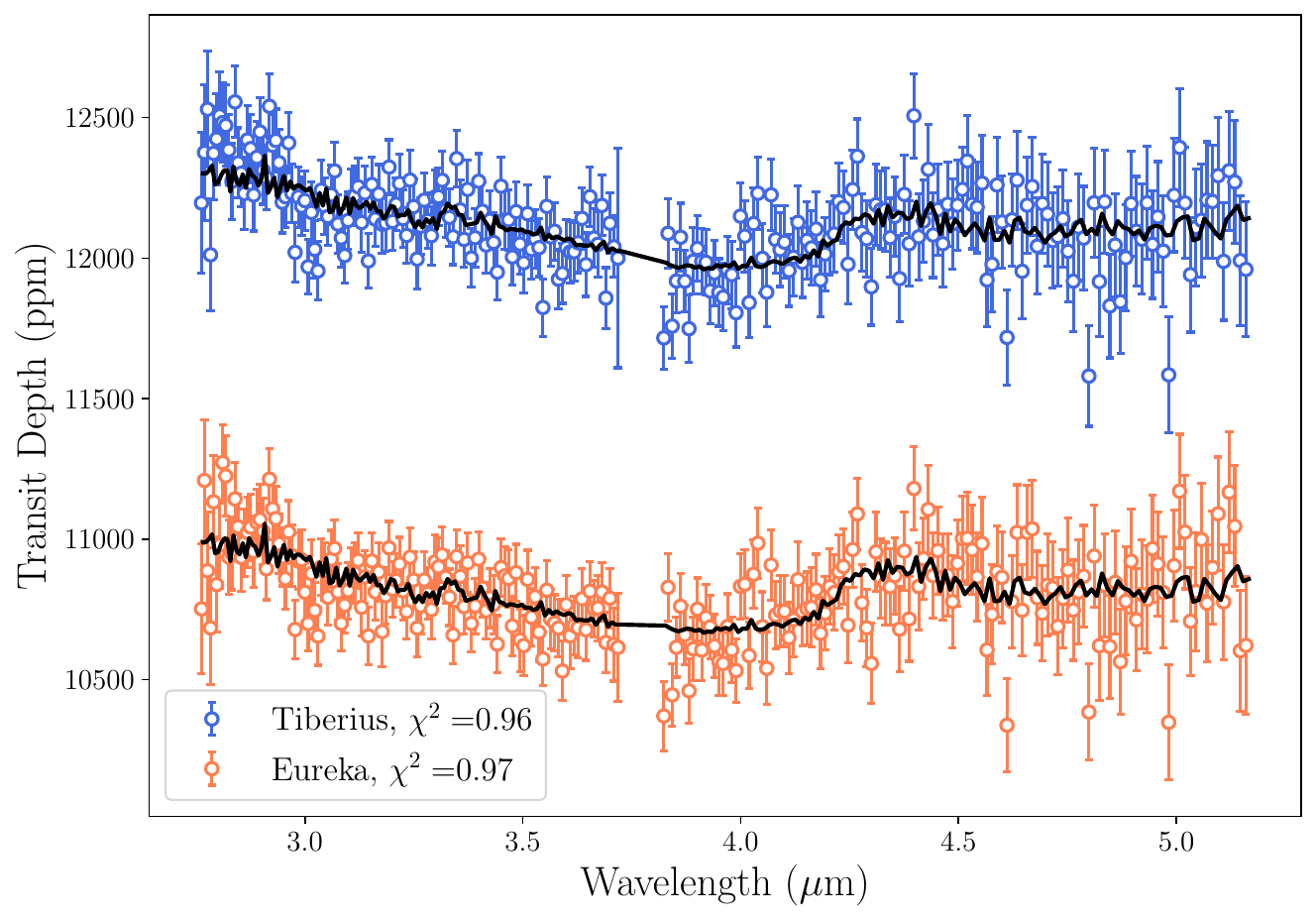}%
    }}
    \caption{Cornerplot showing the posteriors from the BeAR chemical equilibrium retrievals on the \texttt{Tiberius} (blue) and \texttt{Eureka!} (orange) reductions of HAT-P-30 b at $R=400$. The top right insert shows the best-fit models for the \texttt{Tiberius} (blue) and \texttt{Eureka!} (orange) reductions. The \texttt{Eureka!} spectrum is offset by 1300 ppm for visualisation purposes. The legend in the bottom left indicates the reduced $\chi^2$ values for each of the fits.}
    \label{fig:BeAR_cornerplot_R400_chemeq}
\end{figure*}

\begin{figure*}
    \centering
    \settototalheight{\dimen0}{\includegraphics[width=\textwidth]{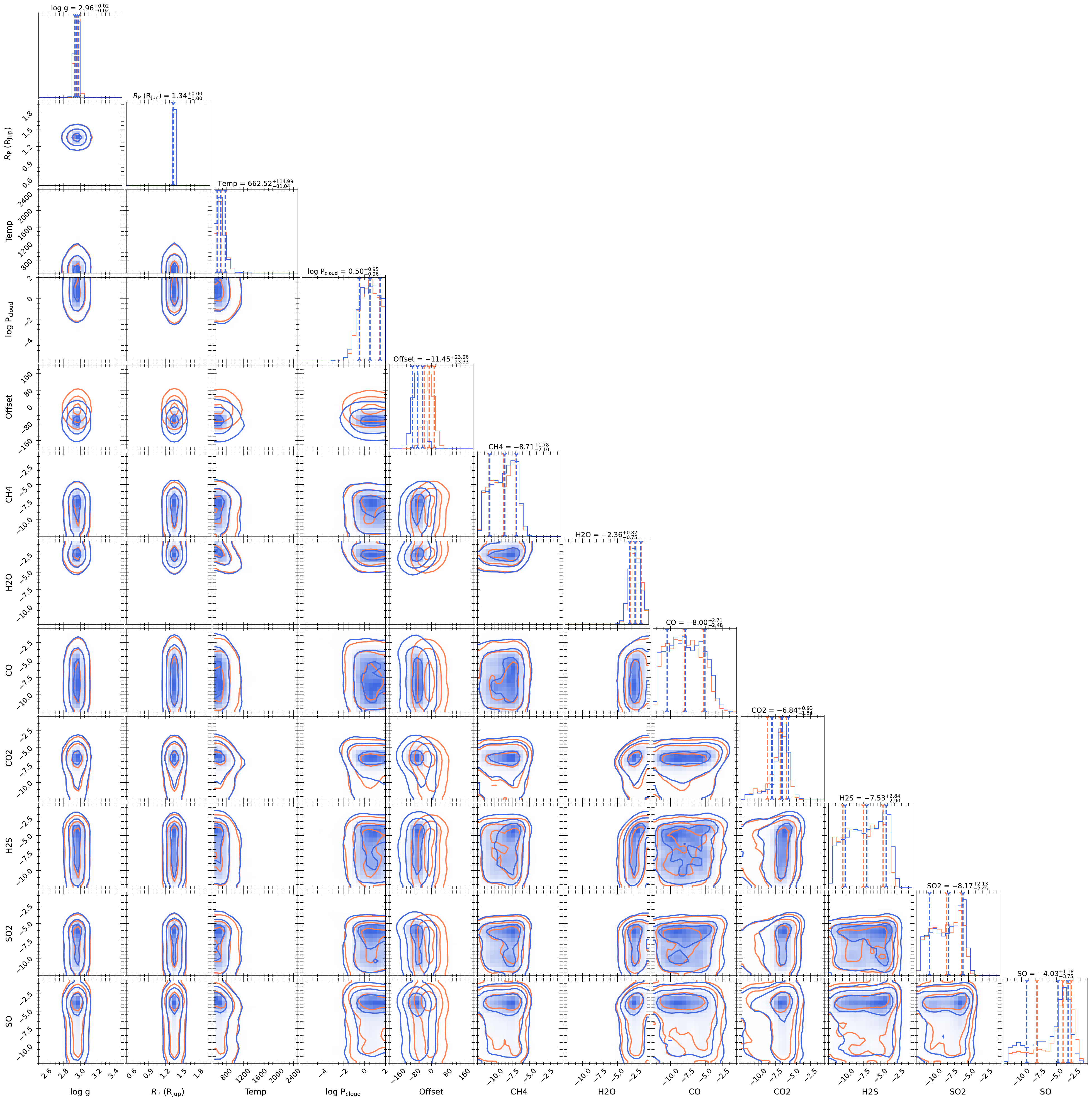}}
    \includegraphics[width=\textwidth]{figures/prt_free_R400_corner.pdf}%
    \llap{\raisebox{\dimen0-4cm}{%
    \includegraphics[height=4cm]{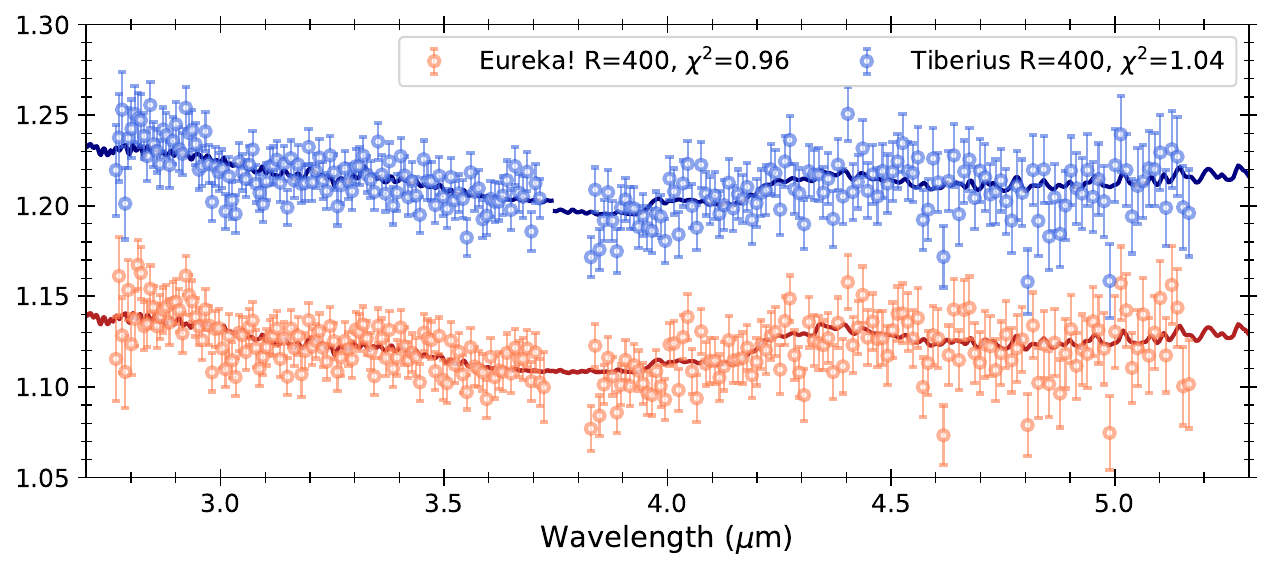}%
    }}
    \caption{Corner plot showing the posterior probability distributions from the \prt free chemistry retrievals on the $R=400$ transmission spectra from \tiberius (blue) and \eureka (orange, from which the numbers are derived). Abundances are given in units of $\log$mass fractions. The top right panel displays the best-fit models from each retrieval, offset by 500 ppm, with the reduced $\chi^2$ of the fit indicated in the legend.}
    \label{fig:prt_free_cornerplot_R100}
\end{figure*}

\begin{figure*}
    \centering
    \settototalheight{\dimen0}{\includegraphics[width=\textwidth]{figures/BeAR_retrieval_comparison_R400.pdf}}
    \includegraphics[width=\textwidth]{figures/BeAR_retrieval_comparison_R400.pdf}%
    \llap{\raisebox{\dimen0-6cm}{%
    \includegraphics[height=6cm]{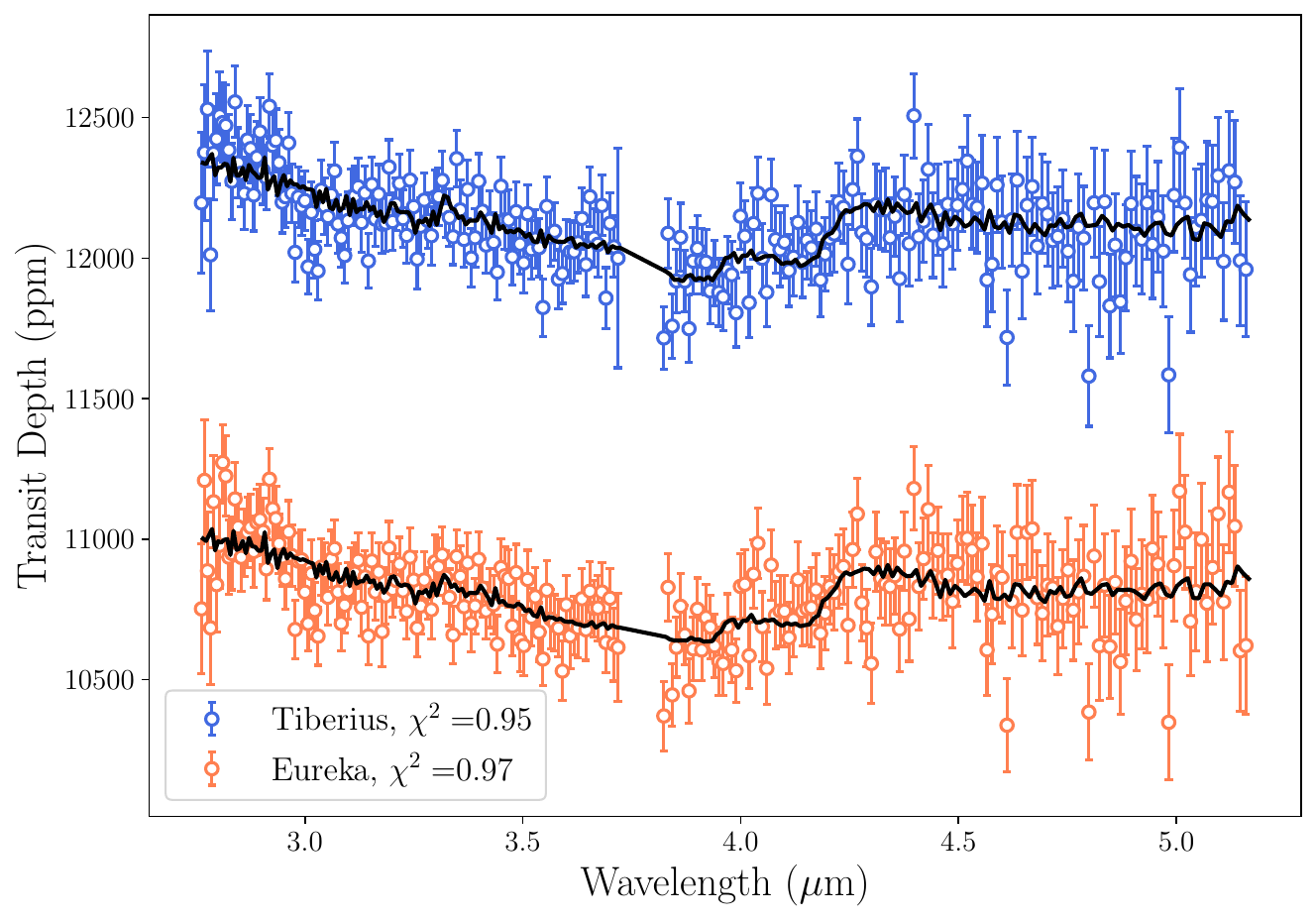}%
    }}
    \caption{Cornerplot showing the posteriors from the BeAR free-chemistry retrievals on the \texttt{Tiberius} (blue) and \texttt{Eureka!} (orange) reductions of HAT-P-30 b at $R=400$. Abundances are given in units of $\log$vertical mixing ratios. The top right insert shows the best-fit models for the \texttt{Tiberius} (blue) and \texttt{Eureka!} (orange) reductions. The \texttt{Eureka!} spectrum is offset by 1300 ppm for visualisation purposes. The legend in the bottom left indicates the reduced $\chi^2$ values for each of the fits.}
    \label{fig:BeAR_cornerplot_R400}
\end{figure*}



\bsp	
\label{lastpage}
\end{document}